%% file: p.tex
\documentclass{jfm}

\usepackage{graphicx}
\usepackage{natbib}
\usepackage{amsmath}
\usepackage{bm}

\newcommand\Rey{\mbox{\textit{Re}}} 
\newcommand{\aver}[1]{\left\langle {#1} \right\rangle}

\title[DNS of turbulent channel flow over porous walls]
{Direct numerical simulation \\ of turbulent channel flow \\ over porous walls}

\author[M.E. Rosti, L. Cortelezzi \& M. Quadrio ]%
{Marco E. Rosti$^{1,2}$,\ns
Luca Cortelezzi$^3$\break
and Maurizio Quadrio$^1$%
  \thanks{Email address for correspondence: maurizio.quadrio@polimi.it}
}

\affiliation{$^1$Department of Aerospace Science and Technology,
Politecnico di Milano, Campus Bovisa, 20136 Milano, Italy
\\[\affilskip]
$^2$Department of Mechanical and Aeronautical Engineering, City University London, \\ EC1V 0HB London, UK
\\[\affilskip]
$^3$Department of Mechanical Engineering, McGill University, Montreal, Quebec H3A 2K6, Canada
}

\begin{document}

\maketitle

\begin{abstract}
We perform direct numerical simulations (DNS) of a turbulent channel flow over porous walls. In the fluid region the flow is governed by the incompressible Navier--Stokes (NS) equations, while in the porous layers the Volume-Averaged Navier--Stokes (VANS) equations are used, which are obtained by volume-averaging the microscopic flow field over a small volume that is larger than the typical dimensions of the pores. In this way the porous medium has a continuum description, and can be specified without the need of a detailed knowledge of the pore microstructure by indipendently assigning permeability and porosity. At the interface between the porous material and the fluid region, momentum-transfer conditions are applied, in which an available coefficient related to the unknown structure of the interface can be used as an error estimate. To set up the numerical problem, the velocity-vorticity formulation of the coupled NS and VANS equations is derived and implemented in a pseudo-spectral DNS solver. Most of the simulations are carried out at $Re_\tau=180$ and consider low-permeability materials; a parameter study is used to describe the role played by permeability, porosity, thickness of the porous material, and the coefficient of the momentum-transfer interface conditions. Among them permeability, even when very small, is shown to play a major role in determining the response of the channel flow to the permeable wall. Turbulence statistics and instantaneous flow fields, in comparative form to the flow over a smooth impermeable wall, are used to understand the main changes introduced by the porous material. A simulations at higher Reynolds number is used to illustrate the main scaling quantities. 
\end{abstract}


\section{Introduction}

Two are the main motivations for this study: to further improve 
the understanding of the effects of porous materials 
on engineering and natural flows, and to gain insight on
the design of novel porous materials. There are, in fact, many 
engineering applications involving fluid flows over or 
through porous materials that will benefit from a better
understanding of the role played by these materials, such as extraction 
of oil from ground reservoirs, management of water ground basins and 
filtration of pollutants through aquifers, transpiration 
cooling, in which the porosity is used to enhance the heat 
exchange capability of the material, filtration processes 
used to separate solid particles from fluids. Examples in 
nature are flows through sedimentary rocks, such as sandstones, 
conglomerates and shales, and water flows over seabeds 
and riverbeds.  Porous media also play a crucial role 
in many biological processes involving fluid and mass transfer 
at the walls of many organic tissues, such as blood vessels, 
lungs and kidneys. The ability to design porous materials
with specific properties (e.g., porosity, permeability, 
momentum transfer at the interface, etc,) could lead
to novel developments in several fields spanning from
aerodynamics to biology, from chemistry to medicine.

The main properties characterizing a porous material are
porosity and permeability. These are obviously average properties
measured on samples of porous materials large with respect to
the characteristic pore size. Porosity (or void fraction), 
$\varepsilon$, is a dimensionless measure of the void spaces 
in a material. It is expressed as a fraction of the volume of 
voids over the total volume and its value varies between 0 and 1. 
Permeability, $K^*$, with dimensions of a length squared, is a measure 
of the ease with which a fluid flows through a porous medium 
(throughout this article, an asterisk denotes a dimensional quantity). 
$K^*=0$ if a medium is impermeable, i.e., if no fluid can flow 
through it, while it becomes infinite if a medium offers no 
resistance to a fluid flow. 
Typical values of porosity for porous materials made of packed 
particles are (see Macdonald et. al., 1979) $0.366 < \varepsilon < 0.64$ 
for spherical glass beads packed in a ``uniformly random'' manner, 
$0.367 < \varepsilon < 0.515$  for spherical marble mixtures, sand 
and gravel mixtures, and ground Blue Metal mixtures, $0.123 < \varepsilon < 0.378$ 
for a variety of consolidated media, $0.32 < \varepsilon < 0.59$
for a wide variety of cylindrical packings, $0.682 < \varepsilon < 0.919$ 
for cylindrical fibers. \cite{beavers-joseph-1967} performed experiments
with three types of foametals $0.78 < \varepsilon < 0.79$ and two types of
aloxites $0.52 < \varepsilon < 0.58$. 

The first empirical law governing Stokes flow through porous media was 
derived by Darcy \citep[see][]{lage-1998} in 1856. More than a century
later, \cite{beavers-joseph-1967} presented the first interface
(jump) condition coupling a porous flow governed by Darcy's
law with an adjacent fully developed laminar channel flow. 
This condition was further developed, with different degrees of 
success, by \cite{neale-nader-1974}, \cite{vafai-thiyagaraja-1987}, 
\cite{vafai-kim-1990} and \cite{hahn-je-choi-2002}, among others.
General porous flow equations, the so-called volume-averaged 
Navier--Stokes equations (VANS), were derived by \cite{whitaker-1996} 
by volume-averaging the Navier--Stokes (NS) equations. The VANS
equations show some analogy with LES equations. 
\cite{ochoatapia-whitaker-1995a} introduced a momentum transfer 
condition, involving a shear stress jump at the porous/fluid interface,
necessary to couple the VANS equations (governing the flow
within the porous material) to the NS equations (governing
the free flowing fluid). Over the years this condition 
was further developed by \cite{alazmi-vafai-2001}, 
\cite{goyeau-etal-2003}, \cite{chandesris-jamet-2006}, 
\cite{valdesparada-goyeau-ochoatapia-2007},
\cite{chandesris-jamet-2007, chandesris-jamet-2009} 
and \cite{valdesparada-etal-2013}. Recently, 
\cite{minale-2014a, minale-2014b} rederived the momentum
transfer condition from physical principles using the
volume-averaging approach introduced by \cite{ochoatapia-whitaker-1995a}.
Minale computed the stress transferred from the free flowing
fluid to both the fluid within the porous medium and to the porous 
matrix, and then imposed a zero stress jump at the interface 
effectively preserving the total stress at it. Although physically sound,
the boundary condition obtained by Minale in practice produces
results similar to those obtained when using the condition derived
by \cite{ochoatapia-whitaker-1995a} and \cite{valdesparada-etal-2013}.
 
The main effects of porous materials on adjacent fluid
flows are the destabilization of laminar flows and the enhancement 
of the Reynolds-shear stresses, with a consequent increase in 
skin-friction drag in turbulent flows. The first results showing
the destabilizing effects of wall permeability were obtained
experimentally by \cite{beavers-sparrow-magnuson-1970}. 
\cite{sparrow-etal-1973} experimentally determined a few critical 
Reynolds numbers in a channel with one porous wall, and performed
a two-dimensional linear stability analysis using Darcy's law 
with the interface condition introduced by \cite{beavers-joseph-1967}. 
Recently, \cite{tilton-cortelezzi-2006, tilton-cortelezzi-2008} 
performed a three-dimensional linear stability analysis of a 
laminar flow in a channel with one or two homogeneous, isotropic, 
porous walls by modelling the flow in the porous walls using
the VANS equations with the interface conditions of 
\cite{ochoatapia-whitaker-1995a}. They reported that wall permeability 
can drastically decrease the stability of fully developed
laminar channel flows.

Scarce is also the literature regarding the effects of permeability
on turbulent flows. Early experiments by \cite{lovera-kennedy-1969} 
on alluvial streams over flat sand beds showed increasing skin friction 
coefficient with increasing Reynolds number. \cite{ruff-gelhar-1972}
investigated turbulent flows in a pipe lined with highly porous
foam. They recognized the importance of the exchange of momentum
across the porous/fluid interface and reported higher skin
friction for porous walls than for solid walls.
Experimentally, \cite{zagni-smith-1976} reported higher skin friction 
for open-channel flows over permeable beds of packed spheres 
than over impermeable walls, and attributed the increase to the 
additional energy dissipation caused by the exchange of momentum 
across the fluid-porous interface. \cite{kong-schetz-1982} measured 
an increase in skin friction in boundary layers over porous walls 
made of sintered metals, bonded screen sheets and perforated titanium 
sheets. In wind-tunnel experiments of boundary layer flows over
a bed of grains, \cite{zippe-graf-1983} reported a rise in skin
friction with respect to impermeable rough walls. 
\cite{shimizu-tsujimoto-nakagawa-1990} investigated the flow induced 
in the transition layer, just below the porous/fluid interface, and 
concluded that permeability enhances momentum flux and Reynolds 
stresses near the interface. 
\cite{hahn-je-choi-2002} performed a DNS of a turbulent
plane channel flow by modelling the presence of 
permeable walls with boundary conditions similar
to those introduced by \cite{beavers-joseph-1967},
i.e. finite tangential (slip) velocity and
zero normal velocity. They reported ``significant 
skin-friction reductions'' at the permeable walls.
This result conflicts with the experimental
evidence and is presumably due to the enforcement
of a zero normal velocity at the porous/fluid
interface, a boundary condition that inhibits the 
correct exchange of momentum across the interface 
\citep[e.g.][]{ruff-gelhar-1972}. It is also interesting to 
note that \cite{perot-moin-1995} performed a DNS of 
a turbulent plane channel flow with infinite
permeability and no-slip boundary conditions in 
order to remove the wall blocking mechanism and 
study the contributon of splashes on wall-turbulence.

In order to describe accurately the mass and momentum transfer between 
a fluid-saturated porous layer and a turbulent flow, one could, in principle, 
perform a DNS by solving the NS equations over the entire domain and
enforcing the no-slip and no-penetration conditions on the highly 
convoluted surface representing the boundary of the porous material. 
In practice, however, this approach is hard to implement because the 
boundary of a porous material has, in general, an extremely complex geometry 
that, often, is not known in full details. Therefore, this approach has
been used only in cases in which the porous medium is highly idealized 
and has a simple geometry. \cite{zhang-prosperetti-2009} and 
\cite{liu-prosperetti-2011} modelled the porous walls of a channel
using disjointed cylinders and spheres in a simple cubic arrangement
and studied the lift, drag and torque generated on the cylindrical 
and spherical particles by a fully developed laminar flow. 
\cite{breugem-boersma-2005} and \cite{chandesris-etal-2013} used a 
three-dimensional Cartesian grid of disjointed cubes. The formers 
investigated the effect of porous walls on the statistics of a fully 
turbulent channel flow, the latters the effects on turbulent heat transfer. 

\cite{breugem-boersma-2005} used two approaches to model the flow
through a porous material and characterize the effects of permeability
on fully turbulent channel flows with a porous wall. In the first
approach, they performed a DNS of a turbulent channel flow over a 
three-dimensional Cartesian grid of disjointed cubes mimicking a 
permeable wall with $\varepsilon = 0.875$, and solved the NS
equations over the entire domain. In the second approach, they 
performed a DNS of the channel flow by solving the NS equations 
in the fluid region and the VANS equations in the porous layers. 
In particular they used the model developed by \cite{irmay-1965} 
to express permeability in terms of porosity, the Burke--Plummer 
equation to express the Forchheimer tensor in terms of porosity 
and a fifth-order polynomial function to model the variation of 
porosity in the thin porous region adjacent to the porous/fluid 
interface, known as transition layer. This comparison is particularly 
meaningful because the VANS equations were obtained by volume 
averaging the NS equations over the Cartesian grid of disjointed 
cubes used in the DNS simulations. The main conclusion of the study 
by \cite{breugem-boersma-2005} was that ``the approach based on 
the VANS equations is capable of an accurate simulation of the 
turbulent flow over and through a permeable wall, even quantitatively.''

One year later, \cite{breugem-boersma-uittenbogaard-2006} leveraged the 
work by \cite{breugem-boersma-2005} to study the influence of a highly 
permeable porous wall, made by a packed bed of particles, on 
turbulent channel flows. To isolate the effect of wall permeability 
from that of wall roughness, they considered highly porous packed beds 
made of particles of small mean diameter, two apparently 
conflicting requirements leading to beds of disjointed particles.
Because of these assumptions, the fluid quickly flows through the
porous medium and the drag force cannot be neglected. The authors modelled 
it by means of the Ergun equation \citep{bird-stewart-lightfoot-2002} 
in combination with their variable-porosity model, thus ensuring
the continuity of both porosity and drag 
force over the interface region. In this approach, the permeability and 
the Forchheimer tensors (responsible for the drag force) can be written in 
terms of porosity and mean particle diameter. The authors presented
the results of four simulations for values of porosity equal to 0, 0.6,
0.8 and 0.95 and classified the permeable walls as highly permeable (near 
which viscous effects are of minor importance, $Re_K \equiv \sqrt{K^*}u^*_{\tau}/\nu^*=9.35$, 
where $u^*_\tau$ is the friction velocity and $\nu^*$ is the kinematic viscosity) 
in the case $\varepsilon=0.95$, partially permeable ($Re_K=1.06$) in the case 
$\varepsilon=0.80$ and effectively impermeable (near which viscous diffusion 
of mean kinetic energy is counterbalanced by viscous dissipation, $Re_K=0.31$) 
in the case $\varepsilon=0.60$. Their results showed that the structure and
dynamics of turbulence above a highly permeable wall, where there are no
low- and high-speed streaks and quasi-streamwise vortices, are very 
different compared to those of a turbulent flow over an effectively 
impermeable wall. Near a highly permeable wall, turbulence is dominated 
by relatively large vortical structures that favor the exchange of 
momentum between the top layer of the porous medium, the transition layer, 
and the channel, and induce a strong increase in the Reynolds-shear 
stresses and, consequently, a strong increase in the skin friction 
compared to an impermeable wall. 

\cite{suga-etal-2010} studied experimentally the effects of 
wall permeability on a turbulent flow in a channel with a 
porous wall. They considered three types of foamed ceramics 
materials whose porosity was almost the same ($\epsilon \approx 0.8$) 
and their permeabilities varied by about a factor four, 
$K^* = 0.020, 0.033$ and $0.087 mm^2$. They observed
that the slip velocity of a flow over a permeable wall 
increases drastically in the range of Reynolds numbers 
where the flow transitions from laminar to turbulent, 
and that transition to turbulence appears at progressively 
lower Reynolds numbers as permeability increases,
consistently with the results of linear stability
analysis \citep[]{tilton-cortelezzi-2006, tilton-cortelezzi-2008}. 
The turbulence statistics of the velocity fluctuations 
showed that the wall-normal component increases as the wall
permeability and/or the Reynolds number increases. The 
authors concluded that permeability weakens the blocking 
effects of a porous wall on the vortex motion, therefore 
contributing to an increase of the wall-shear stress, a 
conclusion consistent with similar observations put forward, 
among others, by \cite{hahn-je-choi-2002}. 
\cite{suga-etal-2010} also performed a numerical 
simulation of the same turbulent flow using the analytic 
wall function at the porous/fluid interface and found their 
results in good agreement with their experimental results 
and the results by \cite{breugem-boersma-uittenbogaard-2006}.

\begin{figure}
\centering
\includegraphics[width=0.49\textwidth]{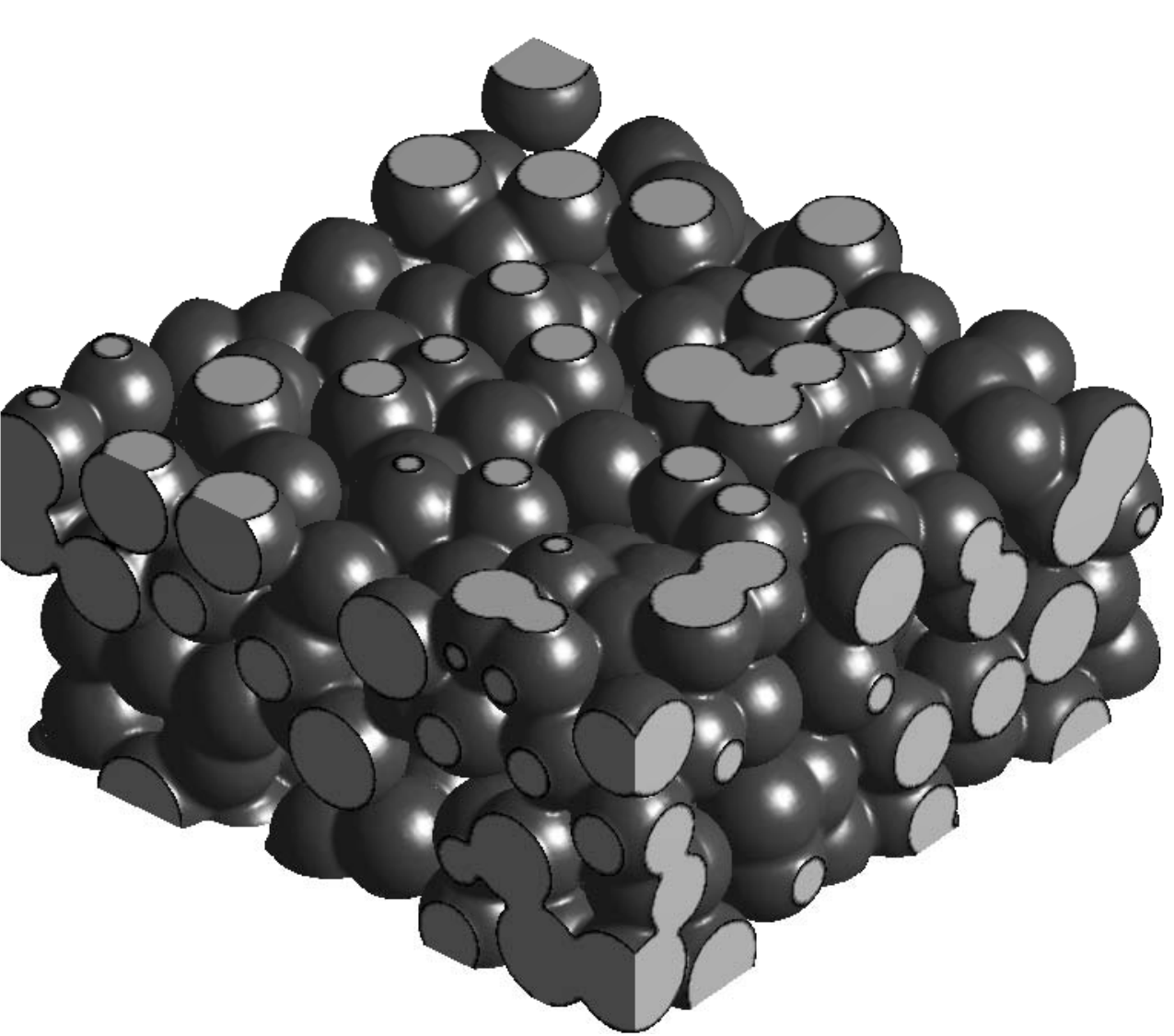}
\includegraphics[width=0.49\textwidth]{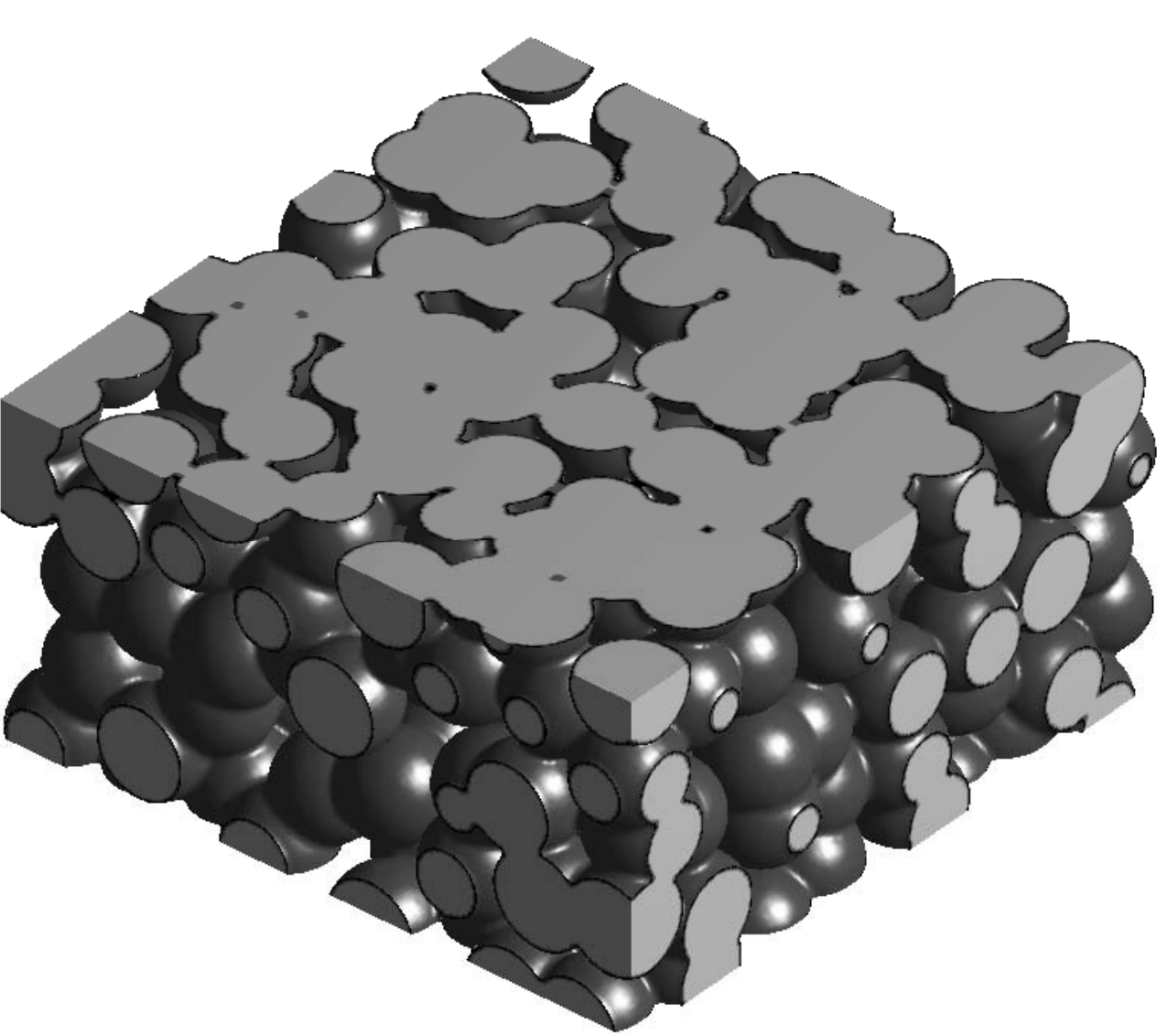}
\caption{Effect of surface machining on the same numerically generated
porous sample: between-particles cut (left) and through-particles 
cut (right) yield highly different interfaces.}
\label{fig:porous-sample}
\end{figure}

The flow at the interface and in the transition region within 
the porous layer depends, for a given porous material, on the 
surface machining of the interface. Figure \ref{fig:porous-sample} 
shows that even using the same (numerically generated) porous 
sample and the same surface machining technique, a totally different 
interfacial geometry is obtained by simply cutting 
the porous material half of a particle diameter deeper. Obviously, 
the flow just above the interface and within the transition region 
for the two samples shown in figure \ref{fig:porous-sample} will be 
noticeably different. Therefore, in general, it is nearly impossible 
to introduce a variable-porosity model that is capable to fit all 
possible interface geometries, even if the porous material used is 
exactly the same. At the contrary, we believe that making a choice 
for a variable-porosity model reduces the generality of the model 
to a particular porous material with a particular interface. 

The complexity of modeling the flow at the porous/fluid interface and within 
the transition region becomes less severe for porous materials of small 
permeability and small mean particle size. In particular, for sufficiently 
low permeabilities, the fluid velocity at the interface is small; consequently 
the convective effects and the drag force experienced by the 
fluid become negligible, because the dense channel-like structures of the 
porous matrix impede motion between layers of fluid. In this case, 
the transition region (of the order of a few pore diameters thick) and
roughness (of the order of one pore diameter high) can be assumed to 
have zero thickness, and porosity and permeability to have a constant 
value up to the interface that has a clearly defined position 
\citep{ochoatapia-whitaker-1995a, ochoatapia-whitaker-1998, valdesparada-goyeau-ochoatapia-2007, valdesparada-etal-2013}.
As a consequence, porosity and permeability are effectively decoupled
because they are assumed to be constant up to the porous/fluid interface. 
The zero-thickness assumption produces a jump in the shear stresses at
the interface, because at the interface the sphere over which the volume averaging 
is performed lies part in the porous region and part in the free fluid
region, see figure \ref{fig:pore}. This jump in stress produces an 
additional boundary condition at the interface, where the magnitude of the jump 
is proportional to a dimensionless parameter $\tau$, the momentum transfer 
coefficient, which is of order one and can be both positive or negative
\citep{ochoatapia-whitaker-1995a, valdesparada-goyeau-ochoatapia-2007, valdesparada-etal-2013}.
The parameter $\tau$ is decoupled from porosity and permeability.
In a recent interpretation by \cite{minale-2014a, minale-2014b}, 
a negative $\tau$ quantifies the amount of stress transfered from the 
free fluid to the porous matrix, while a positive $\tau$ quantifies the 
amount of stress transfered from the porous matrix to the free fluid;
when $\tau=0$ the stress carried by the free fluid is fully transfered 
to the fluid saturating the porous matrix. In this study we use the 
stress jump condition introduced by \cite{ochoatapia-whitaker-1995a}, 
to be consistent with the work of 
\cite{tilton-cortelezzi-2006, tilton-cortelezzi-2008} that 
inspired the present study. Although the relationship between linear stability 
of channel flows and fully turbulent channel flows is not immediate, we will show that the assessment 
of the effects of the porous parameters made by \cite{tilton-cortelezzi-2008}
in their linear stability analysis of pressure driven channel flows 
with porous walls is remarkably accurate for fully turbulent channel 
flows at low Reynolds numbers.  

A porous wall of small permeability $K^*$ is often thought to
behave as an effectively impermeable wall. It is true that, in the 
limit $K^* \rightarrow 0$, a porous wall becomes impermeable and 
behaves as a solid wall. On the other hand, as it has been shown by 
\cite{tilton-cortelezzi-2006, tilton-cortelezzi-2008}, very small 
amounts of permeability have major effects on the stability of 
fully developed laminar flows in channels with one or both porous walls. 
In particular, the critical Reynolds number is most sensitive to small 
permeabilities where it experiences its sharpest drop. Hence, in this 
paper we focus on low-permeability porous materials. There are two main 
reasons for this choice. First, porous materials of small permeabilities 
are common in nature and in industrial applications and, therefore, it 
is of interest to characterize their effects on turbulent flows. Second, 
since porosity, permeability and momentum transfer coefficient are decoupled 
in the model used in the present study, our results could provide insight 
for the design of novel porous materials which target specific engineering 
applications. 

In this work we describe the development and implementation of a computer 
code for the DNS of turbulent channel flows bounded by porous walls. 
The numerical method, based upon a pseudo-spectral strategy, employs 
Fourier discretization in the homogeneous directions and compact finite 
differences in the wall-normal direction and it is used both in the
fluid region and within the porous walls. Within the former the full
NS equations are solved while within the latters the 
VANS equations are solved neglecting the inertial terms. After validating 
the numerical method against linear stability theory, 
the fully turbulent case is addressed. A parametric study is performed by 
varying all the parameters defining the porous medium (permeability, porosity
and thickness) in order to assess their effect on turbulence statistics. 
Furthermore, the effects of the transition layer and interface machining 
are accounted for by varying the parameter controlling the shear stress 
jump at the interface in order to mimic the unavoidable uncertainty related 
to the real interface. Although not the main focus of the present 
work, the effect of varying the channel flow bulk Reynolds number is also shortly 
discussed. 

The paper is organized as follows: in \S\ref{sec:definition}, we discuss 
the channel geometry, governing equations and interface conditions.
We then present the numerical method used to perform the DNSs. 
The code is then validated against results predicted by linear 
stability theory. In \S\ref{sec:results} we describe in statistical 
terms the effects of the porous material on a fully developed 
turbulent flow in a baseline case where the porous medium has 
small permeability, and then discuss
the effects of varying the parameters of the porous material.
Finally, we present a concluding summary in \S\ref{sec:conclusions}.

\section{Problem definition and mathematical formulation}
\label{sec:definition}

\begin{figure}
\centering
\includegraphics[width=0.7\textwidth]{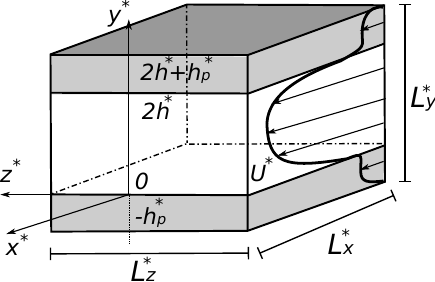}
\caption{Sketch of the channel geometry. The interfaces between the channel region and the porous walls are located at $y^*=0$ and $y^*=2 h^*$; the porous walls have thickness $h_p^*$, so that the two impermeable walls bounding the whole domain are at $y^*=-h_p^*$ and $y^*=2 h^* + h_p^*$.}
\label{fig:channel_with_porous_walls}
\end{figure}
We consider the fully developed flow of an incompressible viscous fluid in 
a channel delimited by two identical, flat, rigid, homogeneous and isotropic, 
infinite porous layers sealed by impermeable walls, see figure \ref{fig:channel_with_porous_walls}. The upper and lower interfaces between the 
fluid and the porous material are located at $y^* = 0$ and $y^*=2h^*$ while the upper and lower impermeable walls are located at $y^* = -h_p^*$ and $y^*= 2h^* + h_p^*$, where $h^*$ is the half-height of the fluid region and $h_p^*$ is the height of the identical porous layers, respectively. 

The Reynolds number of the flow is defined as
\begin{equation}
\label{eq:Re-definition}
\Rey = \frac{U_b^* h^*}{\nu^*},
\end{equation}
where $h^*$ is the characteristic length scale  
and the characteristic velocity is the bulk velocity $U_b^*$, defined 
as the average value of the mean velocity $\overline{u}^*(y^*)$ computed across the whole domain, i.e.
\begin{equation}
\label{eq:Ub}
U_b^* = \dfrac{1}{2 h^*} \int_{-h_p^*}^{2h^* + h_p^*} \overline{u}^* dy^*.
\end{equation}
Note that the integral spans over all three regions (fluid region and porous layers) 
while the weighting, $1/2 h^*$, accounts for the fluid region only. This choice,
dictated by the fact that the mass flux in the porous layers is negligible
in comparison to the total mass flux (especially at low permeabilities), facilitates 
the comparison of the flow in a channel with porous walls to that in a channel bounded 
by impermeable walls. In the next section we present the mathematical formulation 
of the problem made dimensionless by using $h^*$, $U_b^*$ and $h^*/U_b^*$ as characteristic 
length, velocity and time, respectively.

\subsection{Velocity-pressure formulation}
The flow of an incompressible viscous fluid through the domain sketched in figure
\ref{fig:channel_with_porous_walls} is governed by the non-dimensional incompressible 
Navier--Stokes equations
\begin{subequations}  
\label{eq:NS_nondim} 
\begin{align}
         &\dfrac{\partial \bm{u}}{\partial t} + (\bm{u} \cdot \bm{\nabla}) \bm{u} = - \bm{\nabla} p + \dfrac{1}{\Rey} \nabla^2 \bm{u} , \label{eq:NSm_nondim}\\
        &\bm{\nabla} \cdot \bm{u} = 0 . \label{eq:NSc_nondim}
\end{align}
\end{subequations}

\begin{figure}
\centering
\includegraphics[angle=0,width=0.7\textwidth]{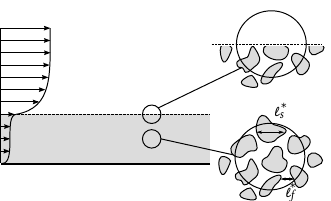}
\caption{Sketch of a porous medium, with $\ell^*_f$ and $\ell^*_s$ the characteristic lengths of the pore and particle diameters of the pore-like structures.}
\label{fig:pore}
\end{figure}
However, it is nearly impossible to apply this model to a flow confined 
by porous layers because porous materials, in general, have very complex 
geometries and are characterized by a wide range of length scales. As 
exemplified in figure \ref{fig:pore}, such scales are bounded by
the smallest scales $\ell^*_f$ (of the fluid phase) and $\ell^*_s$ 
(of the solid phase),  related to the characteristic pore and particle 
diameters of the pore-like structures, and the largest scale $L^*_p$ which 
is the characteristic thickness of the porous layer. To overcome these 
difficulties, \cite{whitaker-1969, whitaker-1986, whitaker-1996} proposed 
to model only the large-scale behaviour of a flow in a porous medium by 
averaging the NS equations over a small sphere, of volume $V^*$ and 
radius $r^*$. This averaging procedure, which is similar to the LES 
decomposition, results in the so-called volume-averaged Navier--Stokes 
(VANS) equations, and relies on the assumption that the length scales 
of the problem are well separated, 
i.e., $\ell^*_s \sim \ell^*_f \ll r^* \ll L^*_p$. 
Under this assumption, the volume-averaged quantities are smooth and 
free of small-scale fluctuations. In other words, the fluid-saturated 
porous medium is described as a continuum, so that fluid quantities, 
like velocity or pressure, are defined at every point in space, 
regardless of their position in the fluid or solid phase.

The first step in the derivation of the VANS equations consists in 
choosing the appropriate averaging method. The \textit{superficial 
volume average} $\aver{\phi^*}^s$, of a generic scalar quantity 
$\phi^*$, is defined \citep{quintard-whitaker-1994b, whitaker-1996} as
\begin{equation} 
\label{eq:average_sup}
\aver{\phi^*}^s = \frac{1}{V^*} \int_{V^*_f} \phi^* dV^*_f,
\end{equation}
where $V^*_f<V^*$ is the volume of fluid contained within the averaging 
volume $V^*$, while the \textit{intrinsic volume average} is defined as
\begin{equation} \label{eq:average_int}
\aver{\phi^*}^f = \frac{1}{V^*_f} \int_{V^*_f} \phi^* dV^*_f.
\end{equation}
These two averages are related as follows
\begin{equation} \label{eq:epsilon}
	\aver{\phi^*}^s = \frac{V^*_f}{V^*} \aver{\phi^*}^f = \varepsilon \aver{\phi^*}^f,
\end{equation}
where $\varepsilon = V^*_f / V^*$ is the porosity, or the 
volume-fraction of fluid contained in $V^*$, which is generally 
a function of the position in a heterogeneous porous medium.

The second step in the derivation of the VANS equations consists 
in defining a relationship between the volume average of a derivative 
of a scalar quantity and the derivative of the volume average of the 
same quantity, both for time and spatial derivatives. The {\em general 
transport theorem} \citep{whitaker-1969}, a generalized formulation 
of the Reynolds transport theorem, provides us with a relationship 
for the volume average of a time derivative, while the {\em spatial 
averaging theorem} \citep{slattery-1967,whitaker-1969} provides us
with a relationship for the volume average of the spatial derivatives.

Finally, assuming the porous material to be homogeneous and isotropic, 
i.e. porosity and permeability remain constant throughout the porous walls, 
and permeability to be sufficiently small to neglect inertial effects,
the dimensionless VANS equations become linear because the drag 
contribution related to the Forchheimer tensor becomes negligible 
compared to the Darcy drag. In their linear-stability study, 
\cite{tilton-cortelezzi-2008} verified that this linearity assumption 
amounts to considering the longitudinal velocity at the porous/fluid interface 
to be lower than about 5\% of the centerline velocity of the channel. Since
the focus of the present work are the effects of porous materials of
small permeabilities on turbulent channel flows, we assume that the 
fluid motion within the porous walls is governed by the following 
dimensionless VANS equations 
\begin{subequations} 
\label{eq:VANS_nondim}
\begin{align}
	 &\dfrac{\partial \aver{\bm{u}}^s}{\partial t} = - \varepsilon \bm{\nabla} \aver{p}^f + \dfrac{1}{\Rey} \nabla^2 \aver{\bm{u}}^s - \dfrac{\varepsilon}{\sigma^2 \Rey} \aver{\bm{u}}^s,
\label{eq:VANSm_nondim} \\
&\bm{\nabla} \cdot \aver{\bm{u}}^s = 0, \label{eq:VANSc_nondim}
\end{align}
\end{subequations}
where $\sigma = \dfrac{\sqrt{K^*}}{h^*}$ is the dimensionless permeability. 
Note that in the momentum equation \eqref{eq:VANSm_nondim} both averages 
are used. The preferred representation of the velocity is the superficial 
volume-average velocity, $\aver{\bm{u}}^s$, because it is always solenoidal, 
while the preferred representation of the pressure is the intrinsic 
volume-averge pressure, $\aver{p}^f$, because it is the pressure measured 
by a probe in an experimental apparatus. 

To simulate accurately a turbulent flow over a porous wall, it is of crucial 
importance to couple correctly the flow in the fluid region, governed by the 
NS equations \eqref{eq:NS_nondim}, with the flow in the porous 
layers, modelled by the VANS equations \eqref{eq:VANS_nondim}. In a real 
system, this coupling takes place in a thin layer (a few pore diameters thick) 
of the porous wall, the so-called transition region, adjacent to the interface 
between fluid and porous regions. Porosity and permeability change in the 
transition region depending on the structure of the porous material and on 
how the surface of the porous layer has been machined. In general, porosity 
and permeability increase rapidly from their value $\varepsilon$ and $\sigma$ 
within the homogeneous porous region to unity and infinity, respectively, 
slightly above the interface. As a consequence of this variation in porosity
and permeability, the fluid velocity increases from the Darcy velocity in the 
homogeneous porous region to its slip value just above the interface. This 
is achieved in the transition region where mass and momentum transfer take 
place. The variations of porosity and permeability in the transition region 
are difficult to model theoretically, and also their measurement is a challenge 
to experimentalists. 

In general, as explained by \cite{minale-2014a,minale-2014b}, at the 
porous/fluid interface, the total stress carried by a fluid freely 
flowing over the interface equals the sum of the stresses 
transferred to the fluid within the porous material and that 
transferred to the porous matrix. In particular, depending on 
the geometry of the porous material and the machining of the interface 
(see figure \ref{fig:porous-sample}), as the fluid flows 
over a saturated porous material, in certain cases part of the 
momentum is transfered from the free flowing fluid to the porous
matrix, and in other cases it is the opposite. 

In the case of porous materials of small permeabilities, 
the difficulty of modeling accurately the flow near the
interface can be reduced by assuming the transition region 
to have zero thickness, and porosity and permeability to 
have a constant values, $\varepsilon$ and $\sigma$ respectively, 
up to the interface. This assumption, however, produces an error 
in the local averaged velocity $\aver{\bm{u}}^s$ and pressure 
$\aver{p}^f$ at the interface because, at the interface, the 
sphere over which the averages are compued lies 
part in the porous region and part in the free fluid, see 
figure \ref{fig:pore}. This error is corrected by means of an 
additional stress jump condition at the porous/fluid interface 
\citep{ochoatapia-whitaker-1995a}, 
that fully couples the NS equations \eqref{eq:NS_nondim} to 
the VANS equations \eqref{eq:VANS_nondim}. Velocity and pressure 
are forced to be continuous at the interface, while the shear 
stresses are in general discontinuous and the magnitude of 
this discontinuity is controlled by a dimensionless parameter $\tau$. 
Over the years, the stress jump condition proposed by
\cite{ochoatapia-whitaker-1995a}
was further developed by \cite{alazmi-vafai-2001}, 
\cite{goyeau-etal-2003}, \cite{chandesris-jamet-2006}, 
\cite{valdesparada-goyeau-ochoatapia-2007},
\cite{chandesris-jamet-2007, chandesris-jamet-2009}, 
\cite{valdesparada-etal-2013}  and \cite{minale-2014a, minale-2014b}.
Although these developments have better explained the
physical mechanisms responsible for mass and momentum 
transfer at the porous/fluid interface, the magnitude of the
corrections with respect to the stress jump condition of
\cite{ochoatapia-whitaker-1995a} are minor.
Therefore, in this study we use the latter condition to be 
consistent with the work of 
\cite{tilton-cortelezzi-2006, tilton-cortelezzi-2008} that 
inspired the present study.

The momentum transfer coefficient $\tau$, as suggested by 
\cite{minale-2014a,minale-2014b}, models the transfer of stress at the 
interface. This dimensionaless parameter is of order one and can be both 
positive or negative, depending on the type of porous material considered
and machining of the interface (see figure \ref{fig:porous-sample}). 
A negative $\tau$ quantifies the amount of stress transfered from the 
free fluid to the porous matrix, while a positive $\tau$ quantifies the 
amount of stress transfered from the porous matrix to the free fluid, 
when $\tau=0$ the stress carried by the free fluid is fully transferred 
to the fluid saturating the porous matrix. \cite{ochoatapia-whitaker-1995b} 
computed the values of $\tau$ that best fitted sets of experimental data
obtained by using as porous materials aloxite and three different
types of foametals. They showed that $\tau$ varies between -1
to +0.7 depending on the type of foametal considered and
reaches a value of +1.47 for aloxite. In order to model a wide
range of porous materials and surface machining, in this work we 
follow \cite{tilton-cortelezzi-2008} and discuss the effects of 
the momentum transfer coefficient as it varies between -1 and +1. 

Based on the hypotheses made above, the momentum-transfer conditions \citep{ochoatapia-whitaker-1995a} at $y=0$ and $y=2$ reduce to 
\begin{subequations} \label{eq:OTW_cond}
\begin{align}
u &= \langle u \rangle^s , \label{eq:OTW_cond_u}\\ 
v &= \langle v \rangle^s , \label{eq:OTW_cond_v}\\
w &= \langle w \rangle^s , \label{eq:OTW_cond_w}\\
p &= \langle p \rangle^f , \label{eq:OTW_cond_p}\\
\displaystyle 
\sigma \left( \dfrac{\partial u}{\partial y} - \dfrac{1}{\varepsilon} \dfrac{\partial \langle u \rangle^s}{\partial y} \right) &= \pm \tau u, \label{eq:OTW_cond_jump_dudy}\\
\displaystyle 
\sigma \left( \dfrac{\partial w}{\partial y} - \dfrac{1}{\varepsilon} \dfrac{\partial \langle w \rangle^s}{\partial y} \right) &= \pm \tau w, \label{eq:OTW_cond_jump_dwdy}
\end{align}
\end{subequations}
where the positive sign in conditions \eqref{eq:OTW_cond_jump_dudy} and
\eqref{eq:OTW_cond_jump_dwdy} applies to the interface located at $y=2$, 
while the negative sign applies to the interface located at $y=0$. Note 
that the $y$-derivative of the velocity components $u$ and $w$ are taken 
within the fluid region while the $y$-derivative of the volume-averaged velocity 
components $\langle u \rangle^s$ and $\langle w \rangle^s$ are taken within 
the porous layers. 

Finally, each porous layer is bounded by an impermeable wall where the 
no-slip and no-penetration conditions apply. Hence, at $y=-h_p$ and 
$y=2+h_p$, we have
\begin{subequations}  \label{eq:BCw_nondim}
\begin{align}
\aver{u}^s &=0, \label{eq:BCw_nondim_u}\\
\aver{v}^s &=0, \label{eq:BCw_nondim_v}\\
\aver{w}^s &=0. \label{eq:BCw_nondim_w}
\end{align}
\end{subequations}

\subsection{Velocity-vorticity formulation}
DNS of a turbulent plane channel flow, when two spatial directions are homogeneous, 
is computationally more efficient when the governing equations are made independent of 
pressure and reformulated, as done for example by \cite{kim-moin-moser-1987}, 
in terms of wall-normal velocity and wall-normal vorticity components. In this 
subsection, we rewrite the problem at hand in terms of the velocity fields, 
$\bm{u}$ and $\aver{\bm{u}}^s$, and the wall-normal vorticity components, $\eta$ 
and $\aver{\eta}^s$.

In the fluid region, the NS equations \eqref{eq:NS_nondim} can be 
reformulated in terms of the wall-normal velocity component $v$ and
wall-normal vorticity component $\eta$ as
\begin{equation} 
\label{eq:NS_v}
\dfrac{\partial \nabla^2 v}{\partial t} = - \dfrac{\partial^2 HU }{\partial y \partial x} - \dfrac{\partial^2 HW}{\partial y \partial z} + \dfrac{\partial^2 HV}{\partial x^2} + \dfrac{\partial^2 HV}{\partial z^2} + \dfrac{1}{\Rey} \nabla^2\nabla^2 v,
\end{equation}
\begin{equation} \label{eq:NS_eta}
\dfrac{\partial \eta}{\partial t} = \dfrac{\partial HU}{\partial z} - \dfrac{\partial HW}{\partial x} + \dfrac{1}{\Rey} \nabla^2 \eta,
\end{equation}
where $\eta$ is 
\begin{equation}
\label{eq:eta_def}
\eta = \dfrac{\partial u}{\partial z} - \dfrac{\partial w}{\partial x},
\end{equation}
and the non-linear terms are defined as
\begin{subequations}
\begin{align}
HU &= - \dfrac{\partial (u u)}{\partial x} - \dfrac{\partial (u v)}{\partial y} - \dfrac{\partial (u w)}{\partial z}, \label{eq:H_U} \\
HV &= - \dfrac{\partial (v u)}{\partial x} - \dfrac{\partial (v v)}{\partial y} - \dfrac{\partial (v w)}{\partial z},
\label{eq:H_V} \\
HW &= - \dfrac{\partial (w u)}{\partial x} - \dfrac{\partial (w v)}{\partial y} - \dfrac{\partial (w w)}{\partial z}.\label{eq:H_W}
\end{align}
\end{subequations}
Equations \eqref{eq:NS_v} and \eqref{eq:NS_eta} together with the continuity equation 
\eqref{eq:NSc_nondim} and the definition of $\eta$ \eqref{eq:eta_def} form a system of 
four equations in the four unknowns $u$, $v$, $w$ and $\eta$. Note that the pressure field
in the fluid region, if needed, can be post-computed by solving the Poisson equation 
obtained by taking the divergence of the momentum equation \eqref{eq:NSm_nondim}, i.e.
\begin{equation} \label{eq:Poisson_fluid}
\nabla^2 p = \dfrac{\partial HU}{\partial x} + \dfrac{\partial HV}{\partial y} + \dfrac{\partial HW}{\partial z}. 
\end{equation}

The VANS equations \eqref{eq:VANS_nondim} can be reformulated following similar steps. 
To derive a pressure-free equation for the volume-averaged wall-normal velocity 
component $\aver{v}^s$, we take the Laplacian of the $y$-component of the VANS 
momentum equation \eqref{eq:VANSm_nondim} and obtain
\begin{equation} \label{eq:VANS_v}
\dfrac{\partial \nabla^2 \aver{v}^s}{\partial t} = \dfrac{1}{\Rey} \nabla^2 \nabla^2 \aver{v}^s - \dfrac{\varepsilon}{\sigma^2 \Rey} \nabla^2 \aver{v}^s.
\end{equation}
The evolution equation for the volume-averaged wall-normal component of the vorticity, 
$\aver{\eta}^s$, can be derived by subtracting the $x$-derivative of the $z$-component of 
the VANS momentum equation \eqref{eq:VANSm_nondim} from the $z$-derivative of the 
$x$-component of the same equation. We obtain
\begin{equation} \label{eq:VANS_eta}
\dfrac{\partial \aver{\eta}^s}{\partial t} = \dfrac{1}{\Rey} \nabla^2 \aver{\eta}^s - \dfrac{\varepsilon}{\sigma^2 \Rey} \aver{\eta}^s,
\end{equation}
where the volume-averaged wall-normal vorticity component is
\begin{equation} \label{eq:eta_p}
\aver{\eta}^s = \dfrac{\partial \aver{u}^s}{\partial z} - \dfrac{\partial \aver{w}^s}{\partial x}.
\end{equation}
Note that equations \eqref{eq:VANS_v} and \eqref{eq:VANS_eta} are linear because we 
have assumed negligible inertial effects in the porous layers, but contain the terms 
$\varepsilon  \nabla^2 \aver{v}^s / \left( \sigma^2 \Rey \right)$ and  
$\varepsilon \aver{\eta}^s / \left(\sigma^2 \Rey \right)$ due to the Darcy drag. 

Owned to the linearity of the above equations, the volume-averaged streamwise, 
$\aver{u}^s$, and spanwise, $\aver{w}^s$, components of the velocity field are 
decoupled from $\aver{v}^s$  and $\aver{\eta}^s$, and  
can be post-computed using the VANS continuity equation \eqref{eq:VANSc_nondim} 
and the definition of volume-averaged wall-normal vorticity component \eqref{eq:eta_p}. 
A Laplace equation for the volume-averaged pressure, $\aver{p}^f$, can be 
derived by taking the divergence of the VANS momentum equation \eqref{eq:VANSm_nondim}.
Therefore, in each porous layer, the volume-averaged pressure can be post-computed 
by solving
\begin{equation} \label{eq:Poisson_porous}
\dfrac{1}{\varepsilon}\nabla^2 \aver{p}^f = 0,
\end{equation}
matching the pressure at the interface and satisfying the boundary conditions. 

The interface \eqref{eq:OTW_cond} and boundary \eqref{eq:BCw_nondim} conditions 
can also be rewritten in the new formulation with simple manipulations. 
The condition \eqref{eq:OTW_cond_v} at $y=0$ and $y=2$ remains obviously 
unchanged in the present formulation, i.e.
\begin{equation} \label{eq:BC_v_0}
v = \aver{v}^s.
\end{equation}
Since the conditions \eqref{eq:OTW_cond} are true everywhere at the interfaces
and all flow quantities are continuous with all derivatives in the  $x$- and 
$z$-directions, then the $x$- and $z$-derivatives of these conditions must also 
be true everywhere at the interfaces. Therefore, adding the $z$-derivative of
\eqref{eq:OTW_cond_w} to the $x$-derivative of \eqref{eq:OTW_cond_u} and using 
the continuity equations \eqref{eq:NSc_nondim} and \eqref{eq:VANSc_nondim}, we 
obtain that the normal derivatives of the normal velocity components in the fluid 
and porous regions are continuous at $y=0$ and $y=2$, i.e.
\begin{equation} \label{eq:BC_v_1}
\dfrac{\partial v}{\partial y} = \dfrac{\partial \aver{v}^s}{\partial y}.
\end{equation}
Subtracting the $x$-derivative of \eqref{eq:OTW_cond_w} from the $z$-derivative 
of \eqref{eq:OTW_cond_u}, we obtain that the normal component of the vorticity 
at $y=0$ and $y=2$ is continuous, i.e.
\begin{equation} \label{eq:BC_eta_0}
\eta = \aver{\eta}^s.
\end{equation}
The shear stresses jump conditions \eqref{eq:OTW_cond_jump_dudy} and
\eqref{eq:OTW_cond_jump_dwdy} at the interfaces must also be reformulated.
Adding the $z$-derivative of \eqref{eq:OTW_cond_jump_dwdy} to 
the $x$-derivative of \eqref{eq:OTW_cond_jump_dudy} and using continuity
\eqref{eq:NSc_nondim} and VANS continuity \eqref{eq:VANSc_nondim} equations,
we obtain a jump condition for the normal derivative of the wall-normal 
velocity component at the interface of the form
\begin{equation} \label{eq:BC_v_2}
\sigma \dfrac{\partial^2 v}{\partial y^2} \mp \tau \dfrac{\partial v}{\partial y} = \dfrac{\sigma}{\varepsilon} \dfrac{\partial^2 \aver{v}^s}{\partial y^2}.
\end{equation}
Subtracting the $x$-derivative of the \eqref{eq:OTW_cond_jump_dwdy} from the 
$z$-derivative of the \eqref{eq:OTW_cond_jump_dudy} and using the definitions
\eqref{eq:eta_def} and \eqref{eq:eta_p}, we obtain a jump condition for the 
wall-normal vorticity component at the interface of the form
\begin{equation} \label{eq:BC_eta_1}
\sigma \dfrac{\partial \eta}{\partial y} \mp \tau \eta = \dfrac{\sigma}{\varepsilon} \dfrac{\partial \aver{\eta}^s}{\partial y}.
\end{equation}
Note that the positive sign in conditions \eqref{eq:BC_v_2} and \eqref{eq:BC_eta_1} 
applies to the interface located at $y=0$, while the negative sign applies to the 
interface located at $y=2$.

Finally, we need to convert the condition on pressure \eqref{eq:OTW_cond_p} at 
the interfaces. Summing the second $x$-derivative of \eqref{eq:OTW_cond_p} to 
the second $z$-derivative of equation \eqref{eq:OTW_cond_p} and using the Poisson 
equations in the fluid \eqref{eq:Poisson_fluid} and porous \eqref{eq:Poisson_porous} 
regions, we have
\begin{equation} \label{eq:ddp}
\dfrac{\partial^2 p}{\partial y^2} - \left( \dfrac{\partial HU}{\partial x} + \dfrac{\partial HV}{\partial y} + \dfrac{\partial HW}{\partial z} \right)= \dfrac{\partial^2 \aver{p}^f}{\partial y^2}.
\end{equation}
To eliminate pressure from the above equation, we take the $y$-derivatives of the 
$y$-component of the momentum equation \eqref{eq:NSm_nondim} to write
\begin{equation} \label{eq:dp}
\dfrac{\partial^2 p}{\partial y^2} = - \dfrac{\partial^2 v}{\partial t \partial y} + \dfrac{1}{\Rey} \nabla^2\dfrac{\partial v}{\partial y} + \dfrac{\partial HV}{\partial y},
\end{equation}
and the $y$-derivatives of the $y$-component of the VANS momentum equation \eqref{eq:VANSm_nondim} to write
\begin{equation} \label{eq:dp_VANS}
\dfrac{\partial^2 \aver{p}^f}{\partial y^2} = - \dfrac{1}{\varepsilon} \dfrac{\partial^2 \aver{v}^s}{\partial t \partial y} + \dfrac{1}{\varepsilon \Rey} \nabla^2 \dfrac{\partial \aver{v}^s}{\partial y} - \dfrac{1}{\sigma^2 \Rey} \dfrac{\partial \aver{v}^s}{\partial y}.
\end{equation}
Substituting \eqref{eq:dp} and \eqref{eq:dp_VANS} into \eqref{eq:ddp}, we obtain
\begin{equation} \label{eq:BC_v_3}
\left( \dfrac{\partial}{\partial t} - \dfrac{1}{\Rey} \nabla^2 \right) \dfrac{\partial v}{\partial y} + \left( \dfrac{\partial HU}{\partial x} + \dfrac{\partial HW}{\partial z} \right) =  \dfrac{1}{\varepsilon} \left( \dfrac{\partial}{\partial t} - \dfrac{1}{\Rey} \nabla^2 \right)  \dfrac{\partial \aver{v}^s}{\partial y} + \dfrac{1}{\sigma^2 \Rey} \dfrac{\partial \aver{v}^s}{\partial y} .
\end{equation}

At the impermeable walls, located at $y=-h_p$ and $y=2+h_p$, the no-penetration 
condition \eqref{eq:BCw_nondim_v} remains unchanged
\begin{equation}
\aver{v}^s =0. 
\label{eq:BCw_nondim_v_eta_v}
\end{equation}
On the other hand, the no-slip condition at the impermeable walls must be reformulated. 
The volume-averaged streamwise and spanwise velocity components, $\aver{u}^s$ and 
$\aver{w}^s$, are zero at the impermeable walls and, therefore, their derivatives
in the $x$- and $z$-directions are also zero. Adding the $x$-derivative of
\eqref{eq:BCw_nondim_u} to the $z$-derivative of \eqref{eq:BCw_nondim_w} and using 
the VANS continuity equation \eqref{eq:VANSc_nondim}, we obtain a boundary condition 
for the wall-normal derivative of $\aver{v}^s$  
\begin{equation} \label{eq:BCw_nondim_v_eta_dv}
\dfrac{\partial \aver{v}^s}{\partial y}=0.
\end{equation}
The last boundary condition can be obtained by subtracting the $x$-derivative of
\eqref{eq:BCw_nondim_w} from the $z$-derivative of \eqref{eq:BCw_nondim_u} and using 
the definition of $\aver{\eta}^s$ \eqref{eq:eta_p}. We obtain that the volume-averaged 
wall-normal component of the vorticity should vanish at the impermeable walls, i.e.
\begin{equation} \label{eq:BCw_nondim_v_eta_eta}
\aver{\eta}^s =0.  
\end{equation}

\subsection{Numerical implementation}
As the VANS governing equations (\ref{eq:VANS_v}) and (\ref{eq:VANS_eta}) have been written into a form that resembles the velocity-vorticity formulation (\ref{eq:NS_v}) and (\ref{eq:NS_eta})
customarily employed for the DNS of the incompressible NS equations, the computer code for their numerical solutions is designed to follow closely what is customary in the field. Except for a few specific differences, noted below, we adopt the general strategy employed by \cite{kim-moin-moser-1987} and many others; for example the choice of Fourier discretization in the wall-parallel homogeneous directions, the pseudo-spectral approach, and the exact removal of aliasing error by expansion of the Fourier velocity modes by a factor (at least) 3/2 before computing the non-linear terms in physical space, and by truncating these additional modes before transferring the non-linear terms back in Fourier space. Our code is derived from that introduced by \cite{luchini-quadrio-2006}, with whom it shares its general architecture. The Fourier discretization in the homogeneous directions is complemented with compact, high-accuracy (formally fourth order, with most operators being sixth-order accurate) explicit finite-differences schemes for the wall-normal direction.

The considered portion of the indefinite channel has lengths $L_x = 2 \pi / \alpha_0$ and $L_z = 2 \pi / \beta_0$ in the streamwise and spanwise direction, and the fundamental wavenumbers $\alpha_0$ and $\beta_0$ are chosen on the basis of physical considerations and with the aim of minimizing truncation effects. The associated number of Fourier modes, $N_x$ and $N_z$, are chosen such that $\Delta x = L_x/N_x$ and $\Delta z = L_z/ N_z$ are small enough to resolve the smallest scales of motion in the turbulent flow. Similar considerations lead to the choice of the number of collocation points along the wall-normal direction. The number of collocation points for the fluid region is denoted by $N_y$, and $N_{y,p}$ refers to the number of points for each porous region.

The evolution equations are advanced in time with a semi-implicit method; a Crank-Nicolson scheme advances the viscous terms, and a third-order, low-storage Runge--Kutta method is used for the non-linear terms. 

The equations of motions have been obtained through a process of spatial derivation along wall-parallel directions, hence information about terms that are uniform in space along these directions is lost, and equations in Fourier space are singular for the null wavenumber. It is useful to introduce a plane-averaging operator $\tilde{\cdot}$ acting along wall-parallel planes, as
\begin{equation}
\tilde{u} \left( y,t \right) = \frac{1}{L_x} \frac{1}{L_z} \int_0^{L_x} \int_0^{L_z} u \left( x,y,z,t \right) \, dx \, dz .
\end{equation}

After noting that $\tilde{v}=0$ everywhere by direct consequence of the continuity equation, two additional equations are required to compute $\tilde{u}$ and $\tilde{w}$ as functions of $y$ and $t$. For the fluid region, they can be obtained by applying the plane-averaging operator 
to the $x$ and $z$ components of the momentum equation \eqref{eq:NSm_nondim},
obtaining
\begin{align}
	\displaystyle \frac{\partial \tilde{u}}{\partial t} &= \frac{1}{\Rey} \frac{\partial^2 \tilde{u}}{\partial y^2} - \frac{\partial \widetilde{uv}}{\partial y} + \tilde{f}_x,  \label{eq:NS_uw_u_mean_flow}\\
	\displaystyle \frac{\partial \tilde{w}}{\partial t} &= \frac{1}{\Rey} \frac{\partial^2 \tilde{w}}{\partial y^2} - \frac{\partial \widetilde{vw}}{\partial y} + \tilde{f}_z,  \label{eq:NS_uw_w_mean_flow}
\end{align}
where $\tilde{f}_x$ and $\tilde{f}_z$ are the forcing terms required to drive the fluid against viscous drag, as discussed below.

The corresponding equations in the porous region, obtained after plane-averaging  \eqref{eq:VANSm_nondim}, are
\begin{align} 
	\displaystyle \frac{\partial \widetilde{\aver{u}^s}}{\partial t} &= \frac{1}{\Rey} \frac{\partial^2 \widetilde{\aver{u}^s}}{\partial y^2} - \frac{\varepsilon}{\sigma^2 Re} \widetilde{\aver{u}^s} + \varepsilon \tilde{f}_x, \label{eq:VANS_uw_u_mean_flow}, \\
	\displaystyle \frac{\partial \widetilde{\aver{w}^s}}{\partial t} &= \frac{1}{\Rey} \frac{\partial^2 \widetilde{\aver{w}^s}}{\partial y^2} - \frac{\varepsilon}{\sigma^2 Re} \widetilde{\aver{w}^s} + \varepsilon \tilde{f}_z . \label{eq:VANS_uw_w_mean_flow}
\end{align}

The flow rates per unit length in the $x$ and $z$ directions are
\begin{align}
Q_x &=  \int_{-h_p}^{0} \widetilde{\aver{u}^s} \, dy + \int_{0}^{2} \tilde{u} \, dy + \int_{2}^{2+h_p} \widetilde{\aver{u}^s} \, dy, \\
Q_z &=  \int_{-h_p}^{0} \widetilde{\aver{w}^s} \, dy + \int_{0}^{2} \tilde{w} \, dy + \int_{2}^{2+h_p} \widetilde{\aver{w}^s} \, dy.
\end{align}

In general, as recently discussed by \cite{hasegawa-quadrio-frohnapfel-2014}, if the pressure gradient is kept constant in time (the Constant Pressure Gradient approach, CPG), the flow rate oscillates in time around a constant value. On the other hand, if the flow rate is kept constant in time (the Constant Flow Rate approach, CFR) the pressure gradient oscillates around a constant value. In the present work, consistently with choosing $U_b^*$ as the characteristic velocity, we opt for enforcing the CFR condition for the streamwise direction; hence the appropriate value for the forcing term $\tilde{f}_x$ (i.e. the instantaneous value of the streamwise pressure gradient) is determined at every time step. For the spanwise direction, on the other hand, the simplest CPG condition with $\tilde{f}_z=0$ is adopted.

\subsection{Validation}
The code is validated against some results obtained via the 
linear stability theory. In particular, we consider the temporal 
evolution of small disturbances to the laminar velocity profile 
\citep{tilton-cortelezzi-2008}, and observe their growth rate. 
Additionally, we have verified that our results converge, as 
permeability goes to zero, to those obtained for a channel with 
impermeable wall. This will be evident in the discussion presented 
in the next Section.

Aside from their wall-normal shape, wavelike velocity perturbations 
to the numerically-computed base flow are defined by their streamwise 
and spanwise real wavenumbers $\alpha$ and $\beta$, and a complex 
streamwise phase speed $c = c_r + i c_i$. Temporal linear stability 
theory predicts that kinetic energy of the perturbations varies 
exponentially with time as $\exp(2 \alpha c_i t)$; furthermore, 
the wall-normal profile of the perturbation can be numerically 
determined.

At $Re=2800$, we consider a porous layer with permeability 
$\sigma = 0.004$, porosity $\epsilon = 0.6$ and momentum transfer 
coefficient $\tau = 0$, as this set of parameters will be thoroughly 
explored in the next Section. The linear stability analysis 
\citep{tilton-cortelezzi-2008, quadrio-etal-2013} establishes 
that a perturbation with $\alpha=1$ and $\beta=0$ may be either 
stable or unstable, depending on the thickness $h_p$ of the porous 
layer. In particular, the flow remains stable at $h_p=0.2$ with 
$c_i=-0.0028958$, whereas it is unstable at $h_p=1$ and $h_p=2$, 
with $c_i=+0.0052738$ and $c_i=+0.0083030$, respectively. 
Figure \ref{fig:validation} compares the predicted growth rates 
with the kinetic energy of the flow computed with the DNS code, 
starting from an initial condition obtained by superposing the 
laminar solution and a perturbation of small enough amplitude to 
preserve linearity. It can be appreciated that stability theory 
and nonlinear numerical simulations produce exponential growth 
rates in very good agreement. For the most unstable case, at 
large times, one notices some small non-linearities in the DNS 
curve, an effect that is obviously absent in the linear case. 
A similar plot is reported in figure \ref{fig:validation_tau} 
where, for validation purposes, two non-zero values of the 
momentum-transfer coefficient $\tau$ are considered for the 
case with $h_p=0.2$: again, the numerical simulations compute 
(negative) growth rates of the perturbation energy which 
perfectly agree with the predictions from the linear stability 
theory.

\begin{figure}
\centering
\input{validation}
\caption{Temporal evolution of the kinetic energy $E$ of disturbances for a channel flow over porous walls at $\Rey=2800$. The porous layers have permeability $\sigma = 0.004$, porosity $\varepsilon = 0.6$ and momentum-transfer coefficient $\tau = 0$. Dashed lines represent the evolution of energy as computed from stability theory, whereas solid lines represent evolution of energy as computed by the DNS code with $N_y=150$ ($N_{y,p}=75$ for $h_p=0.2$ and $N_{y,p}=250$ for $h_p=2$). The three line sets refer to different thicknesses of the porous layer: for $h_p=0.2$ energy decays, for $h_p=1$ energy grows exponentially, and $h_p=2$ shows the largest growth rate.}
\label{fig:validation}
\end{figure}
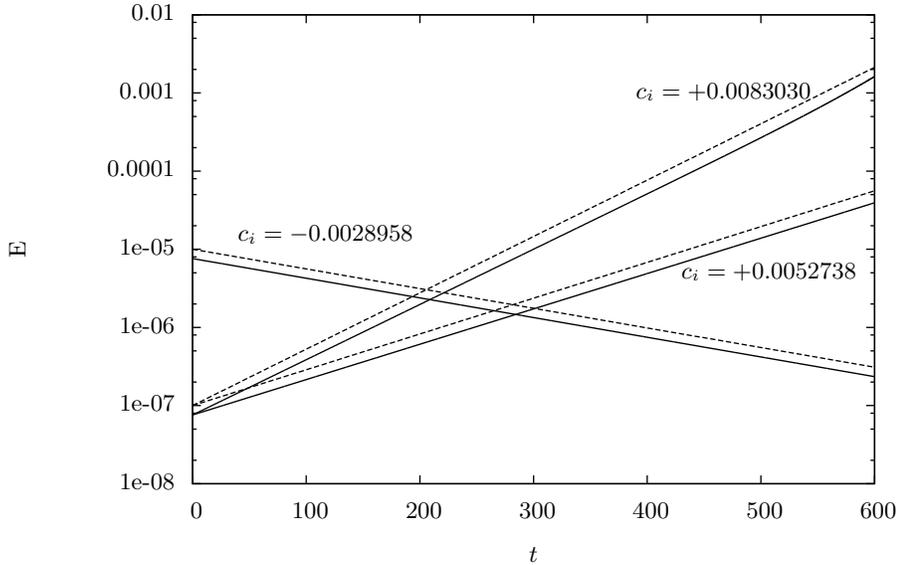

\begin{figure}
\centering
\input{validation_tau}
\caption{Temporal evolution of the kinetic energy $E$ of disturbances for a channel flow over porous walls at $\Rey=2800$. Flow parameters are the same as in figure \ref{fig:validation} ($h_p=0.2$) but for non-zero values of the parameter $\tau$. When $\tau=-1$ the growth rate is $c_i = 0.003118$, and when $\tau=+1$ $c_i=-0.004540$.}
\label{fig:validation_tau}
\end{figure}
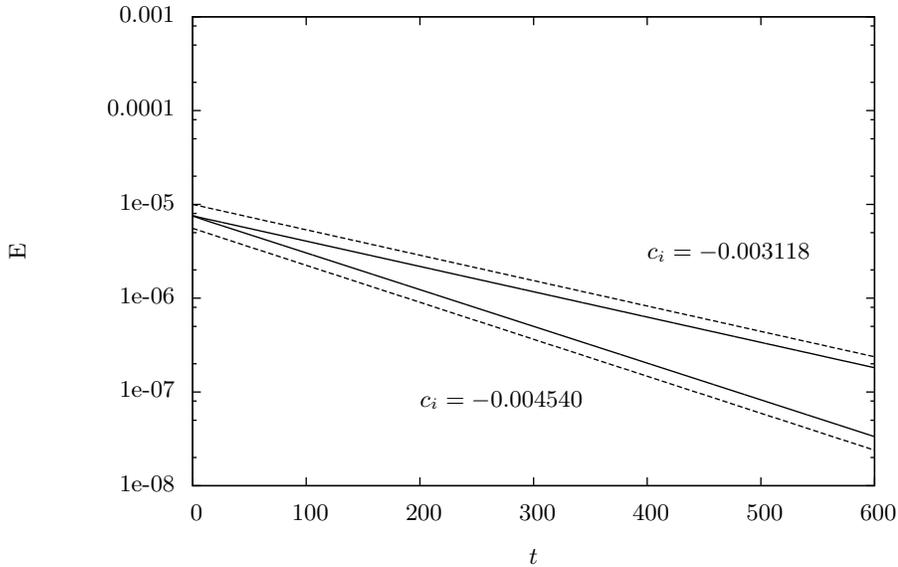

\section{The turbulent flow over a porous layer} 
\label{sec:results}

We consider as a reference for our discussion a turbulent channel flow over impermeable walls, at $Re=2800$, which corresponds to $Re_\tau \approx 180$, the friction Reynolds number chosen by \cite{kim-moin-moser-1987} in their seminal DNS study. Bar discretization issues, the value of $Re$ is all is needed to define the reference simulation of a channel flow bounded by two impermeable walls, whereas porous cases require additional quantities to describe the porous material: the thickness of the porous slab, its permeability and porosity, and the momentum-transfer coefficient that appears in the interface conditions. 

All the simulations are performed at constant flow rate, so that the flow Reynolds number, defined by Eq.~(\ref{eq:Re-definition}), is fixed at $Re=2800$, where the bulk velocity is computed according to (\ref{eq:Ub}). One simulation is carried out at the higher value of $Re=6265$, and is meant to explore how the main interfacial quantities change with Reynolds number. Owing to its increased computational cost, this case has been simulated on a domain of smaller spatial extension. For consistency, two low-$Re$ simulations at different permeabilities have also been recomputed by employing the smaller computational domain. As the focus of this work is the small-permeability regime, we first consider a baseline porosity case with a relatively small value of permeability and study its main turbulent quantities; the parameters of the porous material are then changed, one at a time, to gain a better insight on their relative importance in affecting the turbulent flow. 

In the baseline case, the flow develops over two identical porous walls, 
whose height is $h_p=0.2$; the dimensionless permeability of the porous 
material is $\sigma=0.004$, its porosity is $\varepsilon=0.6$, and the 
coefficient of the momentum-transfer conditions is $\tau=0$. This set 
of parameters is rather classic, having being used in previous studies 
\citep{tilton-cortelezzi-2006, tilton-cortelezzi-2008}, the only peculiarity 
being the small values of $\sigma$ that reflects the present goal of 
investigating low-permeability materials. The value $\sigma=0.004$ 
has been determined in a precursory study \citep{quadrio-etal-2013} as an upper value of 
permeability where the linearity assumption in the porous layer is 
justified. The full set of simulations is reported in Table \ref{tab:cases}, 
where the three cases computed in the 
smaller computational domain are denoted with SD. Table \ref{tab:cases}
also contains some mean quantities like the resulting friction-based 
Reynolds number, the velocity at the interface, and a quantification 
of the wall-normal length scale related to the thickness of the boundary layer 
at the porous/fluid interface. In particular the quantity $\delta$ 
measures how fast the mean velocity profile changes from the interfacial velocity $U_i$ to the inner constant 
Darcy value $U_D$, and is defined as the distance from the 
interface where the difference between the local velocity and $U_D$ 
reduces to $(1 - 1/e)(U_i-U_D)$. On the other hand, the other length-scale $\delta_D$ 
measures the position where the local mean velocity decreases to twice the value of $U_D$.

\begin{table}
\centering
\setlength{\tabcolsep}{10pt}
\begin{tabular}{lrrrrrrrr}
Case &	$\sigma$&	$\varepsilon$&	$\tau$&	$h_p$&	$Re_\tau$ & $U_i$ & $\delta$ & $\delta_D$ \\
\hline \hline
Reference&			-&	-  &	-&	-&   178.5 & - 	& - & - \\
Baseline&			0.004&	0.6&	0&	0.2& 188.3 & 0.0384 & 0.00237 & 0.0272 \\
$\sigma \downarrow$&		0.002&	0.6&	0&	0.2& 182.9 & 0.0180 & 0.00118 & 0.0154 \\
$\sigma \downarrow \downarrow$& 0.001& 0.6&	0&	0.2& 181.8 & 0.0090 & 0.00059 & 0.0086 \\
$\sigma \downarrow \downarrow \downarrow$&	0.0005&	0.6&	0& 0.2& 181.2 & 0.0047 & 0.00031 & 0.0046 \\
$\sigma \downarrow \downarrow \downarrow \downarrow$&	0.00025&	0.6&	0&	0.2&	181.0 & 0.0029 & 0.00020 & 0.0008 \\
$\varepsilon \uparrow$&	0.004&	0.9&	0&	0.2& 187.5 & 0.0464 & 0.00193 & 0.0230 \\
$\varepsilon \downarrow$&	0.004&	0.3&	0&	0.2& 187.6 & 0.0269 & 0.00335 & 0.0359 \\
$\tau \uparrow$&		0.004&	0.6&	$0.5$&	0.2& 185.9 & 0.0609 & 0.00237 & 0.0297 \\
$\tau \uparrow \uparrow$&	0.004&	0.6&	$+1$&	0.2& 177.9 & 0.1516 & 0.00237 & 0.0349 \\
$\tau \downarrow$&		0.004&	0.6&	$-1$&	0.2& 188.3 & 0.0215 & 0.00237 & 0.0242 \\
$h_p$ $\uparrow$&		0.004&	0.6&	0&	2.0& 188.4 & 0.0385 & 0.00237 & 0.0272 \\
Baseline, SD&			0.004&	0.6&	0&	0.2& 188.9 & 0.0381 & 0.00237 & 0.0272 \\
$\sigma \downarrow$, SD	&0.002&	0.6&	0&	0.2& 183.8 & 0.0178 & 0.00118 & 0.0154 \\
$Re \uparrow$, SD		&0.002&	0.6&	0&	0.2& 377.3 & 0.0351 & 0.00119 & 0.0154 \\

\end{tabular}
\caption{Summary of the DNSs performed at different porosity parameters, all at $Re=2800$ except the last one at $Re=6265$. The baseline porosity case has $\sigma=0.004$, $\varepsilon=0.6$ $\tau=0$ and $h_p=0.2$. SD indicates cases computed with a smaller computational domain (see text).}

\label{tab:cases}
\end{table}

For all cases at $Re=2800$, the equations of motion are discretized by using $256  \times 256$  Fourier modes on a computational domain of $4 \pi \times 2 \pi$ in the streamwise and spanwise directions. In the wall-normal direction, $150$ grid points are used in the fluid region, and 75 points discretize the wall-normal derivatives in each porous slab (but $N_{y,p}=250$ in the case with $h_p=2$). The spatial resolution of the numerical simulation is $\Delta x^+ \approx 8.8$ and $\Delta z^+ \approx 4.4$, with a wall-normal resolution $\Delta y^+$ that ranges from $0.16$ near the interface to $4.1$ in the centerline region of the channel. The size of the computational domain and the spatial resolution are comparable to those employed by \cite{kim-moin-moser-1987}. The case at $Re=6265$ clearly has different discretization parameters: with a computational domain of $2 \pi \times \pi$ in the wall-parallel directions, the same level of spatial resolution (in viscous units) is achieved with $N_x=256$, $N_z=256$, and $N_{y,p}=75$, whereas $N_y$ is increased to $N_y=180$.

Viscous units, that have been just used above to express spatial resolution, will be often employed in the following; they are indicated by the superscript $^+$, and are built using the friction velocity $u_\tau$ as the velocity scale and the viscous length $\delta_\nu = \nu / u_\tau$ as the length scale. For a turbulent channel flow with solid walls, the dimensionless friction velocity is defined as
\begin{equation} 
\label{eq:friction_velocity}
u_\tau = \sqrt{\dfrac{1}{\Rey} \left. \dfrac{d \overline{u}}{d y} \right\vert_{y=0}},
\end{equation}
where $\overline{u}$ is the mean velocity, obtained by time averaging $\tilde{u}(y,t)$, 
and the subscript indicates that the derivative is taken at $y=0$, the location of the solid wall. 
When the channel has porous walls, the above definition \eqref{eq:friction_velocity} must 
be modified to account for the turbulent shear stresses that are in general non-zero at the 
porous-fluid interface. We follow \cite{breugem-boersma-2005}, and define
\begin{equation} \label{eq:friction_velocity_total}
u_\tau = \sqrt{\dfrac{1}{\Rey} \left. \dfrac{d \overline{u}}{d y} \right\vert_{y=0} - \left. \overline{u'v'} \right\vert_{y=0}},
\end{equation}
where $\overline{u'v'}$ is the off-diagonal component of the Reynolds stresses tensor, 
and the quantities are evaluated at the porous/fluid interface located at $y=0$. At small 
permeability values like those considered in the present study, as seen in Table \ref{tab:cases} 
the value of the friction Reynolds number changes little from $\Rey_{\tau} \approx 180$ 
that characterizes the reference flow over impermeable walls. 

All the simulations are started from an initial condition that requires some time for the flow to reach a state of statistical equilibrium. After such a state is reached, the calculations are continued for a time interval of 800 time units, during which 160 full flow fields are stored for further statistical analysis. 

\subsection{The baseline case}

\begin{figure}
\centering
\input{fik}
\caption{Wall-normal profile of the integrand of the second term in the FIK identity (\ref{eq:fik}) for the impermeable and porous cases.}
\label{fig:fik}
\end{figure}
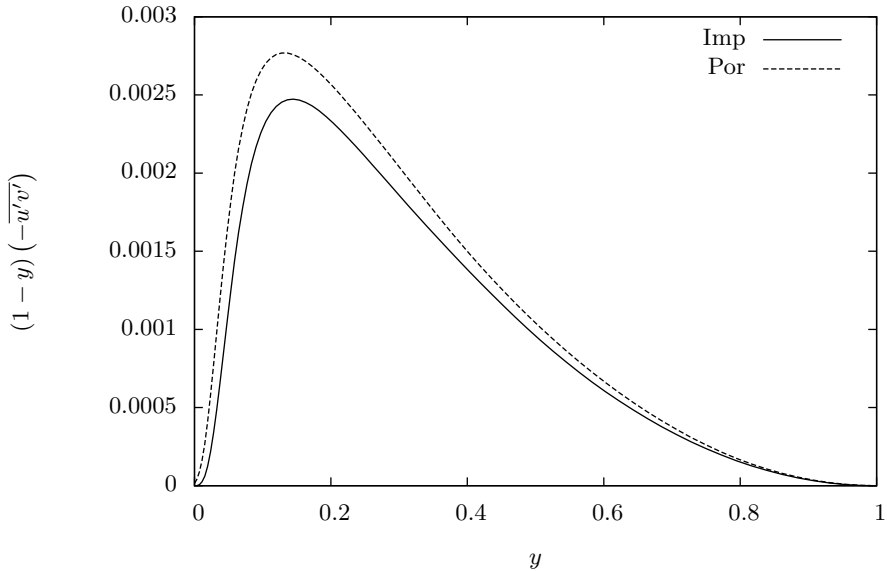


We start the comparison between the turbulent channel flows with porous or impermeable walls by analyzing their mean properties. Small but finite effects of porosity can be appreciated already at a global energetic level: for example the value of $Re_\tau$ increases by approximately 5\% from $Re_\tau=178$ to $Re_\tau=188$, with the skin-friction coefficient $C_f \equiv 2 \tau_w /\rho U_b^2$ going up from $8.14 \times 10^{-3}$ to $9.00 \times 10^{-3}$. Being the simulations carried out under the CFR condition, this implies that the power input to the system, given by the product of flow rate and mean pressure gradient, is correspondingly increased. 

It is known that the skin-friction coefficient in a steady, fully-developed, incompressible, plane channel flow under the CFR condition can be divided into a laminar and a turbulent contribution \citep{fukagata-iwamoto-kasagi-2002}, via the so-called FIK identity. When the fluid flows over porous walls, additional terms arise which indicate how the porous material affects the turbulent friction. To identify these additional terms, we start from the streamwise component of the incompressible momentum equation, that after being averaged in time and along the homogeneous directions becomes
\begin{equation} \label{eq:NSz}
\dfrac{\partial \overline{p}}{\partial x}=
\dfrac{\partial}{\partial y}\left(\dfrac{1}{Re} \dfrac{\partial \overline{u}}{\partial y} 
- \overline{u'v'} \right) .
\end{equation}

Based on the definition (\ref{eq:friction_velocity_total}) of the friction velocity, the friction coefficient is defined as 
\begin{equation} 
\label{eq:cf}
C_f= 2\left(\frac{1}{Re} \left. \frac{\partial \overline{u}}{\partial y} \right|_{y=0}
- \left. \overline{u'v'} \right|_{y=0} \right).
\end{equation}
Integrating \eqref{eq:NSz} from $y=0$ to $y=1$ and noting that both $\partial \overline{u} / \partial y$ and $\overline{u'v'}$ are zero at the centerline for symmetry reasons, a relation between the pressure gradient and the skin-friction coefficient \eqref{eq:cf} is obtained, i.e.
\begin{equation}\label{eq:NSzy}
-\dfrac{\partial \overline{p}}{\partial x}=\dfrac{C_f}{2}.
\end{equation}
Substituting \eqref{eq:NSzy} into \eqref{eq:NSz} we obtain
\begin{equation} \label{eq:NSzyCf}
\dfrac{C_f}{2}=\dfrac{\partial}{\partial y} \left( \overline{u'v'} -\dfrac{1}{Re} \dfrac{\partial \overline{u}}{\partial y} \right) .
\end{equation}
Integrating the above equation \eqref{eq:NSzyCf} twice from $0$ to $y$ leads to 
\begin{equation} \label{eq:NSzyCfyy}
\dfrac{C_f}{2} \left( y-\dfrac{y^2}{2} \right) = \int_0^y - \overline{u'v'} \; dy + \frac{1}{Re} \left( \overline{u} - U_i \right) ,
\end{equation}
where $U_i$ indicates the mean velocity at the interface. A further integration in $y$ from $0$ to $1$ gives
\begin{equation} \label{eq:NSzyCfyyy}
\dfrac{C_f}{6}=\int_0^1 \int_0^y -\overline{u'v'} \; dy \; dy + \dfrac{1}{Re} \left( \int_0^1 \overline{u} \; dy - U_i \right).
\end{equation}
and a subsequent integration by parts gives
\begin{equation}
\int_0^1 \int_0^y -\overline{u'v'} \; dy \; dy = \left[ y \int_0^1 - \overline{u'v'} \; dy \right]_0^1 - \int_0^1 y \left( - \overline{u'v'} \right) \; dy = \int_0^1 - \overline{u'v'} \; dy - \int_0^1 y \left( - \overline{u'v'} \right) \; dy,
\end{equation}
and leads to the following extended FIK identity for a turbulent channel flow with porous wall:
\begin{equation} \label{eq:fik}
C_f= \dfrac{6}{Re} \int_0^1 \overline{u} \; dy + 6 \int_0^1 \left( 1-y \right) \left( - \overline{u'v'} \right) \; dy - \dfrac{6}{Re} U_i.
\end{equation}

This relation states that the skin friction is the sum of three different contributions. The laminar contribution (first term) is identical to the case of impermeable wall. The second term is formally identical to the impermeable one, but the non-zero boundary value of the shear stress at $y=0$ must be remembered. Lastly, the third term accounts for the mean slip velocity at the interface. This extended FIK identity, that reduces to the standard one for impermeable walls, is well verified by the present data, as the value of $C_f=0.0090$ computed via (\ref{eq:fik}) is identical to that obtained by using its definition and the expression (\ref{eq:friction_velocity_total}) for the friction velocity. The laminar, turbulent and interface velocity contributions are $0.0021$, $0.0068$ and $8.2 \times 10^{-5}$. In comparison to the impermeable case, the laminar term is identical, obviously the interfacial velocity term is zero, and the Reynolds stress term is $0.0060$. Figure \ref{fig:fik} portraits the weighted wall-normal profile of the Reynolds shear stress, i.e. the integrand in the second term of (\ref{eq:fik}), in comparative form between impermeable and porous case. The porous integrand is larger than the impermeable wall throughout the whole channel. The case at higher $Re$ has laminar, turbulent and interface velocity contributions 
equal to $0.00096$, $0.00627$ and $0.00003$. 

\begin{figure}
\centering
\input{mean}
\caption{Comparison of the mean velocity profile $\overline{u}$ of turbulent channel 
flow at $\Rey=2800$, over an impermeable wall (reference case, solid line) and a porous 
wall (baseline case, dashed line). The thickness of the porous layer is $h_p=0.2$, the 
permeability is $\sigma=0.004$, the porosity is $\varepsilon=0.6$ and the coefficient 
in the momentum-transfer conditions is $\tau=0$. The inset shows the zoomed-in interfacial 
region where the porous profile exhibits a non-zero slip velocity.}
\label{fig:mean}
\end{figure}
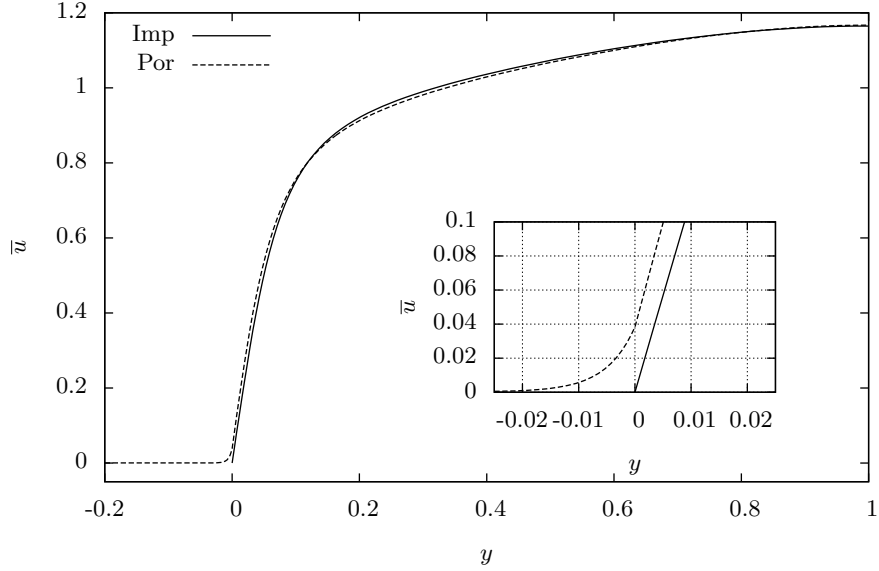

Figure \ref{fig:mean} shows the wall-normal distribution of the streamwise component 
of the mean velocity, in comparative form for the channel flow over impermeable and 
porous walls. The two profiles are quite similar throughout the channel, and reach 
their maximum at the centerline with $\overline{u}(1)=1.165$, with the difference 
that the porous velocity is slightly higher in the region near the interface, and 
consequently lower (owing to the constraint of constant flow rate) in the outer 
region, with the crossover taking place at $y \approx 0.15$. A zoomed view of the 
velocity profile near the interface between the fluid region and the porous layer 
is shown in the inset of figure \ref{fig:mean}, where a slip velocity of about 
$U_i=0.038$ is observed at the interface. The mean velocity within the porous material 
is strictly zero only at the impermeable wall located at $y=-0.2$, and assumes an 
essentially constant value in the bulk of the porous layer. This value equals the 
dimensionless value of the Darcy velocity $U_D^* = - K^* / \mu^*  \,\, d\overline{p^*}/dx^*$, 
which in the present case is $U_D=1.97 \times 10^{-4}$. As the interface at $y=0$ 
is approached, the fluid velocity within the porous material increases further from 
the Darcy value to the interfacial velocity $U_i$, and this process takes place in 
a thin layer, whose thickness $\delta$ turns out to be $0.00237$ when computed as 
the distance from the interface to the location where the local velocity becomes 
$1-1/e \approx 63\%$ of $U_i-U_D$. The velocity profile on the two sides of the 
interface has different slopes, and this, together with a non-zero interfacial 
shear stress for the porous case, explains the different values of the friction 
Reynolds number.

\begin{figure}
\centering
\input{loglaw}
\caption{Comparison of the mean velocity minus the interface velocity, 
$(\overline{u} - U_i)^+$ versus the distance from the interface $y^+$. 
The profiles for an impermeable wall (reference case, solid line) and 
a porous wall (baseline case, dashed line) are compared with the mean 
profile from \cite{kim-moin-moser-1987} (symbols). Dash-dotted lines 
indicate $\overline{u}^+ - U_i^+ = y^+$ and $\overline{u}^+ - U_i^+ = 2.5 \ln y^+ + 5.5$.}
\label{fig:loglaw}
\end{figure}
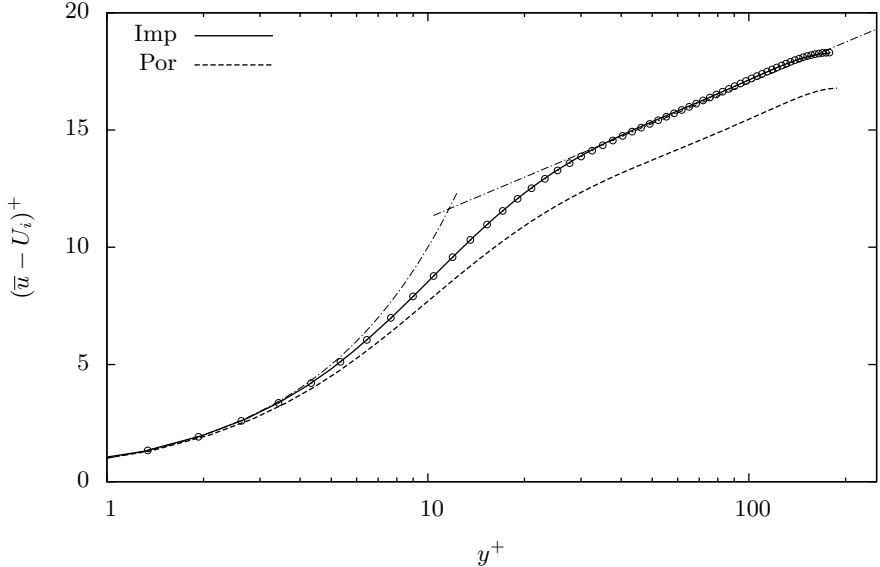

Figure \ref{fig:loglaw} compares the mean velocity profiles in the fluid region only. 
Following \cite{hahn-je-choi-2002}, the profiles are plotted as the difference 
$(\overline{u}-U_i)^+$ versus the logarithm of the distance from the interface $y^+$; 
both quantities are expressed in wall units. The mean velocity profile computed by 
\cite{ kim-moin-moser-1987} is also added, and turns out to be undistinguishable 
from the present results for the impermeable wall. The velocity profile over the 
porous wall possesses the three regions that are present in the standard channel 
flow: the viscous sublayer for $y^+<5$, the logarithmic region for $y^+>30$, and 
the buffer layer in between. In the logarithmic region, 
\begin{equation} \label{eq:log_law}
\left( \overline{u} - U_i \right)^+ = \frac{1}{\kappa} \ln y^+ + B^+,
\end{equation}
where $\kappa$ and $B^+$ are two constants whose values (at this value of $\Rey$) 
are $\kappa=0.4$ and $B^+=5.5$. When plotted in the law-of-the-wall form, the only 
significant difference exhibited by the porous profile is in the logarithmic region, 
where the curve is significantly lower than the case with solid walls (a fit with 
$\kappa=0.4$ yields $B^+=4$). These results are in line with those by 
\cite{hahn-je-choi-2002} and by \cite{breugem-boersma-uittenbogaard-2006}. The 
former mimicked the effects of the porous layer through a suitable set of boundary 
conditions applied at the interface, and found a logarithmic law with unchanged 
slope and decreased constant, indicating a reduced thickness of the viscous sublayer. 
The latter presented a detailed discussion of the mean velocity profile but, as 
mentioned in the Introduction, their method for characterizing the porous material 
is such that a direct, quantitative comparison between their cases and the present results 
is not possible. However, they too conclude that at low permeabilities the slope of 
the velocity profile is unaffected.

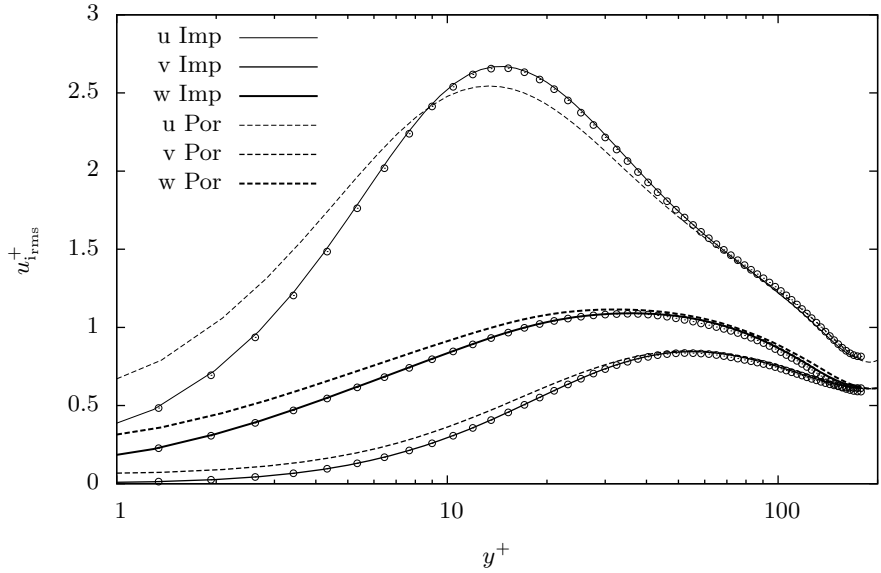
\begin{figure}
\centering
\input{rms}
\caption{R.m.s. values of velocity fluctuations, expressed in wall units, as a function of the distance $y^+$ from the interface, for a turbulent channel flow at $\Rey=2800$. Solid and dashed lines indicate the flow over porous and impermeable walls, respectively. Symbols are for the impermeable case from \cite{kim-moin-moser-1987}.}
\label{fig:rms}
\end{figure}

\begin{figure}
\centering
\input{rms-slab}
\caption{R.m.s. values of velocity fluctuations, expressed in wall units, within the porous slab.}
\label{fig:rms-slab}
\end{figure}
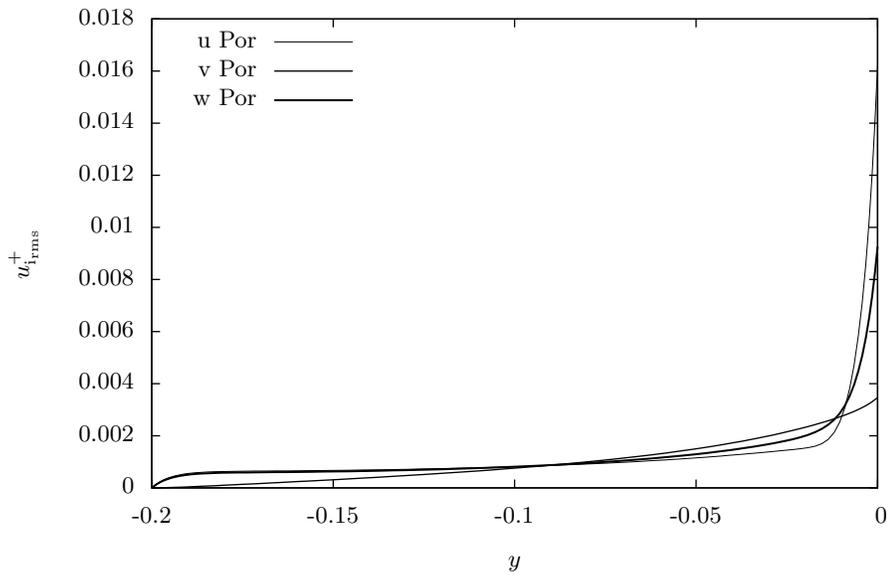

We continue our comparison between the turbulent channel flows with porous or impermeable walls by analyzing the wall-normal distribution of r.m.s. values of velocity fluctuations. Figure \ref{fig:rms} shows their profiles (including data from \cite{kim-moin-moser-1987}, that overlap with the present results for the impermeable wall); they are affected by the porous walls, and the effect is not limited to the interfacial region, but it extends up to $80-100$ wall units. All the components for the porous case present non-null r.m.s. values at $y=0$, where the no-slip condition is not enforced, but they decrease quickly inside the porous layer. In relative terms, the wall-normal and spanwise components are probably the most affected, with the position of their peaks moving towards the interface, a shift attributed by \cite{perot-moin-1995} to the decreased blocking and viscous effects exerted by the porous wall.

Figure \ref{fig:rms-slab} depicts how the r.m.s. intensity of velocity fluctuations change within the porous material from the small but non-zero interfacial value down to zero at the solid wall. At the interface the $v$ component is the smallest and the $u$ component is the largest; inside the porous slab, the wall-parallel components drop to a nearly constant value, and become nearly indistinguishable form each other, decreasing very slowly and eventually dropping to zero across a very thin small wall layer. The $v$ component, on the other hand, presents a markedly lower decay rate and is the largest one for $-0.08 < y < 0.02$. The asymptotic behaviour as the solid wall is approached is linear for the wall-parallel components, and quadratic for the wall-normal one, as required by the no-slip boundary condition and continuity equations. This very presence of velocity fluctuations demonstrate that, strictly speaking, the VANS-filtered flow inside the porous slab is not fully laminar, although the magnitude of the fluctuations is small. The structure of this non-laminar flow, however, is different from turbulence inside the channel, as will be seen in the following.

\begin{figure}
\centering
\input{uv_corrcoeff}
\caption{Correlation coefficient $C_{uv}$ between the streamwise and wall-normal velocity fluctuations for a turbulent flow at $\Rey=2800$. Solid and dashed lines indicate the flow over impermeable and porous walls, respectively. Symbols are for the impermeable case from \cite{kim-moin-moser-1987}.}
\label{fig:uv_corrcoeff}
\end{figure}
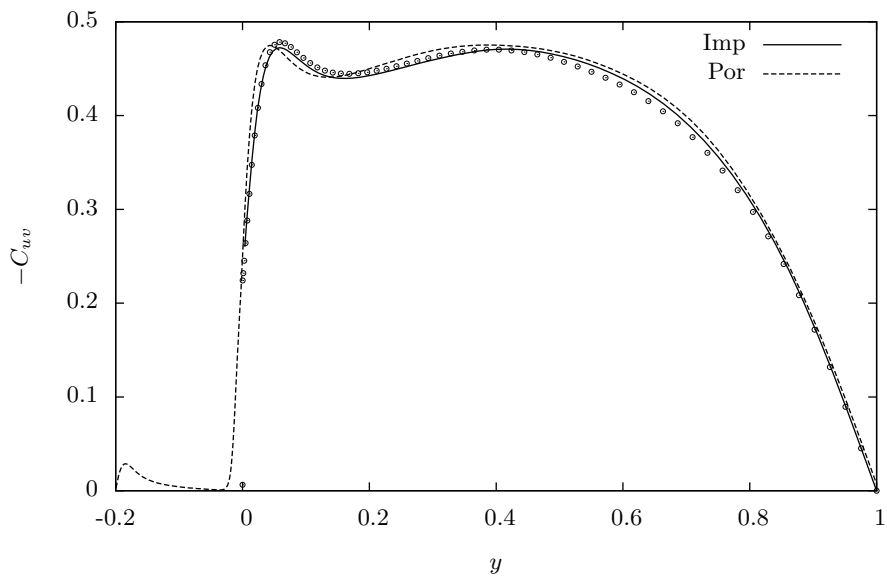

Figure \ref{fig:uv_corrcoeff} effectively illustrates how the structure of 
the (small) turbulent fluctuations changes across the porous layer, by 
plotting the wall-normal profile of the correlation coefficient $-C_{uv}$ 
between the fluctuations of the streamwise and wall-normal velocity components, 
i.e. the off-diagonal component of the Reynolds stress tensor made dimensionless 
by the local $u_{rms}v_{rms}$. The value of $-C_{uv}$ is zero in the isotropic 
case and is also zero at the centerline; it retains a rather large value, 
between 0.4 and 0.5, in most of the channel gap, as in the impermeable case; 
then presents a small local peak near the wall and then decreases to $\approx 0.2$ 
as the wall is approached in the impermeable case. Near the interface with the 
porous medium, on the channel side $-C_{uv}$ is almost unaffected by porosity, 
with only the local peak being slightly moved towards the interface; from the 
interface down $-C_{uv}$ decreases rather quickly to zero. For most of the porous 
layer it remains nearly zero, but when the solid wall is approached $-C_{uv}$ 
progressively increases again, shows a local peak at $y \approx -0.18$, i.e. 
at a distance of $0.02$ from the solid wall, and then eventually drops to zero. 
Hence, regardless of the presence of the porous material and of the limited 
intensity of velocity fluctuations, the peak is due to the presence of the solid wall.

\begin{figure}
\centering
\input{vort}
\caption{R.m.s. values of vorticity fluctuations, expressed in wall units, as a function of the distance $y^+$ from the interface, for a turbulent channel flow at $\Rey=2800$. Solid and dashed lines indicate the flow over impermeable and porous walls, respectively. Symbols are for the impermeable case from \cite{kim-moin-moser-1987}.}
\label{fig:vort}
\end{figure}
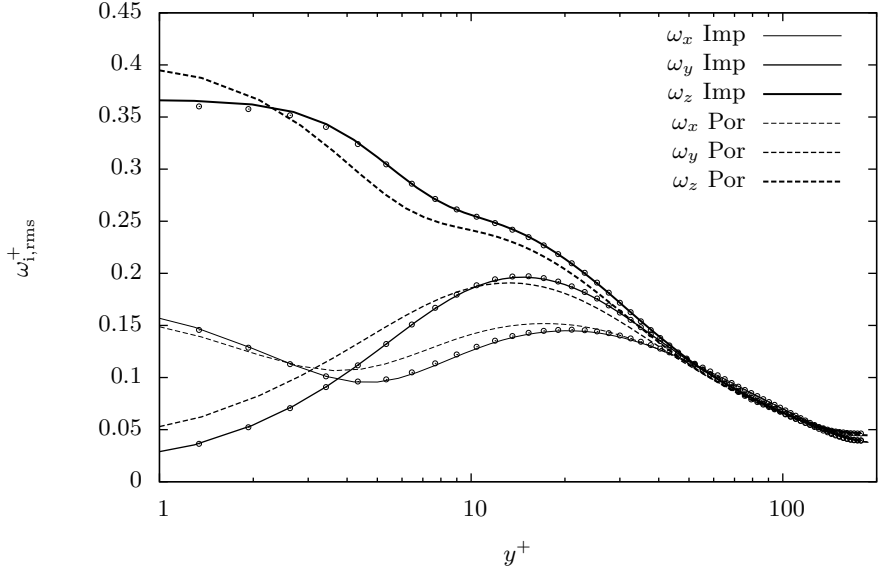

Figure \ref{fig:vort} shows the profiles of the r.m.s. intensities of vorticity 
fluctuations. Again, available data from \cite{kim-moin-moser-1987} are plotted 
for comparison, and overlap very well with the present results for the impermeable 
wall. Similarly to the velocity fluctuations, a non-trivial effect of the porous 
wall upon the vorticity fluctuations can be observed: the spanwise component is 
the largest as in the impermeable case and is increased near the interface, but 
its intensity becomes lower for $4 < y^+ < 50$. The streamwise component is 
slightly decreased for $y^+ < 3$ and increased outwards, while the wall-normal 
component behaves similarly to the spanwise one, being increased for $y^+<10$ 
and slightly decreased for $10 < y^+ < 50$. In particular the local peak of 
$\omega_x$ fluctuations at $y^+ \approx 20$, often associated \citep{kim-moin-moser-1987} 
with the average wall-normal position of quasi-streamwise vortical structures, 
seems to be only slightly affected by the porous material, at least at this level 
of permeability. In agreement with \cite{breugem-boersma-uittenbogaard-2006}, it 
can be inferred that the near-wall structure of the turbulent flow is affected too, 
as this peak is connected with the presence of high- and low-velocity streaks. 

\begin{figure}
\centering
\includegraphics[width=\textwidth]{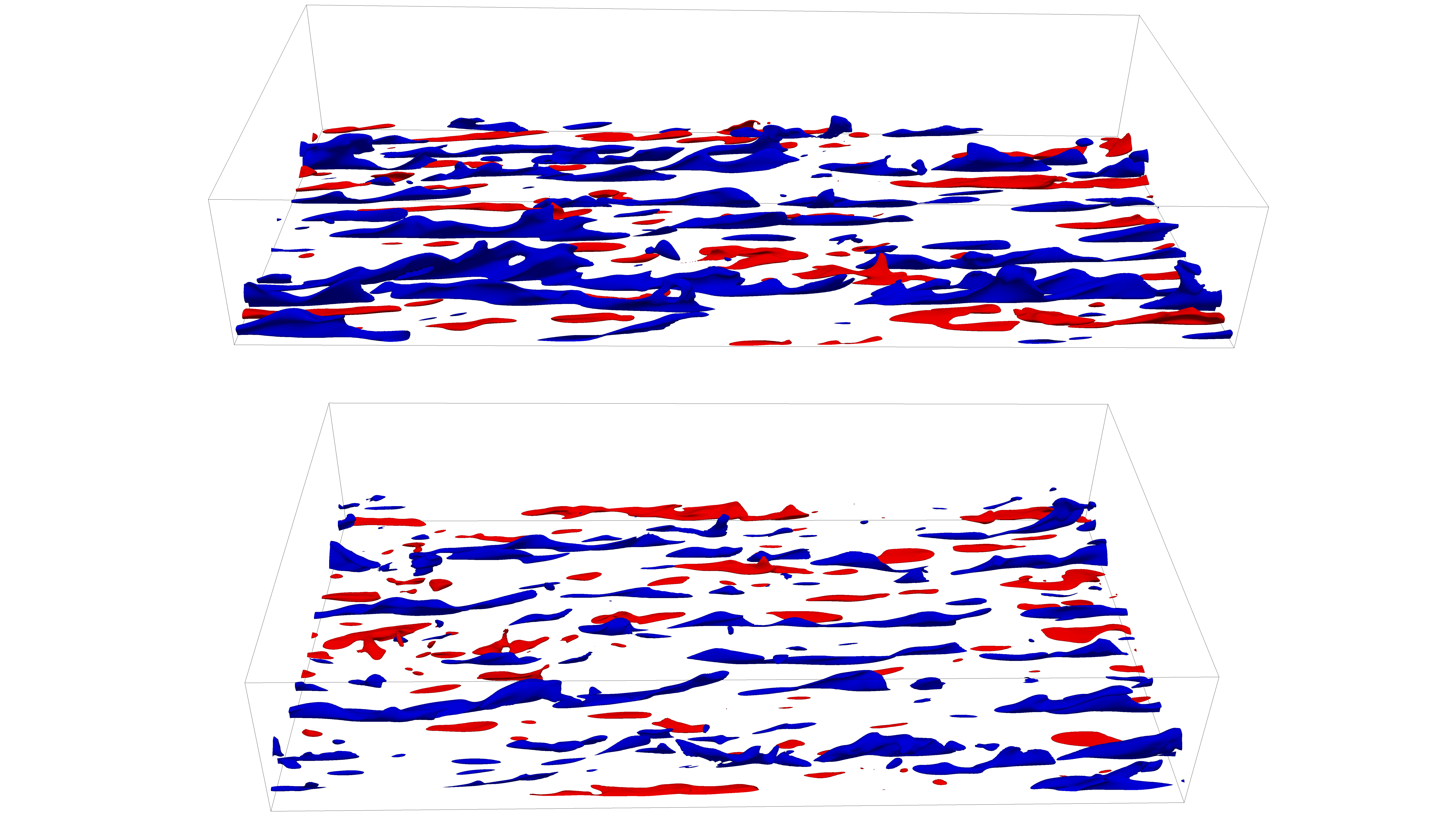}
\caption{Low- and high-velocity near-wall streaks in the turbulent flow over an impermeable wall (top) and over a porous wall (bottom). The streaks are visualized via blue / red isosurfaces for $u'^+=\pm 4$ for the lower half of the channel. The more fragmented nature of the streaky structures over the porous material is evident.}
\label{fig:lss}
\end{figure}

Changes, albeit small, in the near-wall structure of the flow are indeed visually confirmed by figure \ref{fig:lss}, that shows the elongated low- and high-velocity near wall streaks, plotted as isosurfaces where the local velocity assumes the values $u'^+=\pm4$. The small value of permeability notwithstanding, near the porous wall the structures appear as less elongated and more fragmented, especially those parts of the low-speed streaks that are located near the interface. This information will be statistically assessed later. 

\begin{figure}
\centering
\includegraphics[width=\textwidth]{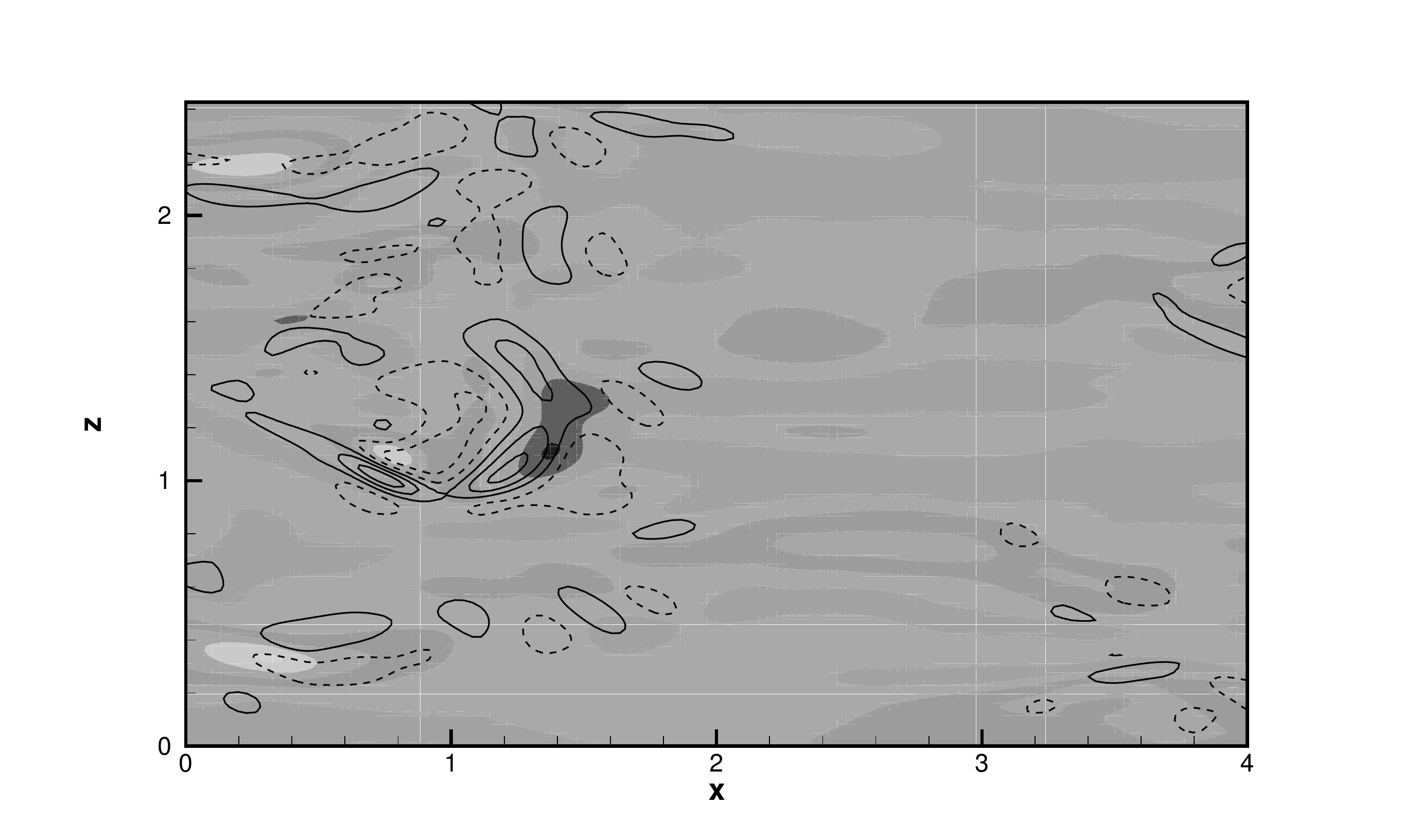}
\caption{Contour levels of the $u$ (colour map) and $v$ (isolines) fluctuating fields at the interface $y=0$ for the baseline porous case. Enlarged view of a limited portion of the wall, with flow from left to right. Colour bands for $u$ are (-0.01,0.02,0.15) from blu/dark to red/light. Contour increments for $v$ are (-0.025,0.01,0.045), with dashed lines for negative contours.}
\label{fig:snap-interface}
\end{figure}

A further look at the interface plane, shown in figure \ref{fig:snap-interface}, amplifies these differences, and demonstrates that the flow is rich of small-scale features, consistent with a picture where the larger coherent structures are being broken into smaller pieces by the porous material. Here the streamwise and wall-normal fluctuating velocity fields are visualized; at the interface both show relatively small fluctuations. The structure of the large-scale low- and high-speed streaks can still be discerned, but the streaks (especially the low-speed ones) are fragmented into smaller structures; at the interface positive fluctuations of $v$ (i.e. related to pockets of fluid being ejected outwards towards the bulk flow) are stronger than negative ones. 

\begin{figure}
\centering
\input{diss}
\caption{Wall-normal profile of the dissipation $\epsilon$ of the 
fluctuating field, for a turbulent channel flow at $\Rey=2800$. 
Solid and dashed line indicate the flow over impermeable and 
porous walls, respectively. The left plot shows quantities scaled 
in outer units, while the right plot employs inner units.}
\label{fig:diss}
\end{figure}
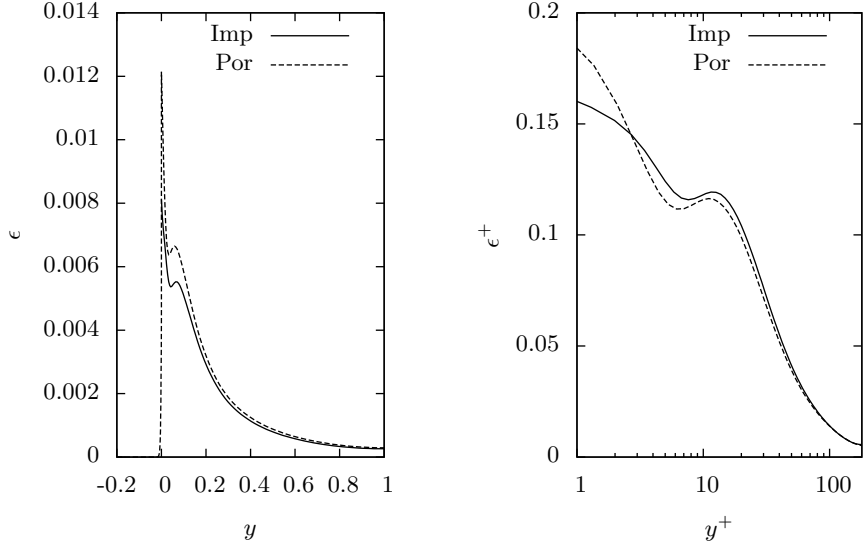

This increased fragmentation is likely to increase viscous dissipation. 
Figure \ref{fig:diss} plots the wall-normal profile of the dissipation 
$\epsilon = \mu \overline{\partial_j u'_i \partial_j u'_i}$ of the 
fluctuating velocity field. While the general aspect of the curve in 
the porous case (figure \ref{fig:diss} left) resembles that of the 
impermeable case, a larger dissipation is evident over a significant 
range of distances from the interface, say $y < 0.4$, and in particular 
the value at the interface is much larger. Dissipation then drops to 
zero very quickly in the porous layer, as velocity fluctuations are 
damped by the porous material, and remains negligible down to the solid 
wall. Most of the large differences, however, simply are a consequence 
of the two flows possessing slightly different values of friction 
Reynolds number $Re_\tau$. The right frame of figure \ref{fig:diss} 
emphasizes the much better match between the two curves in the bulk 
region when inner scaling is adopted to account for the change in 
$Re_\tau$. At the interface with the porous medium, dissipation is 
still observed to assume larger values (although the increase is much 
smaller), but this increase is counterbalanced by a slight decrease 
in the outer part, say for $3 < y^+ < 50$. When expressed in wall 
units and integrated along the wall-normal direction, the turbulent 
dissipation is only slightly reduced in the porous case. 

\cite{quadrio-2011} and \cite{ricco-etal-2012} introduced the concept 
of energy box to illustrate the energy flux through the channel flow 
system, linking energy input per unit time to total energy dissipation, 
i.e. the sum of turbulent dissipation and the dissipation ascribed to 
the mean velocity profile. In the present case, where no active flow 
control device is present, energy enters the system through the pumping
action (product of flow rate and pressure gradient), and in statistical 
equilibrium this energy input must balance the sum of turbulent dissipation 
and direct dissipation from the mean profile. It is known \citep{laadhari-2007} 
that at low values of $Re$ the latter mechanism is a significant one, and 
indeed in our case the mean profile accounts for slightly more than one 
half of the total dissipation. The power input in the baseline porous case, 
however, when expressed in wall units is less than that of the reference 
case, as the flow rate is kept constant in outer units and the friction 
velocity is larger over the porous wall, hence the flow rate in inner 
units decreases. This implies that turbulent dissipation, while slightly 
decreased in wall units and in absolute terms, does in the porous case 
account for a larger share of the entire dissipation.


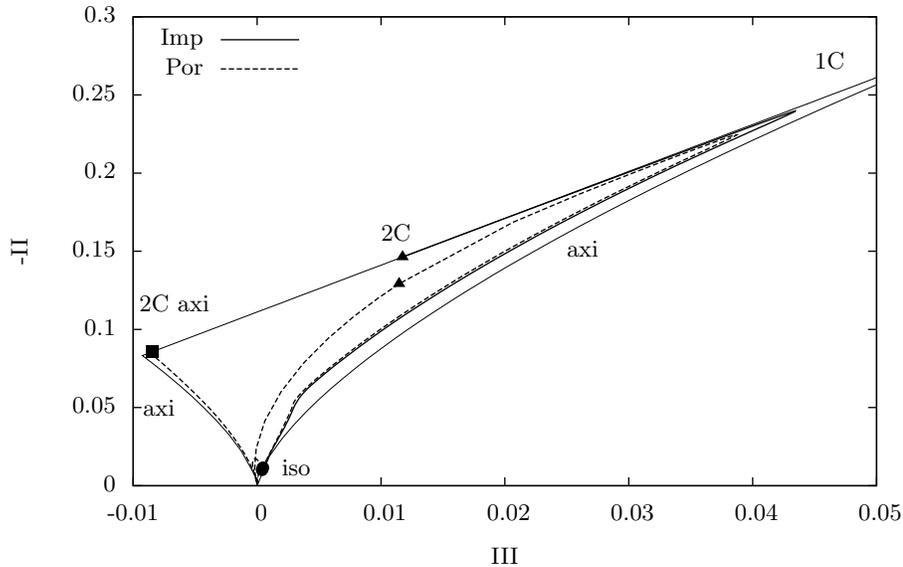
\begin{figure}
\centering
\input{lumley}
\caption{The Lumley triangle, for a turbulent channel flow at $\Rey=2800$. 
$-II$ and $III$ are the second and third invariant of the Reynolds stress 
anisotropy tensor, and the bounding line shows the region of admissible 
turbulence states, amid the extremal states of one-component turbulence 
(1C), two-components turbulence (2C), axisymmetric turbulence (axi), 
two-component axisymmetric turbulence (2C axi) and isotropic turbulence (iso). 
Solid and dashed line indicate the flow over impermeable and porous walls, 
respectively. The triangles correspond to $y=0$ and the square to $y=-h_p$. 
The circle marks the nearly isotropic state, which is visited by the flow 
for $y=1$ and, for the porous case, also in the bulk of the porous layer.}
\label{fig:lumley}
\end{figure}

\begin{figure}
\centering
\input{cor}
\caption{One-dimensional autocorrelation function for the velocity fluctuations for a turbulent channel flow at $\Rey=2800$. The left column is for $y^+ \approx 5$, and the right column for $y^+ \approx 150$; the top row features the streamwise separation, the bottom row the spanwise separation.}
\label{fig:cor}
\end{figure}
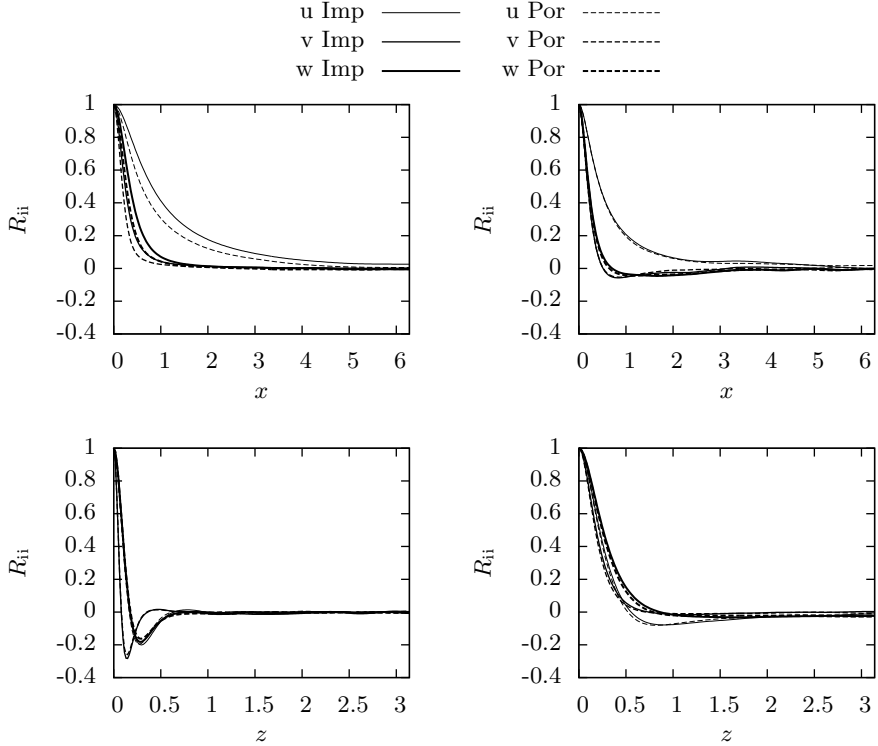

A similar picture emerges from the analysis of the anisotropy invariant 
map (AIM), also called Lumley triangle, shown in figure \ref{fig:lumley}. 
The AIM is a plot on the plane of the second and third invariant of the 
Reynolds stress anisotropy tensor; in this plane, a region is delimited 
by an approximately triangular boundary whose sides correspond to special 
limiting states of the turbulent flow. In the case of impermeable channel 
flow, represented by a solid line, one moves along the line as the wall-normal 
position changes, with the centreline corresponding to an almost isotropic 
state near the origin of the plane, and the very-near-wall region corresponding 
to the part of the curve near the upper boundary of the triangle. Here the 
boundary curve corresponds to a two-components turbulence, as it should be 
in close proximity of a solid wall. The curve for the porous case is nearly 
identical to the reference one for most of the wall-normal positions, with 
the exception of the region very near to the interface. Here the porous curve 
does not follow closely the upper boundary and bends downward, reflecting the 
structural changes induced by the porous material. The differences between 
the two curves, however, are limited to a rather thin layer, and disappear 
for say $y^+ > 5$. Below the interface, the turbulent state further departs 
from the two-components boundary, and quickly reaches the origin of the plane, 
corresponding to the isotropic state. It then remains near the origin until 
the boundary layer over the solid wall is reached: at that point the flow 
state moves from the origin towards the left corner, that represents the 
two-dimensional isotropic state.

These differences, confined within a thin region above the interface, 
possess dynamical consequences that reach out further into the fluid 
region. This was visually evident, for example, already in the instantaneous 
flow fields presented in figures \ref{fig:lss} and \ref{fig:snap-interface}, 
where low-speed and high-speed velocity streaks are shown to be more fragmented 
throughout their whole wall-normal extension. A similar message is conveyed by 
figure \ref{fig:cor}, that shows  the two-points autocorrelation functions for 
the velocity fluctuations, as a function of either the streamwise or the spanwise 
separation. In the outer region of the channel the curves are almost identical, 
but  at $y^+=5$ (left column) the autocorrelations, although qualitatively similar 
between the two flows, distinctly show how the porous wall affects the spatial 
coherency of the structures, reflected by the elongation of the correlation 
functions. Note that at such distance from the interface the structural differences 
just highlighted by the AIM are already vanished. The fact that spanwise 
correlations are marginally affected, but  longitudinal correlations are 
consistently shorter in the near-wall region, confirms similar results 
already obtained by \cite{breugem-boersma-uittenbogaard-2006} and extends 
them to the present, much lower value of permeability $\sigma=0.004$, for 
which the wall is generally believed to be effectively impermeable.

\begin{figure}
\centering
\includegraphics[width=0.8\textwidth]{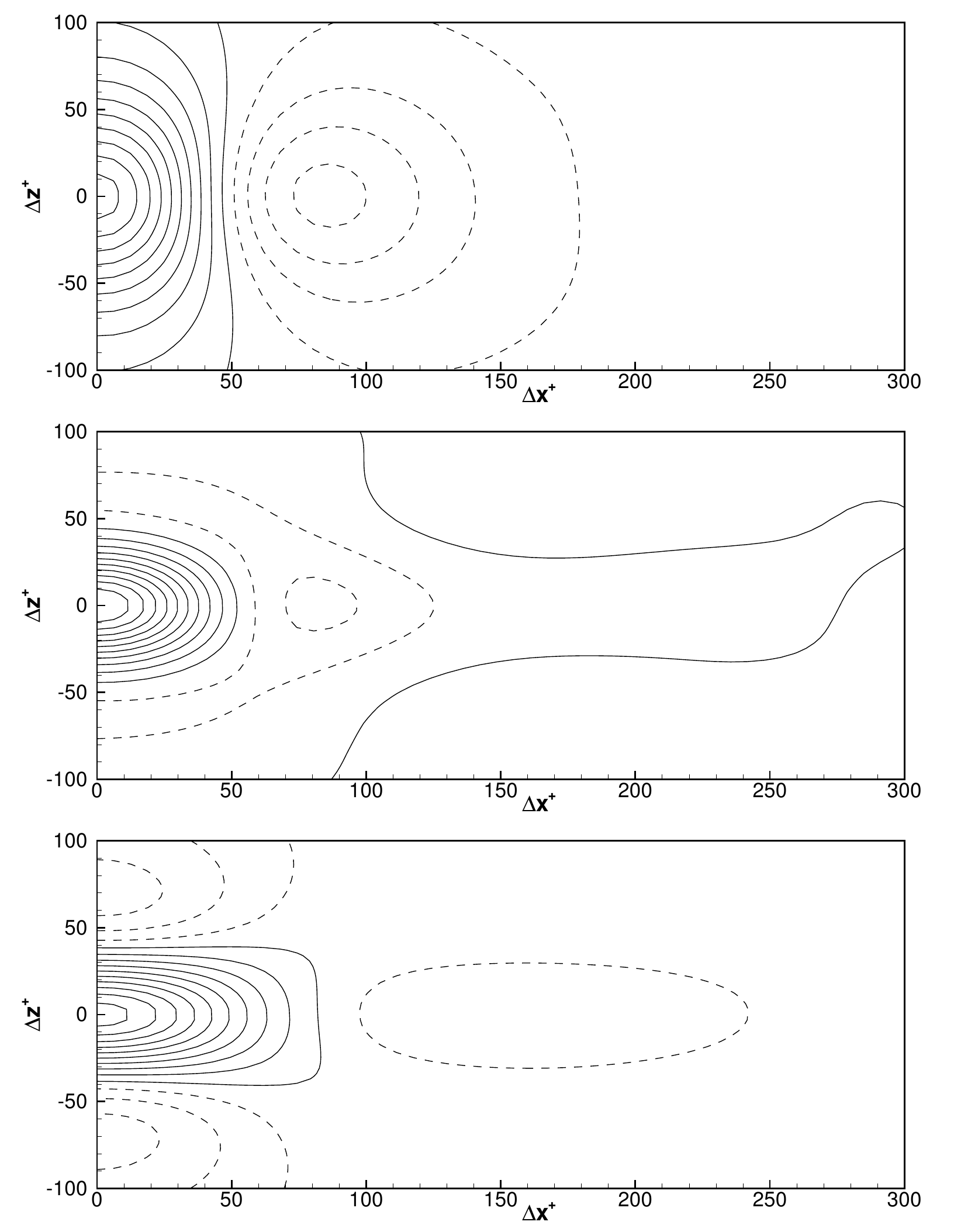}
\caption{Two-dimensional autocorrelation functions for the velocity components $u$ (top), $v$ (middle) and $w$ (bottom), for a turbulent channel flow at $\Rey=2800$, at the center plane $y=-0.1$ of the porous layer. Contours from 0.95 by 0.1 decrements, negative levels are plotted with dashed lines.}
\label{fig:plane_cor}
\end{figure}

The flow inside the porous layer, although characterized by small velocity fluctuations, possesses a well defined structure. We exemplify it with figure \ref{fig:plane_cor}, that features the wall-parallel two-dimensional autocorrelation function for the three velocity components in the middle of the porous layer, i.e. at $y=-0.1$. Although corresponding to small energetic levels, these interesting patterns present relatively large spanwise scales and suggest the presence of a streamwise length scale (distance of the minimum from the origin) of approximatly 90 viscous units (or $x/h \approx 0.5$) for both the streamwise and wall-normal correlations. Even at the present small levels of permeability we thus find a statistical signature of the spanwise vortical structures or rolls that have been already identified over significantly more porous walls \citep{breugem-boersma-uittenbogaard-2006}, rough surfaces \citep{jimenez-etal-2001} and plant canopies \citep{finnigan-2000}, and are suggested \citep{garcia-jimenez-2011-a} to be the main cause of performance loss of drag-reducing riblets when the riblets size is increased above their optimum. The footprints of such structures reach deep into the porous material. 

\begin{figure}
\centering
\includegraphics[width=0.8\textwidth]{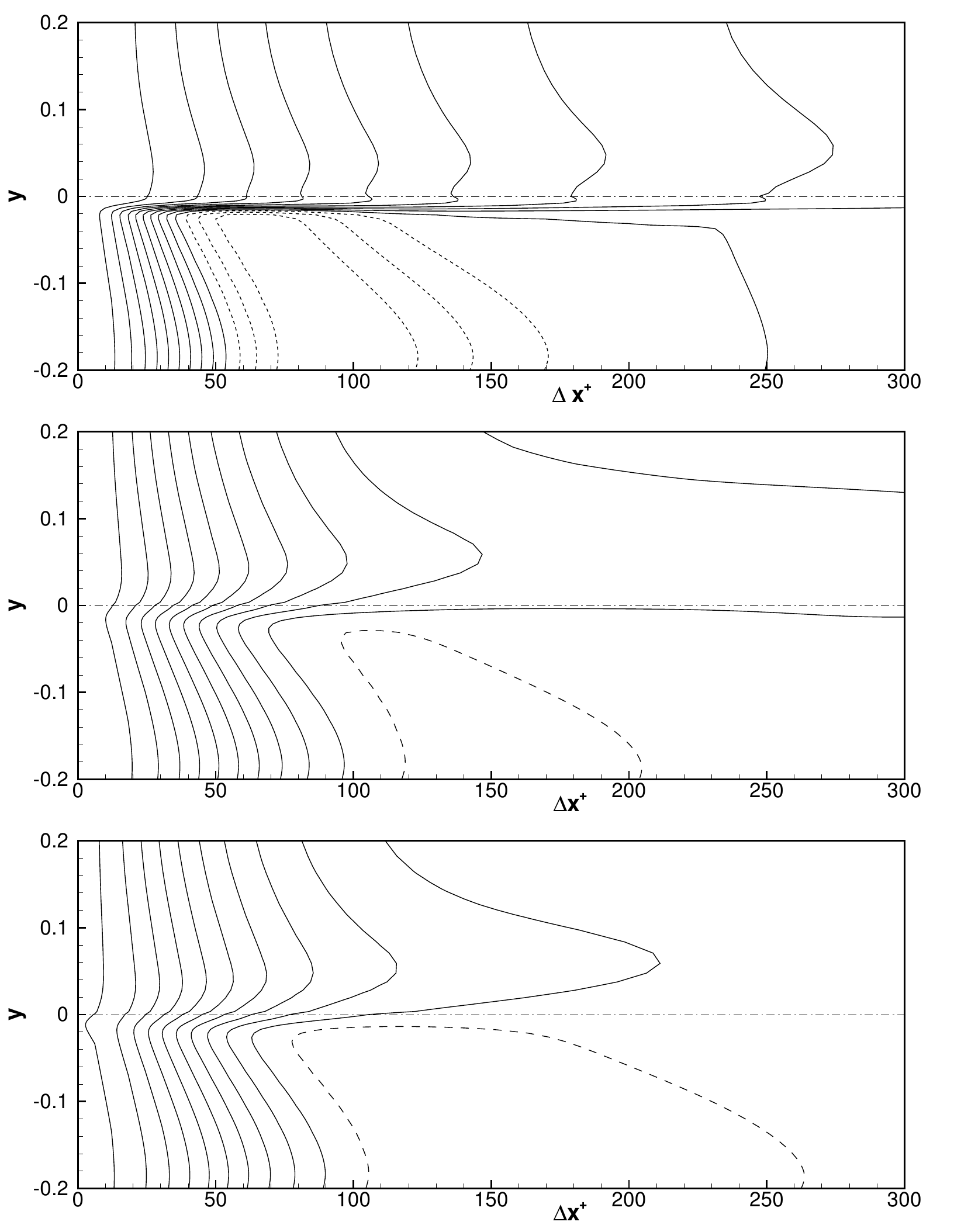}
\caption{One-dimensional autocorrelation function for the velocity components $u$ (top), $v$ (middle) and $w$ (bottom), for a turbulent channel flow at $\Rey=2800$, at zero spanwise and wall-normal separations. The autocorrelations at different $y$ are plotted together, and the dashed-dotted line indicates the position of the interface. Contours from 0.95 by 0.1 decrements, negative levels are plotted with dashed lines.}
\label{fig:x_cor-y}
\end{figure}

The next figure \ref{fig:x_cor-y} illustrates how the structure of the correlation functions change along the wall-normal direction. First, it should be noted explicitly that the figure stacks one-dimensional longitudinal correlations in the $y$ direction. In the channel region the correlations and in particular that for the streamwise component present relatively large lengthscales, related to the elongated streaky structures that characterize the near-wall region, although from figure \ref{fig:snap-interface} one deduces that for the porous case the typical lengthscales are slightly reduced. Then the correlations change rapidly but continuously across the interface, and the lengthscale becomes much shorter for the $u$ component inside the porous material, that cannot support the streak structures, while remaining comparable to the channel lengthscales for the other two components. 
We also recover in the present small-permeability case the result, described by \cite{breugem-boersma-uittenbogaard-2006}, that the correlations present an alternating sequence of progressively weaker negative and positive local peaks. These peaks, although barely noticeable at these levels of permeability, indicate a streamwise periodic organization of the flow structures that \cite{breugem-boersma-uittenbogaard-2006} related to large-scale pressure fluctuations within the porous material.

\subsection{Effect  of porosity parameters}
\label{sec:parameters}

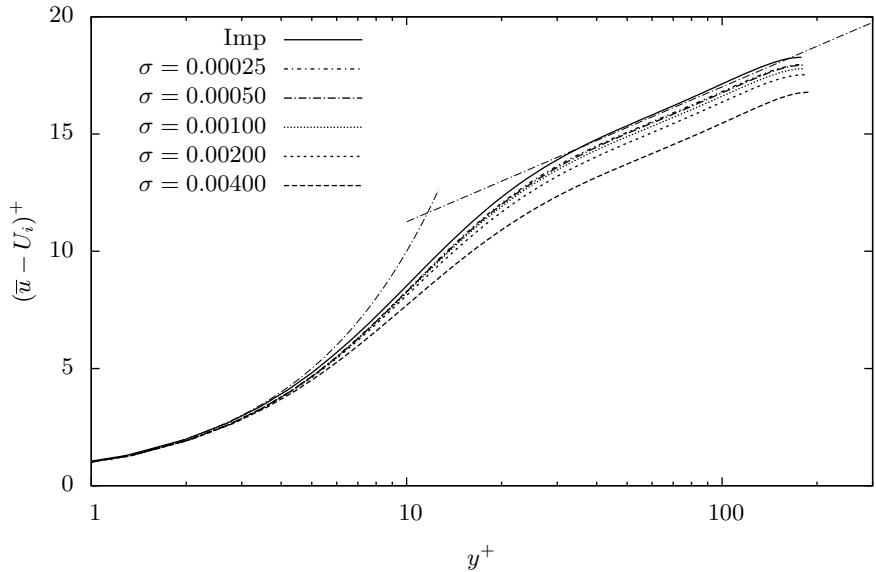
\begin{figure}
\centering
\input{loglaw_sigma}
\caption{Profile of the mean velocity minus the interface velocity, $(\overline{u} - U_i)^+$ plotted versus the distance $y^+$ from the interface. The baseline case with permeability $\sigma=0.004$ (thick dashed line) is compared with decreasing values of $\sigma$ and to the impermeable wall (continuous line).}
\label{fig:loglaw_sigma}
\end{figure}

Permeability $\sigma$ is arguably the most important parameter 
of a porous material. Besides the reference value $\sigma=0.004$, 
four smaller values are considered, as per Table \ref{tab:cases}. 
It can be seen that in all cases the skin-friction (hence the 
value of $Re_\tau$) increases with respect to the reference flow, 
but the increase tends to zero as $\sigma$ becomes smaller. 
Similarly to figure \ref{fig:loglaw}, figure \ref{fig:loglaw_sigma} 
shows the mean velocity profile in law-of-the-wall form, where 
distance $y$ from the interface and mean velocity $(\overline{u} - U_i)$ 
after removal of the interface velocity contribution 
are plotted in wall units. The velocity profile with this choice 
of non-dimensionalization is unchanged in the near-wall portion. 
In the log layer the slope of the profile is unaltered, while the 
intercept increases with decreasing values of $\sigma$ and approaches, 
as expected, $B^+=5.5$ for vanishing $\sigma$. Hence, as the wall 
friction converges towards the impermeable case, the profiles 
monotonically approach the standard wall law that is valid for a 
turbulent channel flow over solid walls. Values of permeability of 
the order of $10^{-4}$ can thus be considered the practical limit 
below which a porous wall behaves like a solid wall in terms of 
mean quantities. 
It is interesting to note that the effects of permeability, at these 
small values, on the mean velocity profile of the turbulent channel 
flow at $Re=2800$ are quantitavely similar to effects of 
permeability on the velocity profile of a laminar pressure driven 
channel flow at $Re=3000$ \citep{tilton-cortelezzi-2008}, where
$\sigma=0.0002$ can be considered the practical limit below which 
a porous wall behaves like a solid wall in terms of linear stability.

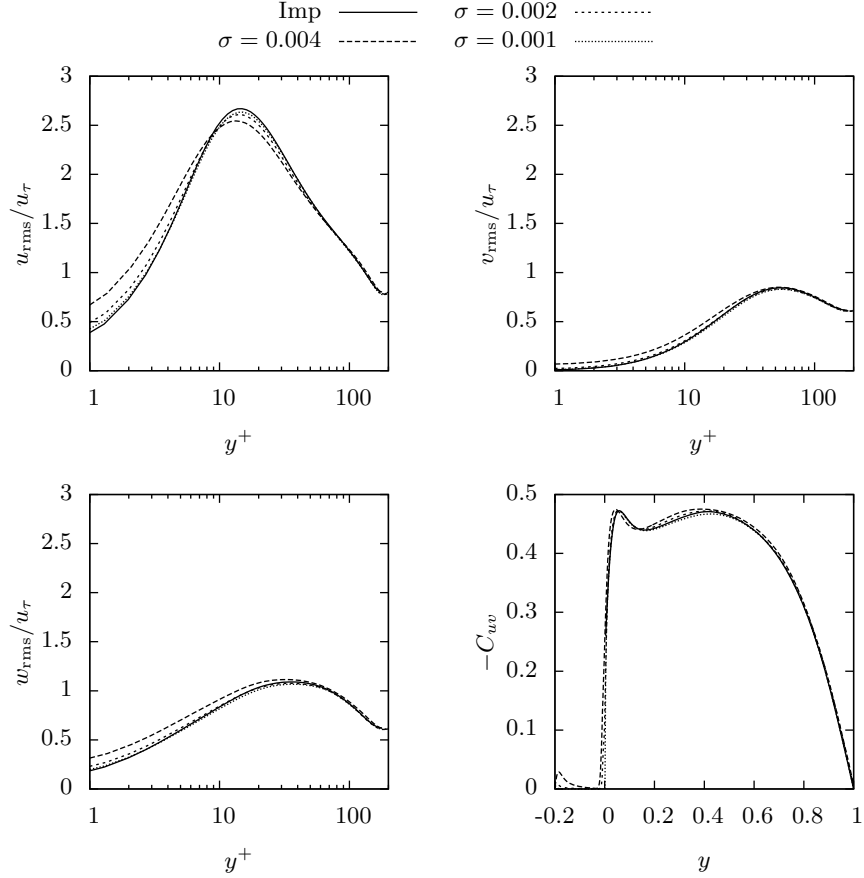
\begin{figure}
\centering
\input{rms_sigma}
\caption{R.m.s. values of velocity fluctuations, expressed in wall 
units, as a function of the distance $y^+$ from the interface, 
and correlation coefficient $C_{uv}$ between the streamwise and 
wall-normal velocity fluctuations. The baseline case with permeability 
$\sigma=0.004$ (thick dashed line) is compared with decreasing values 
of $\sigma$ and to the impermeable wall. The curves for the lowest 
permeability values of $\sigma=0.0005$ and $\sigma=0.00025$ are 
not shown as they overlap the curves for the impermeable case.}
\label{fig:rms_sigma}
\end{figure}

A consistent picture is obtained by plotting the turbulence 
fluctuations, see figure \ref{fig:rms_sigma}. The profiles for the 
r.m.s. values of the velocity components as well as the correlation 
coefficient $-C_{uv}$ confirm that the statistics of the flow do 
converge towards those of the impermeable wall for decreasing values 
of $\sigma$. For these second-order moments, the limiting value of 
permeability below which differences cannot be noticed anymore is 
$\sigma=0.001$, i.e. larger than the limit value for the mean 
velocity profile.

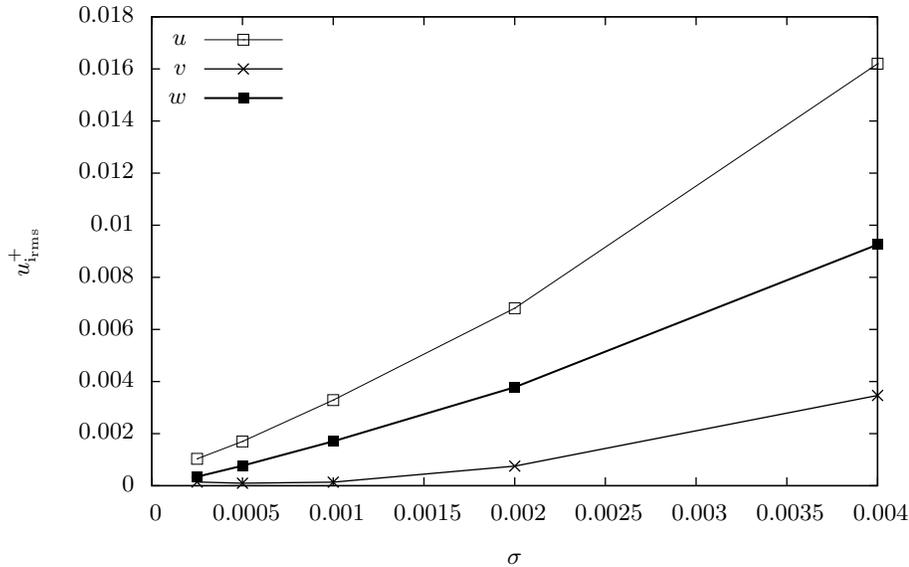
\begin{figure}
\centering
\input{rmsinterface-sigma}
\caption{R.m.s. values of velocity fluctuations at the interface, for different values of permeability $\sigma$.}
\label{fig:rmsinterface-sigma}
\end{figure}

Figure \ref{fig:rmsinterface-sigma} shows, for each component of the velocity vector, how the r.m.s. value of the fluctuations at the interface increases as a function of $\sigma$. A more than linear increase is observed, with the longitudinal component being the largest and the wall-normal component the smallest for a given $\sigma$. In comparing with figures 11 and 12 by \cite{breugem-boersma-uittenbogaard-2006}, similar trends can be observed, but here the fluctuations are two orders of magnitude smaller, owing to the much smaller permeability of the material. 

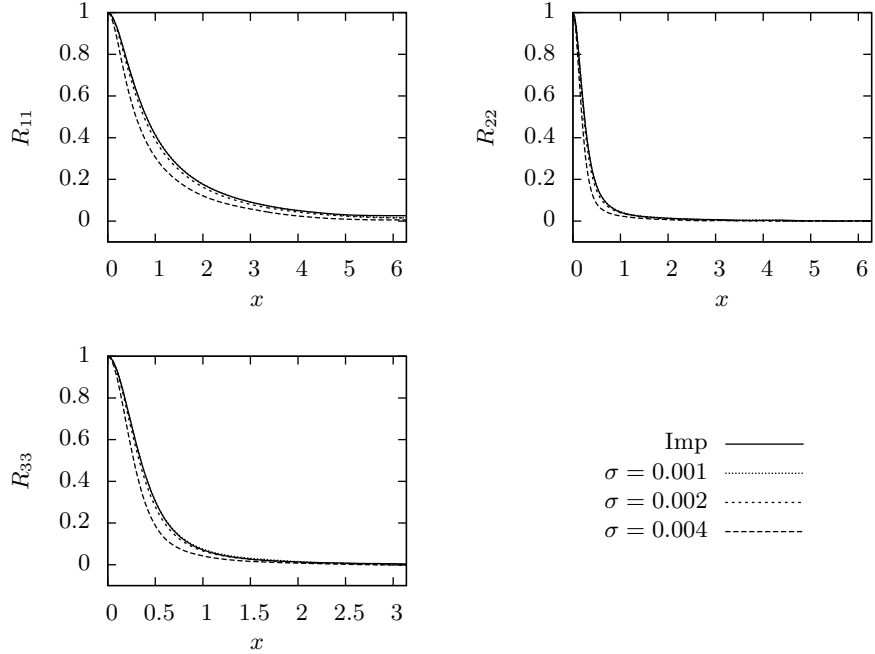
\begin{figure}
\centering
\input{cor_sigma}
\caption{One-dimensional autocorrelation function for the three velocity components, for a turbulent channel flow at $\Rey=2800$, at $y^+ \approx 5$. The baseline case with permeability $\sigma=0.004$ (thick dashed line) is compared to cases with decreasing values of $\sigma$ and to the impermeable wall. The curves for the lowest permeability values of $\sigma=0.0005$ and $\sigma=0.00025$ are not shown as they overlap the curves for the impermeable case.}
\label{fig:cor_sigma}
\end{figure}

The two-point velocity correlations are pictured in 
figure \ref{fig:cor_sigma}. In particular, even the 
longitudinal autocorrelation function for the streamwise 
velocity fluctuations, that is the most affected by 
the porous material as shown in figure \ref{fig:cor},  
becomes indistinguishable from the reference case of a 
channel flow over an impermeable wall for $\sigma<0.001$, 
thus confirming that this can be taken as the practical 
limit below which porosity effects are negligible, as 
far as second moments are concerned.

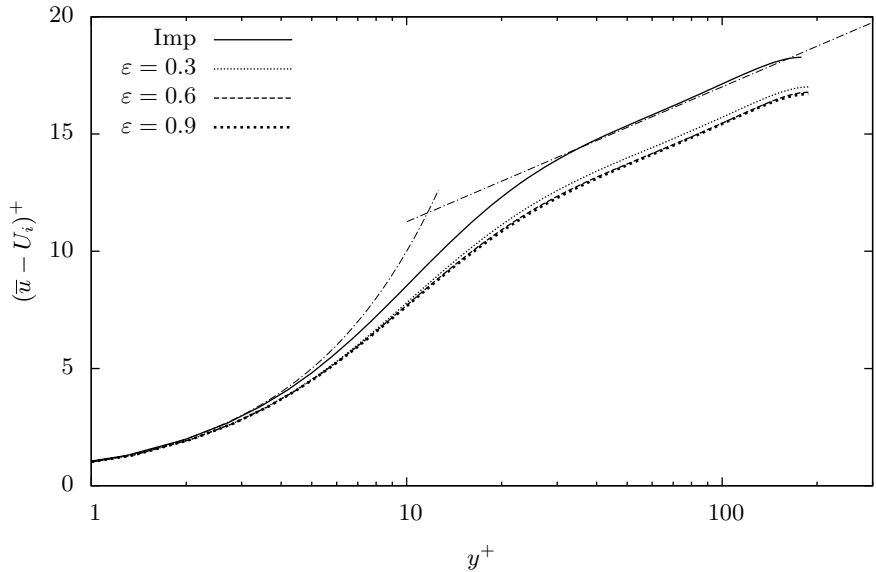
\begin{figure}
\centering
\input{loglaw_epsilon}
\caption{Profile of the mean velocity minus the interface velocity, 
$(\overline{u} - U_i)^+$ plotted versus the distance from the 
interface $y^+$. The baseline case with porosity $\varepsilon=0.6$ 
(dashed line) is compared to larger, 0.9 (thick dotted line), and 
smaller, 0.3 (thin dotted line), values of $\varepsilon$ and to 
the impermeable wall (solid line).}
\label{fig:loglaw_epsilon}
\end{figure}

Aside from permeability, porosity, $\varepsilon$, is the other 
parameter that characterizes a porous material. At odds with 
permeability, the amount of porosity does not affect turbulence 
statistics in a significant way. A quick look at the mean 
velocity profiles in figure \ref{fig:loglaw_epsilon} shows 
that further increasing $\varepsilon$ above the baseline 
value of $\varepsilon=0.6$ has little or no effect on the 
mean profile, whereas halving the porosity value shifts the 
logarithmic part of the velocity profile towards the one for 
the impermeable wall, but the shift is minimal. A similar 
picture emerges after examining other quantities (not shown), 
like the wall-normal profiles of velocity fluctuations, 
correlations and spectra. Overall, this suggests that the 
limiting behaviour of $\varepsilon$ and $\sigma$ are quite 
different. 
The effects of porosity, at these small values of permeability, 
on the mean velocity profile of the turbulent channel 
flow at $Re=2800$ are again quantitavely similar to effects of 
porosity on the velocity profile of a laminar pressure driven 
channel flow at $Re=3000$ \citep{tilton-cortelezzi-2008}.

The effects of the momentum-transfer coefficient $\tau$ on
the turbulence statistics are now considered.
To understand the importance of the role played by
the parameter $\tau$, one should envision the changes
in the values of $\sigma$ and $\epsilon$ as one moves
from deep down in the porous material, up to the
fluid/porous interface and into the fluid. Deep inside the
porous materials the values of $\sigma$ and $\epsilon$
are the bulk values representative of the porous
material chosen. Approaching the interface the
values of $\sigma$ and $\epsilon$ change rapidly
from their bulk values to $\sigma=\infty$ and
$\epsilon=1$ in the fluid just above the interface.
The change takes place in the porous material in
a thin transition layer adjacent to the interface
and the functional variation of $\sigma$ and $\epsilon$
depends mainly on the porosity of the material and on
how the interface has been machined (see figure \ref{fig:porous-sample}).

In this study, however, we focus on porous materials of small 
permeabilities and small mean particle sizes. In particular, for sufficiently 
low permeabilities, the fluid velocity at the porous/fluid interface 
is small; the convective effects, and consequently the drag force 
experienced by the fluid, become negligible, because the dense channel-like 
structures of the porous matrix impede motion between layers of fluid. In this 
case, the transition region (of the order of a few pore diameters thick) and
roughness (of the order of one pore diameter high) can be assumed to 
have zero thickness, and porosity and permeability to have a constant 
value up to the interface that has a clearly defined position 
\citep{ochoatapia-whitaker-1995a, ochoatapia-whitaker-1998, valdesparada-goyeau-ochoatapia-2007, valdesparada-etal-2013}.
Therefore, following \cite{ochoatapia-whitaker-1995a}, we model
the effects of the variation of $\sigma$ and $\epsilon$ in
the transition layer using the momentum-transfer coefficient $\tau$
(see the interface conditions \eqref{eq:BC_v_2} and
\eqref{eq:BC_eta_1}). The parameter $\tau$, as suggested by 
\cite{minale-2014a,minale-2014b}, models the transfer of stress at the 
porous/fluid interface. This dimensionless parameter is of order one 
and can be both positive or negative, depending on the type of porous 
material considered and the type of machining of the interface 
\citep{ochoatapia-whitaker-1998}. In the case when $\tau=0$, 
the stress carried by the free flowing  fluid is fully transferred to the 
fluid saturating the porous matrix, a negative $\tau$ quantifies the amount 
of stress transfered from the free fluid to the porous matrix, while 
a positive $\tau$ quantifies the amount of stress transfered from the 
porous matrix to the free fluid. Note that, in our model, the 
momentum transfer coefficient is decoupled from both porosity
and permeability and, therefore, its effects can be assessed
independently. In order to model a wide range of porous materials 
and surface machining, we follow \cite{tilton-cortelezzi-2008}
and discuss the effects of the momentum transfer coefficient as 
it varies between -1 and +1. In particular, we consider four
values of $\tau=$ -1, 0, 0.5 and 1, because \cite{tilton-cortelezzi-2008}
showed that positive values of $\tau$ have a strong nonlinear effect on 
the linear stability of a pressure driven flow in a channel with porous
wall. 

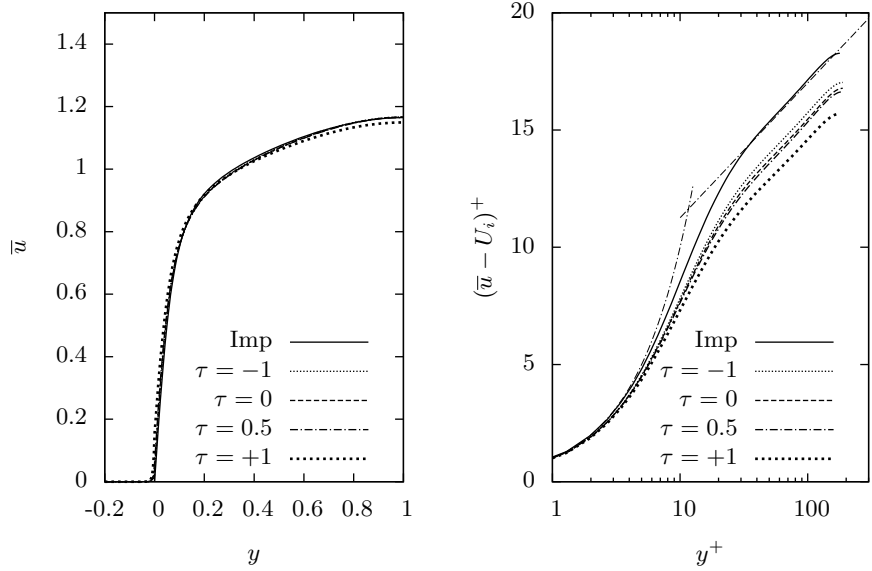
\begin{figure}
\centering
\input{loglaw_tau}
\caption{Profile of the mean velocity minus the interface velocity, 
$(\overline{u} - U_i)^+$ plotted versus the distance $y^+$ from 
the interface. The baseline case with momentum-transfer coefficient 
$\tau=0$ (dashed line) is compared to to the impermeable wall 
(solid line), and to the cases with $\tau=-1$, $\tau=0.5$ and $\tau=+1$.}
\label{fig:loglaw_tau}
\end{figure}

Figure \ref{fig:loglaw_tau} presents the mean velocity profiles 
for the four considered values of the momentum transfer coefficient. 
The nonlinear effects of $\tau$ on the mean velocity profile are 
quite evident: while the mean velocity profiles for the cases
$\tau=$ -1, 0, and 0.5 are grouped together, the case $\tau=1$ 
induces a significant increase of the mean profile at the interface, 
and a consequent decrease in the central region of the channel owing 
to the constant flow rate constraint. The effects of $\tau$, at these 
small values of permeability, on the mean velocity profile of the 
turbulent channel flow at $Re=2800$ are again quantitavely similar 
to effects of $\tau$ on the velocity profile of a laminar pressure 
driven channel flow at $Re=3000$ \citep{tilton-cortelezzi-2008}.
Also the wall-normal derivatives 
of the mean velocity profile at the interface are significantly and
nonlinearly affected by the value of $\tau$, with the friction velocity 
$u_\tau$ being larger when $\tau=-1$ and smaller when $\tau=0.5$
and much smaller when $\tau=1$ compared to the baseline case ($\tau=0$). 
As reported in Table \ref{tab:cases}, $\tau=1$ yields a 
wall friction which is down to the value of the impermeable wall.
This unexpected result is supported by further information presented
below. Note, however, that for $\tau=1$ the interfacial velocity 
is about four times larger than the baseline case, and this could hinder 
the linearity assumption in the porous layer, therefore, quantitative 
considerations related to this case should be regarded with some 
caution.

The right frame of figure \ref{fig:loglaw_tau} shows the mean 
velocities profiles, in wall units, on a logarithmic scale. 
When $\tau=0.5$ and $1$ the velocity profile in the logarithmic 
region decreases nonlinearly, while the negative $\tau$ raises it. 
It must be observed that interpreting the downward shift of the 
log law as an indication of increased friction drag is at odds with 
the values of $Re_\tau$ reported in Table \ref{tab:cases}, 
where one would expect for example the case with $\tau=+1$ 
and $Re_\tau=177.9$ being nearer to the reference flow with 
$Re_\tau=178.5$ than the cases with $\tau=-1$ and $Re_\tau=188.3$. 
This is not the case in figure \ref{fig:loglaw_tau} because 
of the interface velocty $U_i$ being subtracted from the mean 
velocity profile.

\begin{figure}
\centering
\input{rms_tau}
\caption{R.m.s. values of velocity fluctuations, expressed in wall units, as a function of the distance from the interface $y^+$, and correlation coefficient $C_{uv}$ between the streamwise and wall-normal velocity fluctuations. The baseline case with momentum-transfer coefficient $\tau=0$ (dashed line) is compared to $\tau=1$ (thick dotted line) and $\tau=-1$ (thin dotted line) and to the impermeable wall (solid line).}
\label{fig:rms_tau}
\end{figure}
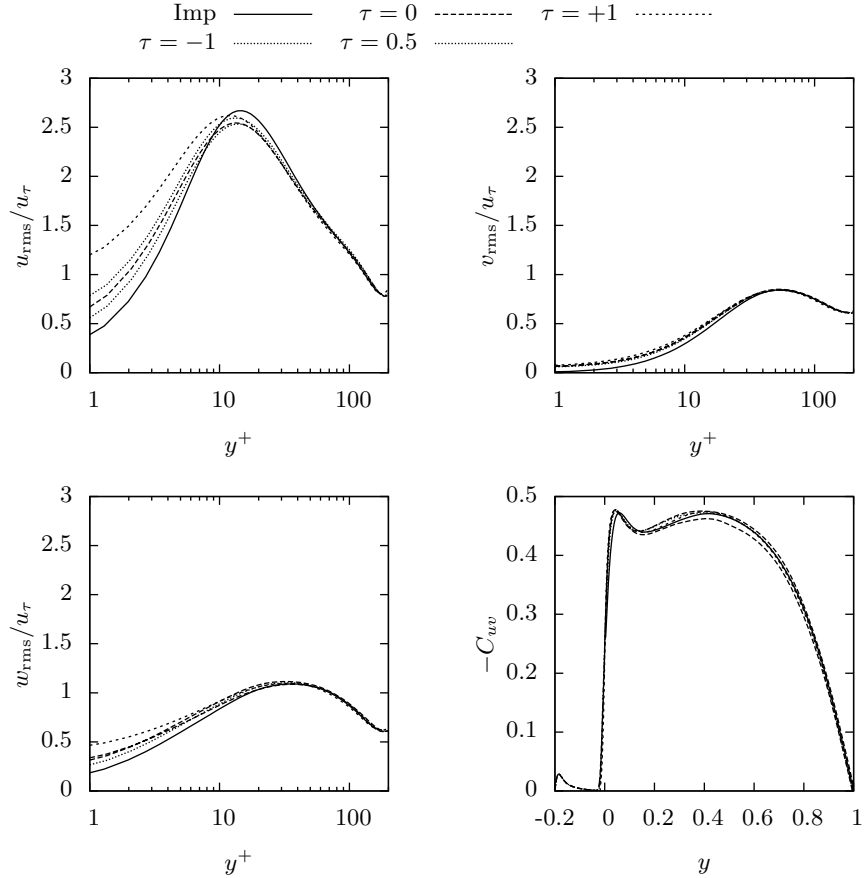

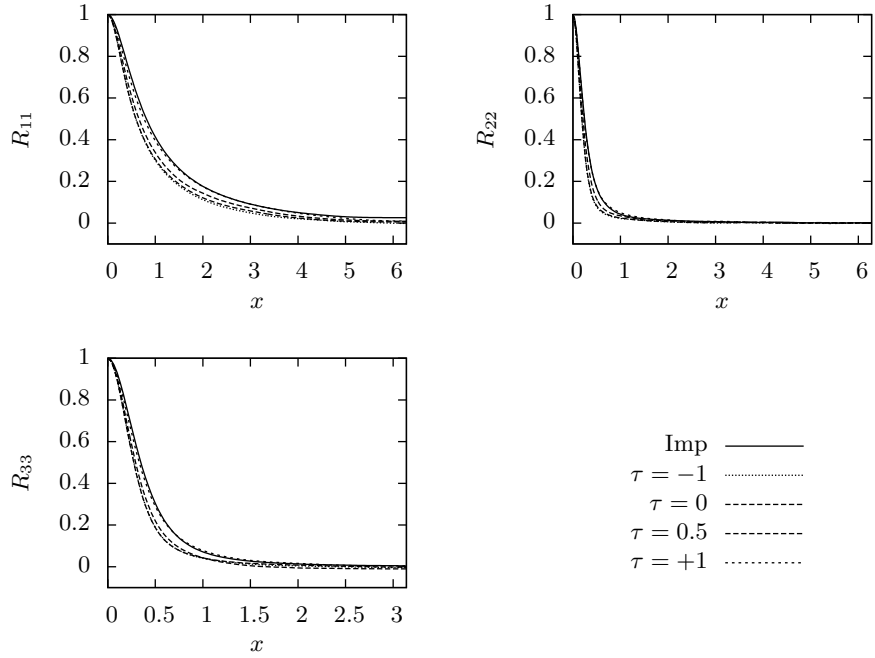
\begin{figure}
\centering
\input{cor_tau}
\caption{One-dimensional autocorrelation function for the three velocity components, for a turbulent channel flow at $\Rey=2800$, at $y^+ \approx 5$. The baseline case with $\tau=0$ (thick dashed line) is compared to $\tau=-1$ (dotted line) and $\tau=1$ (thin dashed line) to the impermeable wall (continuous line). The case with $\tau=1$ is almost identical to the impermeable case.}
\label{fig:cor_tau}
\end{figure}

Figure \ref{fig:rms_tau} shows how the Reynolds stresses are 
affected by the momentum transfer coefficient. We observe that the 
positive values of $\tau$ induce a significant and nonlinear 
increase of the $u'$ and $w'$ profiles with respect to the 
baseline case, whereas $v'$ changes little and in the opposite 
direction. The negative value of $\tau$ is observed to have an 
opposite but weaker effect. The case with $\tau=1$ is also notable 
as the streamwise correlations, shown in figure \ref{fig:cor_tau}, 
turn out to be more elongated along the streamwise direction, and 
to overlap the curves of the impermeable case.

\begin{figure}
\includegraphics[width=\textwidth]{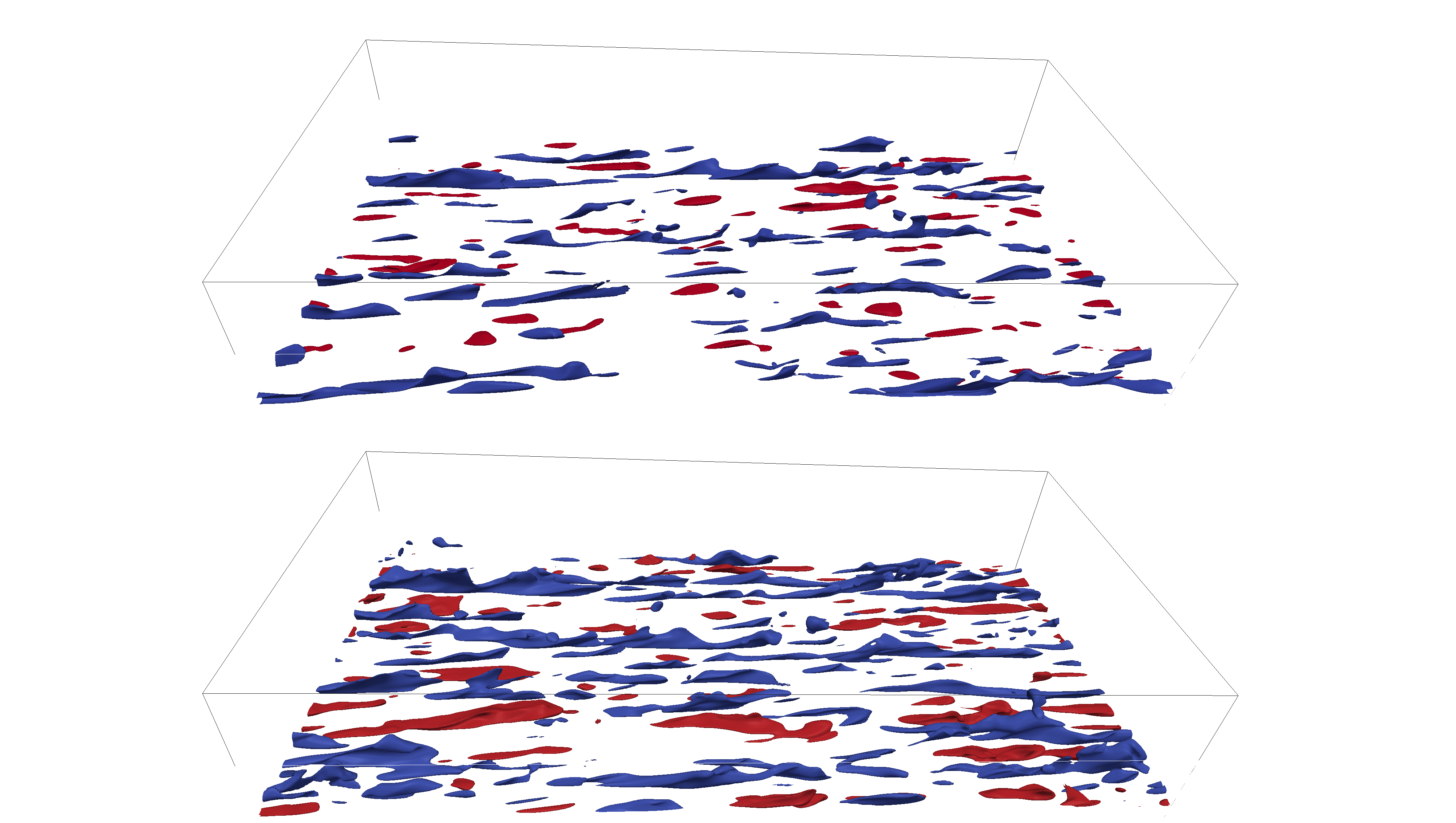}
\caption{Low- and high-velocity near-wall streaks for the case with $\tau=-1$ (top) and $\tau=+1$ (bottom). The streaks are visualized via blue / red isosurfaces for $u'^+=\pm 4$ for the lower half of the channel. Streaks are shorter and more fragmented for $\tau=-1$.}
\label{fig:lss_tau}
\end{figure}

Figure \ref{fig:lss_tau} presents instantaneous snapshots 
of the low- and high-speed streaks for the cases $\tau=-1$ and 
$\tau=+1$. The comparison of the streaks for the cases $\tau=0$
(figure \ref{fig:lss}) and $\tau=-1$  confirms that 
$\tau=-1$ produces more fragmented structures in the near-wall 
region, leading to shorter correlation lengths. Recall that the 
case $\tau=0$, our baseline case shown in figure \ref{fig:lss}, 
models the situation where the total stress carried by the fluid 
freely flowing over the interface is fully transferred to the 
fluid saturating the porous matrix while no stress is transfered 
to or from the porous matrix. In this case, the streak structures 
are shorter and more fragmented than in the impermeable case, 
especially those parts of the low-speed streaks that are located 
near the porous/fluid interface. In the case 
$\tau = -1$, part of the stress carried by the fluid freely 
flowing over the porous/fluid interface is transferred to 
the porous matrix descreasing the average interface velocity, $U_i$, 
from 0.0384 ($\tau=0$) to 0.0215. The apparent effect on 
near-wall turbulence is to increase further the 
fragmentation of the near wall streaks. In the case $\tau = +1$, instead,
part of the stress accumulated in the porous matrix is transferred 
to the fluid freely flowing over the porous/fluid interface 
increasing the average interface velocity, $U_i$, 
from 0.0384 ($\tau=0$) to 0.1516. The structure of the near 
wall streaks apprears to be very similar to the structure 
of the impermeable case (shown in figure \ref{fig:lss}). 
Apparently, at these small values of permeability, the high 
interface velocity inhibits the effect of permeability restoring 
the near wall streaks structure typical of the impermeable case. 
However, considering the high interface velocity, the case 
$\tau=+1$ seems to mimick the flow over a porous ``hydrophobic''
material. 

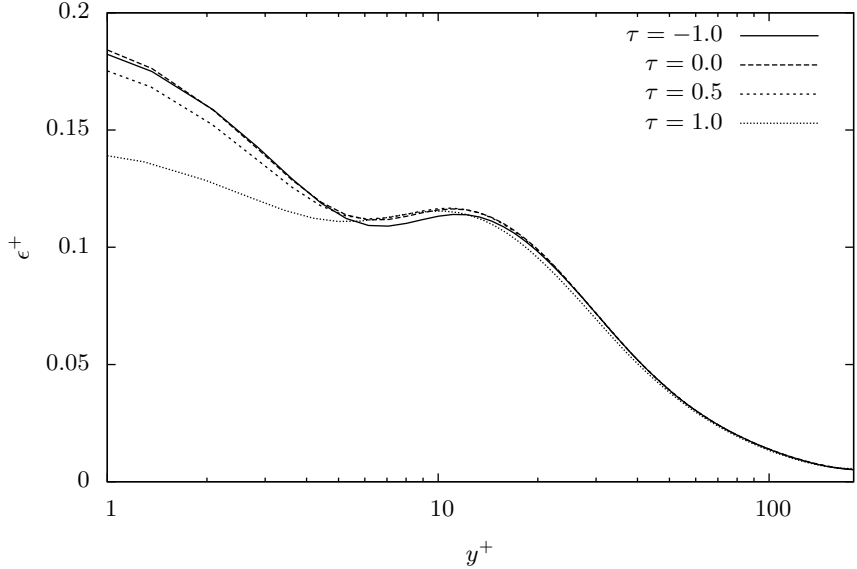
\begin{figure}
\centering
\input{diss_tau} 
\caption{Wall-normal profile in wall units of the 
dissipation $\epsilon^+$ of the fluctuating field, for a turbulent 
channel flow at $\Rey=2800$ with porous walls having the same
porosity ($\varepsilon=0.6$) and permeability ($\sigma=0.004$) 
but characterized by different values of the momentum transfer 
coefficient: $\tau=-1$ (solid line), 0 (long-dashed line),
0.5 (short-dashed line) and 1 (dotted line).}
\label{fig:diss_tau}
\end{figure}

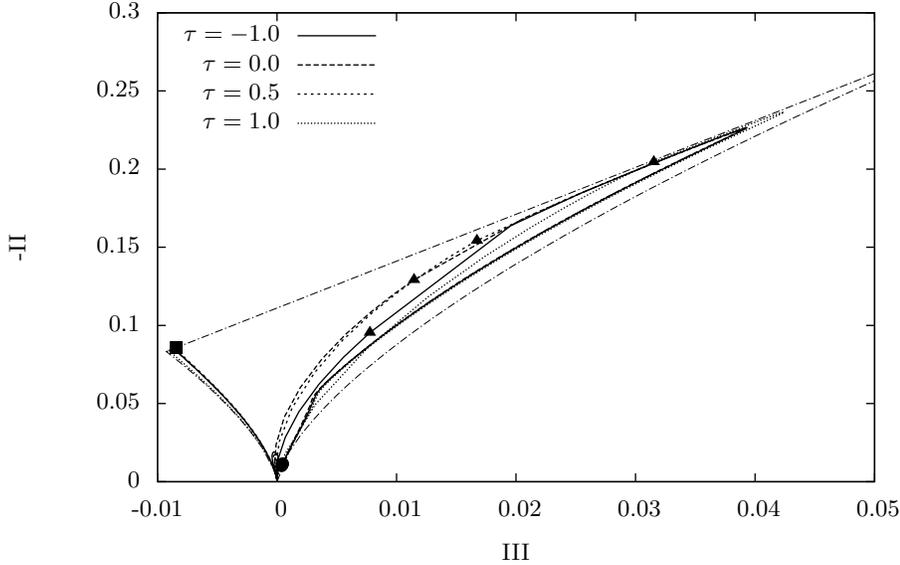
\begin{figure}
\centering
\input{lumley_tau}
\caption{The Lumley triangle, for a turbulent channel flow at $\Rey=2800$
with porous walls having the same porosity ($\varepsilon=0.6$) and 
permeability ($\sigma=0.004$) but characterized by different values 
of the momentum transfer coefficient: $\tau=-1$ (solid line), 
0 (long-dashed line), +0.5 (short-dashed line) and +1 (dotted line). 
$-II$ and $III$ are the second and third invariant of the Reynolds stress 
anisotropy tensor, and the bounding line shows the region of admissible 
turbulence states, amid the extremal states of one-component turbulence 
(1C), two-components turbulence (2C), axisymmetric turbulence (axi), 
two-component axisymmetric turbulence (2C axi) and isotropic turbulence (iso). 
The triangles correspond to the porous/fluid interface ($y=0$) and 
the square to the impermeable wall ($y=-h_p$). The circle marks the 
nearly isotropic state, which is visited by the flow at the center
of the channel ($y=1$) and, for the porous case, in the bulk of the 
porous layer.}
\label{fig:lumley_tau}
\end{figure}

Figure \ref{fig:diss_tau} presents, in wall units, the wall-normal profile 
of the dissipation $\epsilon = \mu \overline{\partial_j u'_i \partial_j u'_i}$ 
of the fluctuating velocity field for the cases $\tau=-1$, $0$, $+0.5$ and 1.
Although the overall trend of the curves is similar and resembles that 
of the impermeable case (figure \ref{fig:diss} left), figure \ref{fig:diss_tau}
shows substantial differences, for positive values of $\tau$, in near-wall 
dissipation up to about $y^+=5$, where all the curves tend to group together 
and stay together up to the center of the channel where they finally merge. 
The curves for the cases $\tau=-1$ and $0$ are almost identical, with
the case $\tau=0$ showing a slightly higher dissipation near the interface.
Apparently, the higher fragmentation observed in the low- and high-speed 
streaks does not have an impact on the near-wall viscous dissipation.
The dissipation tends to decrease for positive values of $\tau$.
For $\tau=+0.5$, the dissipation at the interface is about 3\% lower than
the baseline case ($\tau=0$), while for $\tau=+1$ it is about 28\% 
lower, underlining a strong nonlinear effect of the positive values of $\tau$. 

An overall picture of the differences in the structure of
the turbulent fields for the four different values of $\tau$
(-1, 0, +0.5, +1) is provided by the AIM, or Lumley triangle, 
shown in figure \ref{fig:lumley_tau}. All four curves start from 
a nearly isotropic state indicated  by the solid circles (corresponding 
to the center of the channel) located near the origin of the AIM. 
As the wall-normal position moves from the center of the channel 
toward the porous/fluid interface, the curves move all together 
along the lower side of the triangle corresponding to the axisymmetric 
state ($III>0$), or rod-like turbulence. As the curves approach 
the right vertex of the triangle, the curve for $\tau=+1$ more 
than the others, they turn and start moving along the upper side 
of the triangle corresponding to two-component turbulence. As 
the wall-normal position approaches the porous/fluid interface, 
the turbulence structure in the four cases becomes substantially 
different, as indicated by the well separated locations of the 
solid triangle symbols representing the turbulent states at the 
interface. As the wall-normal position approaches and crosses
the interface and moves into the porous region, the curves follow
different trajectories as they visit again the isotropic state
(corresponding to the center of the porous layer) located at 
the origin of the AIM. The curve for the case $\tau=+1$ is the only 
one that crosses itself, at around $III=0.1$, and overlaps itself 
as it approches the isotropic state indicating that some turbulent 
states are realized both in the channel and within the porous layer. 
The curves for the cases $\tau=+0.5$ and $0$ appear to be almost 
identical. The curve for the case $\tau=-1$, is the one 
that encounters the interface closest to the origin of the AIM
(see solid triangle located at about $III=0.08$). As the curves 
revisit and depart from the isotropic state they all form a small 
loop, the largest for the case $\tau=-1$ the smallest for $\tau=+1$. 
Finally, as the wall-normal position approaches the impermeable wall 
that seals the porous layer, all the curves follow the right side 
of the triangle corresponding to the axysimmetric state ($III<0$), 
or disk-like turbulence, and end at the left vertex of the triangle, 
corresponding to the two-component axysimmetric state.

Figures \ref{fig:lss_tau}--\ref{fig:lumley_tau} provide concrete evidence 
of the substantial and nonlinear impact of the momentum transfer coefficient
$\tau$ on near-wall turbulence. These results emphasize the impact
of the machining of the interface (see figure \ref{fig:porous-sample})
on the transport phenomena in the transition layer and on the structure
of near-wall turbulence.

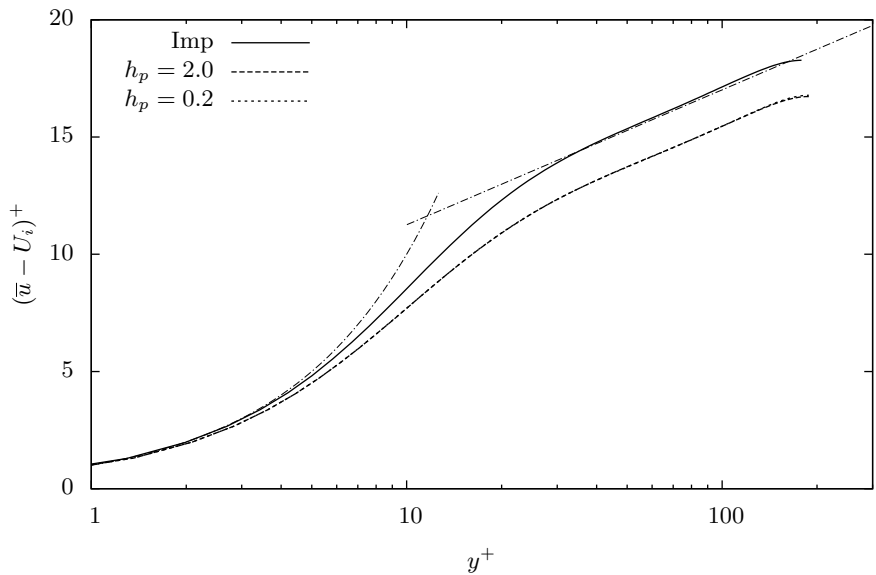
\begin{figure}
\centering
\input{loglaw_hp}

\caption{Profile of the mean velocity minus the interface velocity, $(\overline{u} -U_i)^+$ plotted versus the distance from the interface $y^+$. The baseline case with thickness $h_p=0.2$ (dashed line) is compared to $h_p=2$ (dotted line) and to the impermeable wall (solid line).}
\label{fig:loglaw_hp}
\end{figure}

The last geometric parameter that we consider is the 
thickness $h_p$ of the porous layer. In the baseline 
case the thickness of the porous walls is $h_p=0.2$,
i.e. 10\% of the channel height, a thickness which is
sufficient to avoid the interaction between the two 
boundary layers created in the transition layer 
and at the impermeable wall \citep{quadrio-etal-2013}. Hence, 
for $h_p=0.2$, the velocity profile within the 
porous material has a portion where it remains 
essentially constant at the Darcy value (see figure \ref{fig:mean}). 
In order to asses the effect of $h_p$ on our turbulent
flow, we compare the results of the baseline case with
the results of a DNS where the thickness of the porous
walls is 10 times higher, i.e. $h_p=2$.
Figure \ref{fig:loglaw_hp} presents the mean velocity profiles
for the case $h_p=0.2$ (baseline case, dashed line) and 2 
(dotted line) and clearly shows that a much larger thickness 
does not bring further changes in the turbulent flow 
within the channel and to its statistics. This observation 
is supported by the values of various flow parameters reported 
in Table \ref{tab:cases}, as well as by the higher-order 
flow statistics (not shown). Note that this result appears 
to be at odds with results from linear stability theory. Our 
validation cases, for example, plotted in figure \ref{fig:validation}, 
shows the drastic change in the stability properties of the 
flow, for the considered parameters, when passing from $h_p=0.2$ 
(stable) to $h_p=1$ and $h_p=2$ (unstable). 
These results are consistent with those reported by 
\cite{tilton-cortelezzi-2008} in their linear stability
analysis of a pressure driven channel flow.
They show that increasing the thickness of a porous wall,
thick enough to avoid the interaction between the two
boundary layers, does not affect the laminar velocity
profile. On the other hand, they showed that a thicker
porous wall allows for a larger penetration in the porous
layer of the wall-normal velocity perturbation making
the flow more linearly unstable.

\begin{figure}
\centering
\input{loglaw_re}
\caption{Profile of the mean velocity minus the interface velocity, 
$(\overline{u} - U_i)^+$ plotted versus the distance $y^+$ from the 
interface. All the simulations carried out in the smaller domain 
are considered: the two cases at $Re=2800$ with permeabilities 
$\sigma=0.004$ and $\sigma=0.002$ are compared to the case at 
$Re=6250$ and $\sigma=0.002$. The values of $Re_K$ are $0.753$, 
$0.366$ and $0.754$ respectively.}
\label{fig:loglaw_re}
\end{figure}
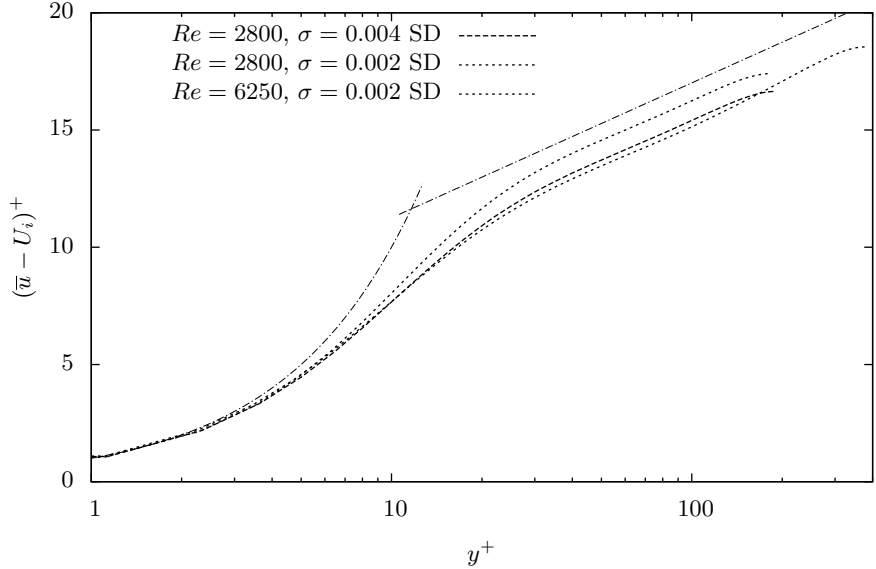

This Section ends with a brief discussion of the effects of another 
key parameter, namely the Reynolds number of the flow. Unfortunately, 
the numerical tool developed within the present work, although computationally 
efficient, only possesses shared-memory parallel capabilities. Thus, 
massive simulations cannot be easily afforded. However, we have been 
able to simulate one case at approximately twice the value of $Re_\tau$ 
of the baseline case, with a view to identify the main scaling laws. 
Since we had to compromise on the size of the computational domain 
to keep the computing time manageable, a pair of low-$Re$ cases, 
including the baseline one, have been recomputed in the smaller 
computational domain to enable a meaningful comparison. The main 
outcome of this study is that the dominant parameter to describe 
the effects of the porous material is the Reynolds number $Re_K$
\citep{breugem-boersma-uittenbogaard-2006}, defined as
\[
Re_K = \frac{\sqrt{K^*} u_\tau^*}{\nu^*}.
\]
In other words, two porous materials show the same behaviour once 
the porosity-related length scale $\sqrt{K^*}$, expressed in wall 
units, is constant. Indeed, figure \ref{fig:loglaw_re} shows how the 
mean velocity profile at higher $Re$ is quite far from the low-$Re$ 
case at the same value of $\sigma$ (which by definition is the 
lengthscale $\sqrt{K^*}$ normalized by $h^*$) but is very near to 
the low-$Re$ case at similar $Re_K$. Similar results hold for the 
interfacial velocity $U_i$. From Table \ref{tab:cases} one notices 
how its values in outer units differ, but they collapse when expressed 
in inner units and for cases with the same $Re_K$. The two cases with 
$Re_K=0.75$ have $U_i^+=0.57$ and $U_i^+=0.58$ respectively, and  
also the lengthscale $\delta$ appears to scale well in wall units, with 
$\delta^+=0.48$ and $\delta^+=0.45$ respectively.

\section{Concluding summary}
\label{sec:conclusions}

We have carried out a number of direct numerical simulations of 
turbulent channel flow over two porous walls. The flow inside 
the fluid region is described by the Navier--Stokes equations, 
while the Volume-Averaged Naviers--Stokes (VANS) equations are 
solved inside the porous layers. The two sets of equations are 
coupled at the interface between fluid and porous material via 
the momentum transfer conditions, featuring a coefficient $\tau$ 
that accounts for the effects of machining the interface. As the 
study is motivated by our interest in porous material with 
relatively small permeability, inertial effects are neglected 
in the porous material, resulting in considerable simplifications 
to the VANS equations and to the interface conditions.

Our formulation allows treating porosity $\varepsilon$ and 
permeability $\sigma$ as two indipendent parameters. The 
efficiency of our computer code has allowed us to run several 
simulations and to collect well-converged statistics by varying 
the values of all the parameters involved and assessing the 
sensitivity of the flow statistics to each of them. 

We have found that, once permeability and porosity are 
decoupled, which is made possible in our formulation, the 
permeability $\sigma$ emerges as the key parameter, although 
the uncertainty in the machining of the interface, expressed 
by the momentum transfer coefficient $\tau$, is noticeably 
important too, and is related to the porosity of the material. 
The turbulent flow in the channel is affected by the porous 
wall even at very low values of permeability, where non-zero 
values of vertical velocity at the interface can still influence 
the flow dynamics. The penetration depth of the turbulent motions 
within the porous slab does not appear to primarily depend upon 
$\sigma$, hence the porous layer is permeated by a non-trivial 
flow with a structure that we have given a statistical description. 

Our results provide insight about the design of novel porous
materials of small permeabilities and small mean particle size
for industrial applications. Two are the main aspects to
keep into account when designing a novel porous material:
its permeability and its mechanical response to the
stress generated by the free flowing fluid over a porous/fluid
interface. Since the effects of permeability are dominant 
with respect to the effects of porosity, in designing a 
porous material, one should focus on obtaining the correct 
permeability while having substantial freedom in choosing 
the porosity. This freedom allows one to choose the composition
and structure of the porous material that better fits the
mechanical requirement of the potential application. In
other words, depending on the application, one can design
materials that absorb stress from the free flowing fluid
at the interface or materials that behave in the opposite way. 
Great care should be put in machining the interface because
it can contribute to the transfer of stress from and to
the porous matrix and enhance or mitigate the efficiency 
of the transport phenomena at the interface.

\section*{Acknowledgments}
One of the authors, L.C., acknowledges the financial support 
provided by NSERC under Contract No. RGPIN217169, some enlighting
discussion with Prof. Nils Tilton, and the kind 
hospitality of the Department of Aerospace Science and Technologies 
of the Politecnico di Milano where he spent a recent sabbatical leave 
and where this study has been conceived and performed. Another
author, M.E.R., acknowledges the computational facilities made available by the Center
for Intelligent Machines of McGill University for performing the direct numerical 
simulations presented in this study.  

\bibliographystyle{jfm}

\end{document}

%% file: validation.tex
\begingroup
  \makeatletter
  \providecommand\color[2][]{%
    \GenericError{(gnuplot) \space\space\space\@spaces}{%
      Package color not loaded in conjunction with
      terminal option `colourtext'%
    }{See the gnuplot documentation for explanation.%
    }{Either use 'blacktext' in gnuplot or load the package
      color.sty in LaTeX.}%
    \renewcommand\color[2][]{}%
  }%
  \providecommand\includegraphics[2][]{%
    \GenericError{(gnuplot) \space\space\space\@spaces}{%
      Package graphicx or graphics not loaded%
    }{See the gnuplot documentation for explanation.%
    }{The gnuplot epslatex terminal needs graphicx.sty or graphics.sty.}%
    \renewcommand\includegraphics[2][]{}%
  }%
  \providecommand\rotatebox[2]{#2}%
  \@ifundefined{ifGPcolor}{%
    \newif\ifGPcolor
    \GPcolorfalse
  }{}%
  \@ifundefined{ifGPblacktext}{%
    \newif\ifGPblacktext
    \GPblacktexttrue
  }{}%
  \let\gplgaddtomacro\g@addto@macro
  \gdef\gplbacktext{}%
  \gdef\gplfronttext{}%
  \makeatother
  \ifGPblacktext
    \def\colorrgb#1{}%
    \def\colorgray#1{}%
  \else
    \ifGPcolor
      \def\colorrgb#1{\color[rgb]{#1}}%
      \def\colorgray#1{\color[gray]{#1}}%
      \expandafter\def\csname LTw\endcsname{\color{white}}%
      \expandafter\def\csname LTb\endcsname{\color{black}}%
      \expandafter\def\csname LTa\endcsname{\color{black}}%
      \expandafter\def\csname LT0\endcsname{\color[rgb]{1,0,0}}%
      \expandafter\def\csname LT1\endcsname{\color[rgb]{0,1,0}}%
      \expandafter\def\csname LT2\endcsname{\color[rgb]{0,0,1}}%
      \expandafter\def\csname LT3\endcsname{\color[rgb]{1,0,1}}%
      \expandafter\def\csname LT4\endcsname{\color[rgb]{0,1,1}}%
      \expandafter\def\csname LT5\endcsname{\color[rgb]{1,1,0}}%
      \expandafter\def\csname LT6\endcsname{\color[rgb]{0,0,0}}%
      \expandafter\def\csname LT7\endcsname{\color[rgb]{1,0.3,0}}%
      \expandafter\def\csname LT8\endcsname{\color[rgb]{0.5,0.5,0.5}}%
    \else
      \def\colorrgb#1{\color{black}}%
      \def\colorgray#1{\color[gray]{#1}}%
      \expandafter\def\csname LTw\endcsname{\color{white}}%
      \expandafter\def\csname LTb\endcsname{\color{black}}%
      \expandafter\def\csname LTa\endcsname{\color{black}}%
      \expandafter\def\csname LT0\endcsname{\color{black}}%
      \expandafter\def\csname LT1\endcsname{\color{black}}%
      \expandafter\def\csname LT2\endcsname{\color{black}}%
      \expandafter\def\csname LT3\endcsname{\color{black}}%
      \expandafter\def\csname LT4\endcsname{\color{black}}%
      \expandafter\def\csname LT5\endcsname{\color{black}}%
      \expandafter\def\csname LT6\endcsname{\color{black}}%
      \expandafter\def\csname LT7\endcsname{\color{black}}%
      \expandafter\def\csname LT8\endcsname{\color{black}}%
    \fi
  \fi
  \setlength{\unitlength}{0.0500bp}%
  \begin{picture}(7012.00,4506.00)%
    \gplgaddtomacro\gplbacktext{%
      \csname LTb\endcsname%
      \put(1342,704){\makebox(0,0)[r]{\strut{} 1e-08}}%
      \put(1342,1294){\makebox(0,0)[r]{\strut{} 1e-07}}%
      \put(1342,1883){\makebox(0,0)[r]{\strut{} 1e-06}}%
      \put(1342,2473){\makebox(0,0)[r]{\strut{} 1e-05}}%
      \put(1342,3062){\makebox(0,0)[r]{\strut{} 0.0001}}%
      \put(1342,3652){\makebox(0,0)[r]{\strut{} 0.001}}%
      \put(1342,4241){\makebox(0,0)[r]{\strut{} 0.01}}%
      \put(1474,484){\makebox(0,0){\strut{} 0}}%
      \put(2331,484){\makebox(0,0){\strut{} 100}}%
      \put(3188,484){\makebox(0,0){\strut{} 200}}%
      \put(4045,484){\makebox(0,0){\strut{} 300}}%
      \put(4901,484){\makebox(0,0){\strut{} 400}}%
      \put(5758,484){\makebox(0,0){\strut{} 500}}%
      \put(6615,484){\makebox(0,0){\strut{} 600}}%
      \put(176,2472){\rotatebox{-270}{\makebox(0,0){\strut{}E}}}%
      \put(4044,154){\makebox(0,0){\strut{}$t$}}%
      \put(1817,2576){\makebox(0,0)[l]{\strut{}$c_i=-0.0028958$}}%
      \put(5158,2295){\makebox(0,0)[l]{\strut{}$c_i=+0.0052738$}}%
      \put(4816,3652){\makebox(0,0)[l]{\strut{}$c_i=+0.0083030$}}%
    }%
    \gplgaddtomacro\gplfronttext{%
    }%
    \gplbacktext
    \put(0,0){\includegraphics{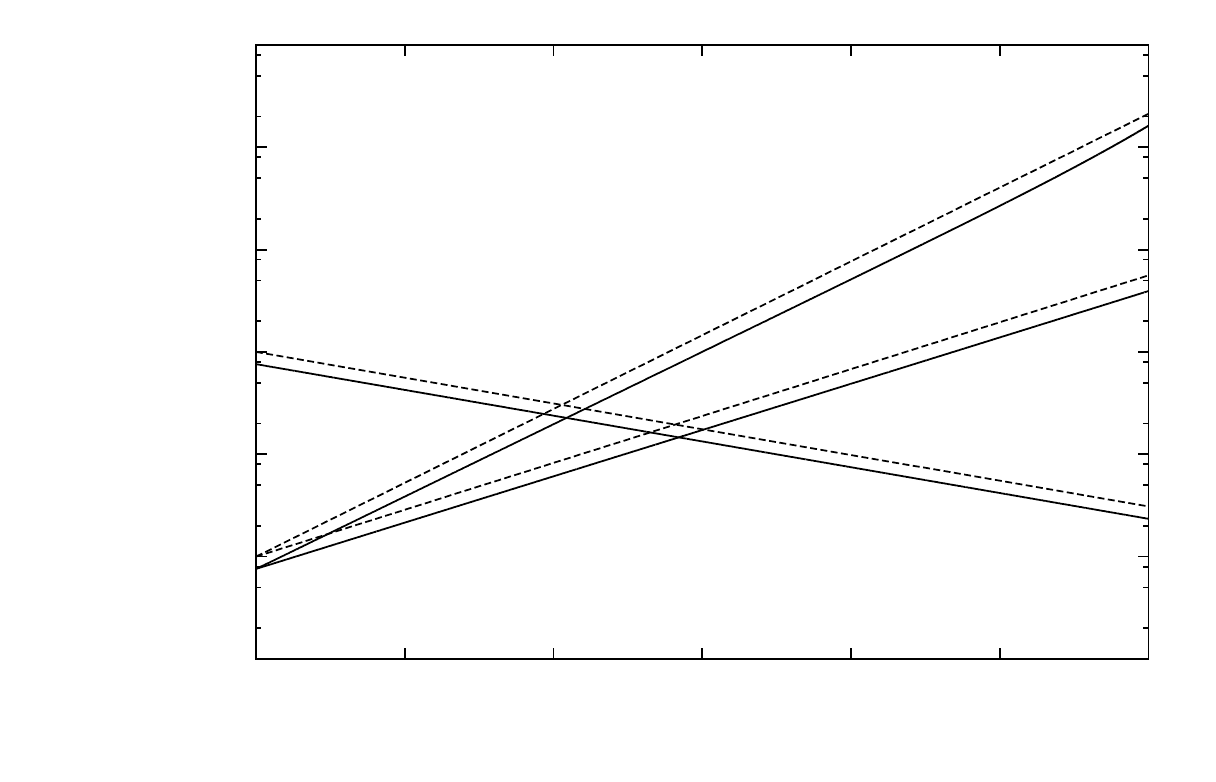}}%
    \gplfronttext
  \end{picture}%
\endgroup

%% file: validation_tau.tex
\begingroup
  \makeatletter
  \providecommand\color[2][]{%
    \GenericError{(gnuplot) \space\space\space\@spaces}{%
      Package color not loaded in conjunction with
      terminal option `colourtext'%
    }{See the gnuplot documentation for explanation.%
    }{Either use 'blacktext' in gnuplot or load the package
      color.sty in LaTeX.}%
    \renewcommand\color[2][]{}%
  }%
  \providecommand\includegraphics[2][]{%
    \GenericError{(gnuplot) \space\space\space\@spaces}{%
      Package graphicx or graphics not loaded%
    }{See the gnuplot documentation for explanation.%
    }{The gnuplot epslatex terminal needs graphicx.sty or graphics.sty.}%
    \renewcommand\includegraphics[2][]{}%
  }%
  \providecommand\rotatebox[2]{#2}%
  \@ifundefined{ifGPcolor}{%
    \newif\ifGPcolor
    \GPcolorfalse
  }{}%
  \@ifundefined{ifGPblacktext}{%
    \newif\ifGPblacktext
    \GPblacktexttrue
  }{}%
  \let\gplgaddtomacro\g@addto@macro
  \gdef\gplbacktext{}%
  \gdef\gplfronttext{}%
  \makeatother
  \ifGPblacktext
    \def\colorrgb#1{}%
    \def\colorgray#1{}%
  \else
    \ifGPcolor
      \def\colorrgb#1{\color[rgb]{#1}}%
      \def\colorgray#1{\color[gray]{#1}}%
      \expandafter\def\csname LTw\endcsname{\color{white}}%
      \expandafter\def\csname LTb\endcsname{\color{black}}%
      \expandafter\def\csname LTa\endcsname{\color{black}}%
      \expandafter\def\csname LT0\endcsname{\color[rgb]{1,0,0}}%
      \expandafter\def\csname LT1\endcsname{\color[rgb]{0,1,0}}%
      \expandafter\def\csname LT2\endcsname{\color[rgb]{0,0,1}}%
      \expandafter\def\csname LT3\endcsname{\color[rgb]{1,0,1}}%
      \expandafter\def\csname LT4\endcsname{\color[rgb]{0,1,1}}%
      \expandafter\def\csname LT5\endcsname{\color[rgb]{1,1,0}}%
      \expandafter\def\csname LT6\endcsname{\color[rgb]{0,0,0}}%
      \expandafter\def\csname LT7\endcsname{\color[rgb]{1,0.3,0}}%
      \expandafter\def\csname LT8\endcsname{\color[rgb]{0.5,0.5,0.5}}%
    \else
      \def\colorrgb#1{\color{black}}%
      \def\colorgray#1{\color[gray]{#1}}%
      \expandafter\def\csname LTw\endcsname{\color{white}}%
      \expandafter\def\csname LTb\endcsname{\color{black}}%
      \expandafter\def\csname LTa\endcsname{\color{black}}%
      \expandafter\def\csname LT0\endcsname{\color{black}}%
      \expandafter\def\csname LT1\endcsname{\color{black}}%
      \expandafter\def\csname LT2\endcsname{\color{black}}%
      \expandafter\def\csname LT3\endcsname{\color{black}}%
      \expandafter\def\csname LT4\endcsname{\color{black}}%
      \expandafter\def\csname LT5\endcsname{\color{black}}%
      \expandafter\def\csname LT6\endcsname{\color{black}}%
      \expandafter\def\csname LT7\endcsname{\color{black}}%
      \expandafter\def\csname LT8\endcsname{\color{black}}%
    \fi
  \fi
  \setlength{\unitlength}{0.0500bp}%
  \begin{picture}(7012.00,4506.00)%
    \gplgaddtomacro\gplbacktext{%
      \csname LTb\endcsname%
      \put(1342,704){\makebox(0,0)[r]{\strut{} 1e-08}}%
      \put(1342,1411){\makebox(0,0)[r]{\strut{} 1e-07}}%
      \put(1342,2119){\makebox(0,0)[r]{\strut{} 1e-06}}%
      \put(1342,2826){\makebox(0,0)[r]{\strut{} 1e-05}}%
      \put(1342,3534){\makebox(0,0)[r]{\strut{} 0.0001}}%
      \put(1342,4241){\makebox(0,0)[r]{\strut{} 0.001}}%
      \put(1474,484){\makebox(0,0){\strut{} 0}}%
      \put(2331,484){\makebox(0,0){\strut{} 100}}%
      \put(3188,484){\makebox(0,0){\strut{} 200}}%
      \put(4045,484){\makebox(0,0){\strut{} 300}}%
      \put(4901,484){\makebox(0,0){\strut{} 400}}%
      \put(5758,484){\makebox(0,0){\strut{} 500}}%
      \put(6615,484){\makebox(0,0){\strut{} 600}}%
      \put(176,2472){\rotatebox{-270}{\makebox(0,0){\strut{}E}}}%
      \put(4044,154){\makebox(0,0){\strut{}$t$}}%
      \put(4901,2456){\makebox(0,0)[l]{\strut{}$c_i=-0.003118$}}%
      \put(3188,1343){\makebox(0,0)[l]{\strut{}$c_i=-0.004540$}}%
    }%
    \gplgaddtomacro\gplfronttext{%
    }%
    \gplbacktext
    \put(0,0){\includegraphics{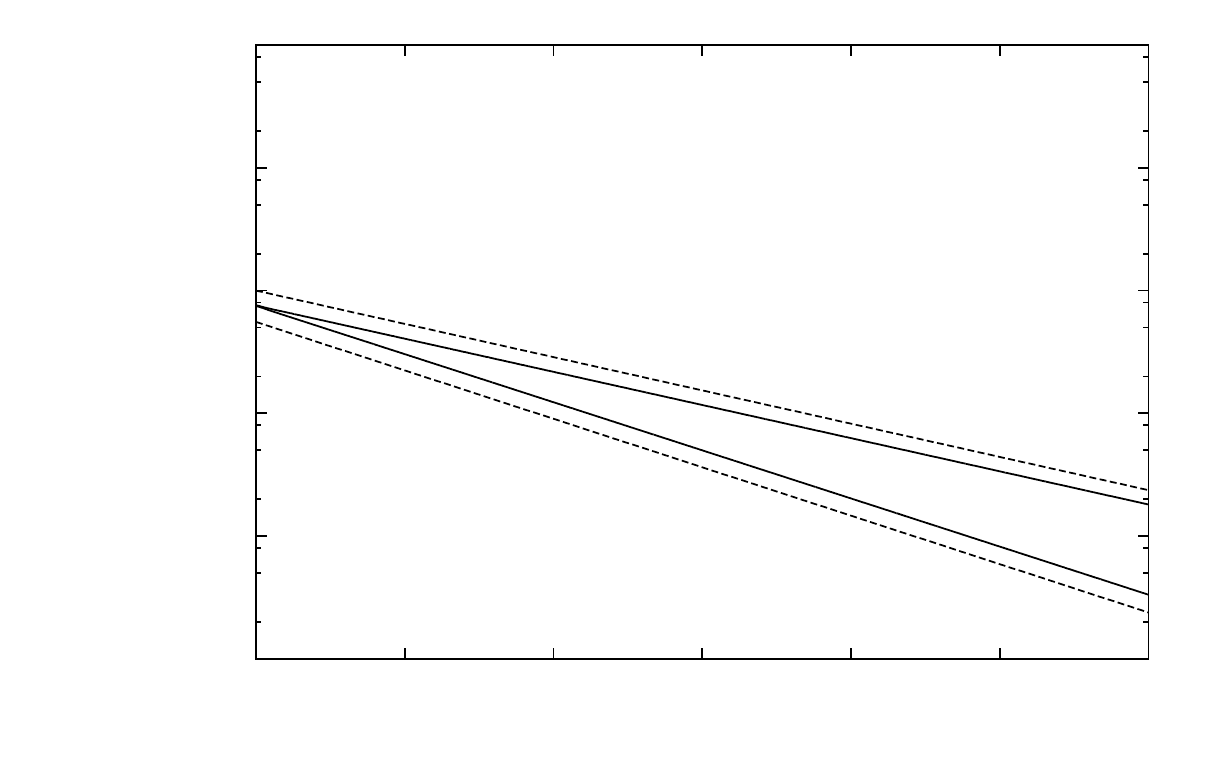}}%
    \gplfronttext
  \end{picture}%
\endgroup

%% file: fik.tex
\begingroup
  \makeatletter
  \providecommand\color[2][]{%
    \GenericError{(gnuplot) \space\space\space\@spaces}{%
      Package color not loaded in conjunction with
      terminal option `colourtext'%
    }{See the gnuplot documentation for explanation.%
    }{Either use 'blacktext' in gnuplot or load the package
      color.sty in LaTeX.}%
    \renewcommand\color[2][]{}%
  }%
  \providecommand\includegraphics[2][]{%
    \GenericError{(gnuplot) \space\space\space\@spaces}{%
      Package graphicx or graphics not loaded%
    }{See the gnuplot documentation for explanation.%
    }{The gnuplot epslatex terminal needs graphicx.sty or graphics.sty.}%
    \renewcommand\includegraphics[2][]{}%
  }%
  \providecommand\rotatebox[2]{#2}%
  \@ifundefined{ifGPcolor}{%
    \newif\ifGPcolor
    \GPcolorfalse
  }{}%
  \@ifundefined{ifGPblacktext}{%
    \newif\ifGPblacktext
    \GPblacktexttrue
  }{}%
  \let\gplgaddtomacro\g@addto@macro
  \gdef\gplbacktext{}%
  \gdef\gplfronttext{}%
  \makeatother
  \ifGPblacktext
    \def\colorrgb#1{}%
    \def\colorgray#1{}%
  \else
    \ifGPcolor
      \def\colorrgb#1{\color[rgb]{#1}}%
      \def\colorgray#1{\color[gray]{#1}}%
      \expandafter\def\csname LTw\endcsname{\color{white}}%
      \expandafter\def\csname LTb\endcsname{\color{black}}%
      \expandafter\def\csname LTa\endcsname{\color{black}}%
      \expandafter\def\csname LT0\endcsname{\color[rgb]{1,0,0}}%
      \expandafter\def\csname LT1\endcsname{\color[rgb]{0,1,0}}%
      \expandafter\def\csname LT2\endcsname{\color[rgb]{0,0,1}}%
      \expandafter\def\csname LT3\endcsname{\color[rgb]{1,0,1}}%
      \expandafter\def\csname LT4\endcsname{\color[rgb]{0,1,1}}%
      \expandafter\def\csname LT5\endcsname{\color[rgb]{1,1,0}}%
      \expandafter\def\csname LT6\endcsname{\color[rgb]{0,0,0}}%
      \expandafter\def\csname LT7\endcsname{\color[rgb]{1,0.3,0}}%
      \expandafter\def\csname LT8\endcsname{\color[rgb]{0.5,0.5,0.5}}%
    \else
      \def\colorrgb#1{\color{black}}%
      \def\colorgray#1{\color[gray]{#1}}%
      \expandafter\def\csname LTw\endcsname{\color{white}}%
      \expandafter\def\csname LTb\endcsname{\color{black}}%
      \expandafter\def\csname LTa\endcsname{\color{black}}%
      \expandafter\def\csname LT0\endcsname{\color{black}}%
      \expandafter\def\csname LT1\endcsname{\color{black}}%
      \expandafter\def\csname LT2\endcsname{\color{black}}%
      \expandafter\def\csname LT3\endcsname{\color{black}}%
      \expandafter\def\csname LT4\endcsname{\color{black}}%
      \expandafter\def\csname LT5\endcsname{\color{black}}%
      \expandafter\def\csname LT6\endcsname{\color{black}}%
      \expandafter\def\csname LT7\endcsname{\color{black}}%
      \expandafter\def\csname LT8\endcsname{\color{black}}%
    \fi
  \fi
  \setlength{\unitlength}{0.0500bp}%
  \begin{picture}(7012.00,4506.00)%
    \gplgaddtomacro\gplbacktext{%
      \csname LTb\endcsname%
      \put(1342,704){\makebox(0,0)[r]{\strut{} 0}}%
      \put(1342,1294){\makebox(0,0)[r]{\strut{} 0.0005}}%
      \put(1342,1883){\makebox(0,0)[r]{\strut{} 0.001}}%
      \put(1342,2473){\makebox(0,0)[r]{\strut{} 0.0015}}%
      \put(1342,3062){\makebox(0,0)[r]{\strut{} 0.002}}%
      \put(1342,3652){\makebox(0,0)[r]{\strut{} 0.0025}}%
      \put(1342,4241){\makebox(0,0)[r]{\strut{} 0.003}}%
      \put(1474,484){\makebox(0,0){\strut{} 0}}%
      \put(2502,484){\makebox(0,0){\strut{} 0.2}}%
      \put(3530,484){\makebox(0,0){\strut{} 0.4}}%
      \put(4559,484){\makebox(0,0){\strut{} 0.6}}%
      \put(5587,484){\makebox(0,0){\strut{} 0.8}}%
      \put(6615,484){\makebox(0,0){\strut{} 1}}%
      \put(176,2472){\rotatebox{-270}{\makebox(0,0){\strut{}$\left( 1-y \right) \left( - \overline{u'v'} \right)$}}}%
      \put(4044,154){\makebox(0,0){\strut{}$y$}}%
    }%
    \gplgaddtomacro\gplfronttext{%
      \csname LTb\endcsname%
      \put(5628,4068){\makebox(0,0)[r]{\strut{}Imp}}%
      \csname LTb\endcsname%
      \put(5628,3848){\makebox(0,0)[r]{\strut{}Por}}%
    }%
    \gplbacktext
    \put(0,0){\includegraphics{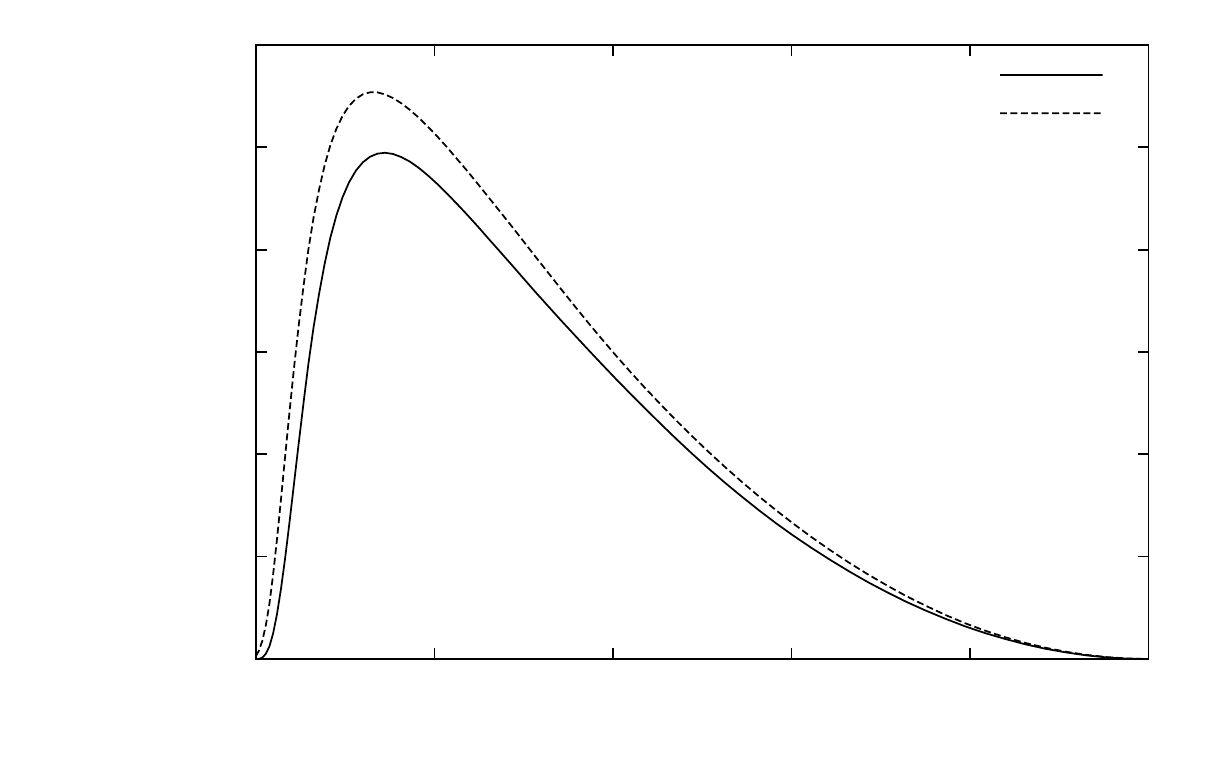}}%
    \gplfronttext
  \end{picture}%
\endgroup

%% file: mean.tex
\begingroup
  \makeatletter
  \providecommand\color[2][]{%
    \GenericError{(gnuplot) \space\space\space\@spaces}{%
      Package color not loaded in conjunction with
      terminal option `colourtext'%
    }{See the gnuplot documentation for explanation.%
    }{Either use 'blacktext' in gnuplot or load the package
      color.sty in LaTeX.}%
    \renewcommand\color[2][]{}%
  }%
  \providecommand\includegraphics[2][]{%
    \GenericError{(gnuplot) \space\space\space\@spaces}{%
      Package graphicx or graphics not loaded%
    }{See the gnuplot documentation for explanation.%
    }{The gnuplot epslatex terminal needs graphicx.sty or graphics.sty.}%
    \renewcommand\includegraphics[2][]{}%
  }%
  \providecommand\rotatebox[2]{#2}%
  \@ifundefined{ifGPcolor}{%
    \newif\ifGPcolor
    \GPcolorfalse
  }{}%
  \@ifundefined{ifGPblacktext}{%
    \newif\ifGPblacktext
    \GPblacktexttrue
  }{}%
  \let\gplgaddtomacro\g@addto@macro
  \gdef\gplbacktext{}%
  \gdef\gplfronttext{}%
  \makeatother
  \ifGPblacktext
    \def\colorrgb#1{}%
    \def\colorgray#1{}%
  \else
    \ifGPcolor
      \def\colorrgb#1{\color[rgb]{#1}}%
      \def\colorgray#1{\color[gray]{#1}}%
      \expandafter\def\csname LTw\endcsname{\color{white}}%
      \expandafter\def\csname LTb\endcsname{\color{black}}%
      \expandafter\def\csname LTa\endcsname{\color{black}}%
      \expandafter\def\csname LT0\endcsname{\color[rgb]{1,0,0}}%
      \expandafter\def\csname LT1\endcsname{\color[rgb]{0,1,0}}%
      \expandafter\def\csname LT2\endcsname{\color[rgb]{0,0,1}}%
      \expandafter\def\csname LT3\endcsname{\color[rgb]{1,0,1}}%
      \expandafter\def\csname LT4\endcsname{\color[rgb]{0,1,1}}%
      \expandafter\def\csname LT5\endcsname{\color[rgb]{1,1,0}}%
      \expandafter\def\csname LT6\endcsname{\color[rgb]{0,0,0}}%
      \expandafter\def\csname LT7\endcsname{\color[rgb]{1,0.3,0}}%
      \expandafter\def\csname LT8\endcsname{\color[rgb]{0.5,0.5,0.5}}%
    \else
      \def\colorrgb#1{\color{black}}%
      \def\colorgray#1{\color[gray]{#1}}%
      \expandafter\def\csname LTw\endcsname{\color{white}}%
      \expandafter\def\csname LTb\endcsname{\color{black}}%
      \expandafter\def\csname LTa\endcsname{\color{black}}%
      \expandafter\def\csname LT0\endcsname{\color{black}}%
      \expandafter\def\csname LT1\endcsname{\color{black}}%
      \expandafter\def\csname LT2\endcsname{\color{black}}%
      \expandafter\def\csname LT3\endcsname{\color{black}}%
      \expandafter\def\csname LT4\endcsname{\color{black}}%
      \expandafter\def\csname LT5\endcsname{\color{black}}%
      \expandafter\def\csname LT6\endcsname{\color{black}}%
      \expandafter\def\csname LT7\endcsname{\color{black}}%
      \expandafter\def\csname LT8\endcsname{\color{black}}%
    \fi
  \fi
  \setlength{\unitlength}{0.0500bp}%
  \begin{picture}(7012.00,4506.00)%
    \gplgaddtomacro\gplbacktext{%
      \csname LTb\endcsname%
      \put(726,845){\makebox(0,0)[r]{\strut{} 0}}%
      \put(726,1411){\makebox(0,0)[r]{\strut{} 0.2}}%
      \put(726,1977){\makebox(0,0)[r]{\strut{} 0.4}}%
      \put(726,2543){\makebox(0,0)[r]{\strut{} 0.6}}%
      \put(726,3109){\makebox(0,0)[r]{\strut{} 0.8}}%
      \put(726,3675){\makebox(0,0)[r]{\strut{} 1}}%
      \put(726,4241){\makebox(0,0)[r]{\strut{} 1.2}}%
      \put(858,484){\makebox(0,0){\strut{}-0.2}}%
      \put(1818,484){\makebox(0,0){\strut{} 0}}%
      \put(2777,484){\makebox(0,0){\strut{} 0.2}}%
      \put(3737,484){\makebox(0,0){\strut{} 0.4}}%
      \put(4696,484){\makebox(0,0){\strut{} 0.6}}%
      \put(5656,484){\makebox(0,0){\strut{} 0.8}}%
      \put(6615,484){\makebox(0,0){\strut{} 1}}%
      \put(220,2472){\rotatebox{-270}{\makebox(0,0){\strut{}$\overline{u}$}}}%
      \put(3736,154){\makebox(0,0){\strut{}$y$}}%
    }%
    \gplgaddtomacro\gplfronttext{%
      \csname LTb\endcsname%
      \put(1386,4068){\makebox(0,0)[r]{\strut{}Imp}}%
      \csname LTb\endcsname%
      \put(1386,3848){\makebox(0,0)[r]{\strut{}Por}}%
    }%
    \gplgaddtomacro\gplbacktext{%
      \csname LTb\endcsname%
      \put(3662,1379){\makebox(0,0)[r]{\strut{} 0}}%
      \csname LTb\endcsname%
      \put(3662,1636){\makebox(0,0)[r]{\strut{} 0.02}}%
      \csname LTb\endcsname%
      \put(3662,1893){\makebox(0,0)[r]{\strut{} 0.04}}%
      \csname LTb\endcsname%
      \put(3662,2150){\makebox(0,0)[r]{\strut{} 0.06}}%
      \csname LTb\endcsname%
      \put(3662,2407){\makebox(0,0)[r]{\strut{} 0.08}}%
      \csname LTb\endcsname%
      \put(3662,2664){\makebox(0,0)[r]{\strut{} 0.1}}%
      \csname LTb\endcsname%
      \put(4006,1159){\makebox(0,0){\strut{}-0.02}}%
      \csname LTb\endcsname%
      \put(4430,1159){\makebox(0,0){\strut{}-0.01}}%
      \csname LTb\endcsname%
      \put(4854,1159){\makebox(0,0){\strut{} 0}}%
      \csname LTb\endcsname%
      \put(5277,1159){\makebox(0,0){\strut{} 0.01}}%
      \csname LTb\endcsname%
      \put(5701,1159){\makebox(0,0){\strut{} 0.02}}%
      \put(3156,2021){\rotatebox{-270}{\makebox(0,0){\strut{}$\overline{u}$}}}%
      \put(4853,829){\makebox(0,0){\strut{}$y$}}%
    }%
    \gplgaddtomacro\gplfronttext{%
    }%
    \gplbacktext
    \put(0,0){\includegraphics{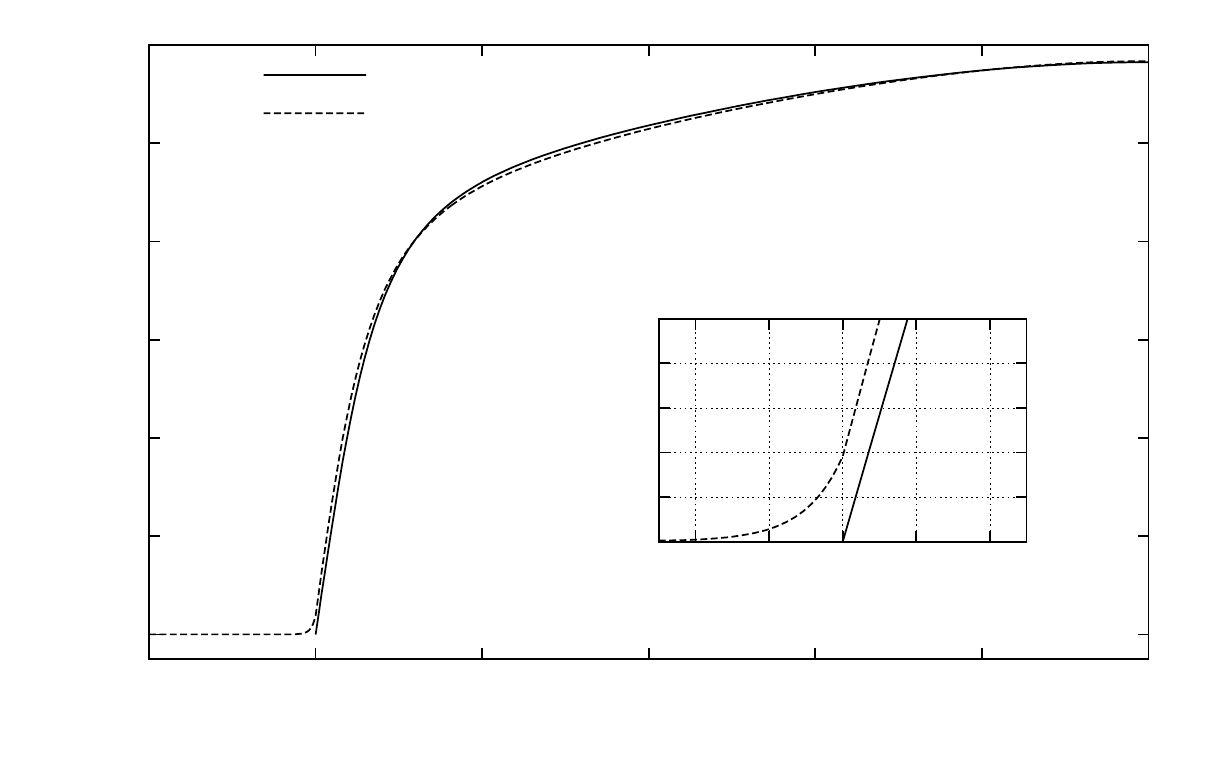}}%
    \gplfronttext
  \end{picture}%
\endgroup

%% file: loglaw.tex
\begingroup
  \makeatletter
  \providecommand\color[2][]{%
    \GenericError{(gnuplot) \space\space\space\@spaces}{%
      Package color not loaded in conjunction with
      terminal option `colourtext'%
    }{See the gnuplot documentation for explanation.%
    }{Either use 'blacktext' in gnuplot or load the package
      color.sty in LaTeX.}%
    \renewcommand\color[2][]{}%
  }%
  \providecommand\includegraphics[2][]{%
    \GenericError{(gnuplot) \space\space\space\@spaces}{%
      Package graphicx or graphics not loaded%
    }{See the gnuplot documentation for explanation.%
    }{The gnuplot epslatex terminal needs graphicx.sty or graphics.sty.}%
    \renewcommand\includegraphics[2][]{}%
  }%
  \providecommand\rotatebox[2]{#2}%
  \@ifundefined{ifGPcolor}{%
    \newif\ifGPcolor
    \GPcolorfalse
  }{}%
  \@ifundefined{ifGPblacktext}{%
    \newif\ifGPblacktext
    \GPblacktexttrue
  }{}%
  \let\gplgaddtomacro\g@addto@macro
  \gdef\gplbacktext{}%
  \gdef\gplfronttext{}%
  \makeatother
  \ifGPblacktext
    \def\colorrgb#1{}%
    \def\colorgray#1{}%
  \else
    \ifGPcolor
      \def\colorrgb#1{\color[rgb]{#1}}%
      \def\colorgray#1{\color[gray]{#1}}%
      \expandafter\def\csname LTw\endcsname{\color{white}}%
      \expandafter\def\csname LTb\endcsname{\color{black}}%
      \expandafter\def\csname LTa\endcsname{\color{black}}%
      \expandafter\def\csname LT0\endcsname{\color[rgb]{1,0,0}}%
      \expandafter\def\csname LT1\endcsname{\color[rgb]{0,1,0}}%
      \expandafter\def\csname LT2\endcsname{\color[rgb]{0,0,1}}%
      \expandafter\def\csname LT3\endcsname{\color[rgb]{1,0,1}}%
      \expandafter\def\csname LT4\endcsname{\color[rgb]{0,1,1}}%
      \expandafter\def\csname LT5\endcsname{\color[rgb]{1,1,0}}%
      \expandafter\def\csname LT6\endcsname{\color[rgb]{0,0,0}}%
      \expandafter\def\csname LT7\endcsname{\color[rgb]{1,0.3,0}}%
      \expandafter\def\csname LT8\endcsname{\color[rgb]{0.5,0.5,0.5}}%
    \else
      \def\colorrgb#1{\color{black}}%
      \def\colorgray#1{\color[gray]{#1}}%
      \expandafter\def\csname LTw\endcsname{\color{white}}%
      \expandafter\def\csname LTb\endcsname{\color{black}}%
      \expandafter\def\csname LTa\endcsname{\color{black}}%
      \expandafter\def\csname LT0\endcsname{\color{black}}%
      \expandafter\def\csname LT1\endcsname{\color{black}}%
      \expandafter\def\csname LT2\endcsname{\color{black}}%
      \expandafter\def\csname LT3\endcsname{\color{black}}%
      \expandafter\def\csname LT4\endcsname{\color{black}}%
      \expandafter\def\csname LT5\endcsname{\color{black}}%
      \expandafter\def\csname LT6\endcsname{\color{black}}%
      \expandafter\def\csname LT7\endcsname{\color{black}}%
      \expandafter\def\csname LT8\endcsname{\color{black}}%
    \fi
  \fi
  \setlength{\unitlength}{0.0500bp}%
  \begin{picture}(7012.00,4506.00)%
    \gplgaddtomacro\gplbacktext{%
      \csname LTb\endcsname%
      \put(682,704){\makebox(0,0)[r]{\strut{} 0}}%
      \put(682,1588){\makebox(0,0)[r]{\strut{} 5}}%
      \put(682,2473){\makebox(0,0)[r]{\strut{} 10}}%
      \put(682,3357){\makebox(0,0)[r]{\strut{} 15}}%
      \put(682,4241){\makebox(0,0)[r]{\strut{} 20}}%
      \put(814,484){\makebox(0,0){\strut{} 1}}%
      \put(3233,484){\makebox(0,0){\strut{} 10}}%
      \put(5652,484){\makebox(0,0){\strut{} 100}}%
      \put(176,2472){\rotatebox{-270}{\makebox(0,0){\strut{}$(\overline{u}-U_i)^+$}}}%
      \put(3714,154){\makebox(0,0){\strut{}$y^+$}}%
    }%
    \gplgaddtomacro\gplfronttext{%
      \csname LTb\endcsname%
      \put(1342,4068){\makebox(0,0)[r]{\strut{}Imp}}%
      \csname LTb\endcsname%
      \put(1342,3848){\makebox(0,0)[r]{\strut{}Por}}%
    }%
    \gplbacktext
    \put(0,0){\includegraphics{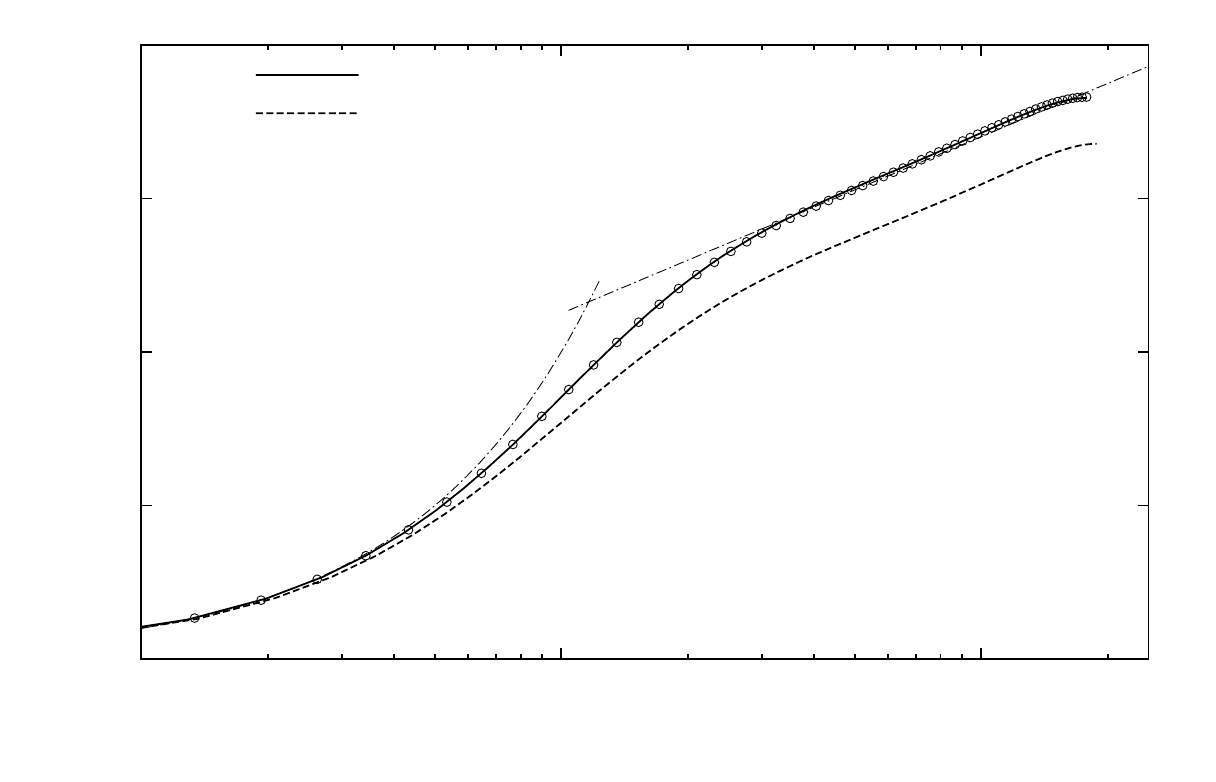}}%
    \gplfronttext
  \end{picture}%
\endgroup

%% file: rms.tex
\begingroup
  \makeatletter
  \providecommand\color[2][]{%
    \GenericError{(gnuplot) \space\space\space\@spaces}{%
      Package color not loaded in conjunction with
      terminal option `colourtext'%
    }{See the gnuplot documentation for explanation.%
    }{Either use 'blacktext' in gnuplot or load the package
      color.sty in LaTeX.}%
    \renewcommand\color[2][]{}%
  }%
  \providecommand\includegraphics[2][]{%
    \GenericError{(gnuplot) \space\space\space\@spaces}{%
      Package graphicx or graphics not loaded%
    }{See the gnuplot documentation for explanation.%
    }{The gnuplot epslatex terminal needs graphicx.sty or graphics.sty.}%
    \renewcommand\includegraphics[2][]{}%
  }%
  \providecommand\rotatebox[2]{#2}%
  \@ifundefined{ifGPcolor}{%
    \newif\ifGPcolor
    \GPcolorfalse
  }{}%
  \@ifundefined{ifGPblacktext}{%
    \newif\ifGPblacktext
    \GPblacktexttrue
  }{}%
  \let\gplgaddtomacro\g@addto@macro
  \gdef\gplbacktext{}%
  \gdef\gplfronttext{}%
  \makeatother
  \ifGPblacktext
    \def\colorrgb#1{}%
    \def\colorgray#1{}%
  \else
    \ifGPcolor
      \def\colorrgb#1{\color[rgb]{#1}}%
      \def\colorgray#1{\color[gray]{#1}}%
      \expandafter\def\csname LTw\endcsname{\color{white}}%
      \expandafter\def\csname LTb\endcsname{\color{black}}%
      \expandafter\def\csname LTa\endcsname{\color{black}}%
      \expandafter\def\csname LT0\endcsname{\color[rgb]{1,0,0}}%
      \expandafter\def\csname LT1\endcsname{\color[rgb]{0,1,0}}%
      \expandafter\def\csname LT2\endcsname{\color[rgb]{0,0,1}}%
      \expandafter\def\csname LT3\endcsname{\color[rgb]{1,0,1}}%
      \expandafter\def\csname LT4\endcsname{\color[rgb]{0,1,1}}%
      \expandafter\def\csname LT5\endcsname{\color[rgb]{1,1,0}}%
      \expandafter\def\csname LT6\endcsname{\color[rgb]{0,0,0}}%
      \expandafter\def\csname LT7\endcsname{\color[rgb]{1,0.3,0}}%
      \expandafter\def\csname LT8\endcsname{\color[rgb]{0.5,0.5,0.5}}%
    \else
      \def\colorrgb#1{\color{black}}%
      \def\colorgray#1{\color[gray]{#1}}%
      \expandafter\def\csname LTw\endcsname{\color{white}}%
      \expandafter\def\csname LTb\endcsname{\color{black}}%
      \expandafter\def\csname LTa\endcsname{\color{black}}%
      \expandafter\def\csname LT0\endcsname{\color{black}}%
      \expandafter\def\csname LT1\endcsname{\color{black}}%
      \expandafter\def\csname LT2\endcsname{\color{black}}%
      \expandafter\def\csname LT3\endcsname{\color{black}}%
      \expandafter\def\csname LT4\endcsname{\color{black}}%
      \expandafter\def\csname LT5\endcsname{\color{black}}%
      \expandafter\def\csname LT6\endcsname{\color{black}}%
      \expandafter\def\csname LT7\endcsname{\color{black}}%
      \expandafter\def\csname LT8\endcsname{\color{black}}%
    \fi
  \fi
  \setlength{\unitlength}{0.0500bp}%
  \begin{picture}(7012.00,4506.00)%
    \gplgaddtomacro\gplbacktext{%
      \csname LTb\endcsname%
      \put(748,704){\makebox(0,0)[r]{\strut{} 0}}%
      \put(748,1294){\makebox(0,0)[r]{\strut{} 0.5}}%
      \put(748,1883){\makebox(0,0)[r]{\strut{} 1}}%
      \put(748,2473){\makebox(0,0)[r]{\strut{} 1.5}}%
      \put(748,3062){\makebox(0,0)[r]{\strut{} 2}}%
      \put(748,3652){\makebox(0,0)[r]{\strut{} 2.5}}%
      \put(748,4241){\makebox(0,0)[r]{\strut{} 3}}%
      \put(880,484){\makebox(0,0){\strut{} 1}}%
      \put(3372,484){\makebox(0,0){\strut{} 10}}%
      \put(5865,484){\makebox(0,0){\strut{} 100}}%
      \put(176,2472){\rotatebox{-270}{\makebox(0,0){\strut{}$u_{\textrm{i}_\textrm{rms}}^+$}}}%
      \put(3747,154){\makebox(0,0){\strut{}$y^+$}}%
    }%
    \gplgaddtomacro\gplfronttext{%
      \csname LTb\endcsname%
      \put(1672,4068){\makebox(0,0)[r]{\strut{}u Imp}}%
      \csname LTb\endcsname%
      \put(1672,3848){\makebox(0,0)[r]{\strut{}v Imp}}%
      \csname LTb\endcsname%
      \put(1672,3628){\makebox(0,0)[r]{\strut{}w Imp}}%
      \csname LTb\endcsname%
      \put(1672,3408){\makebox(0,0)[r]{\strut{}u Por}}%
      \csname LTb\endcsname%
      \put(1672,3188){\makebox(0,0)[r]{\strut{}v Por}}%
      \csname LTb\endcsname%
      \put(1672,2968){\makebox(0,0)[r]{\strut{}w Por}}%
    }%
    \gplbacktext
    \put(0,0){\includegraphics{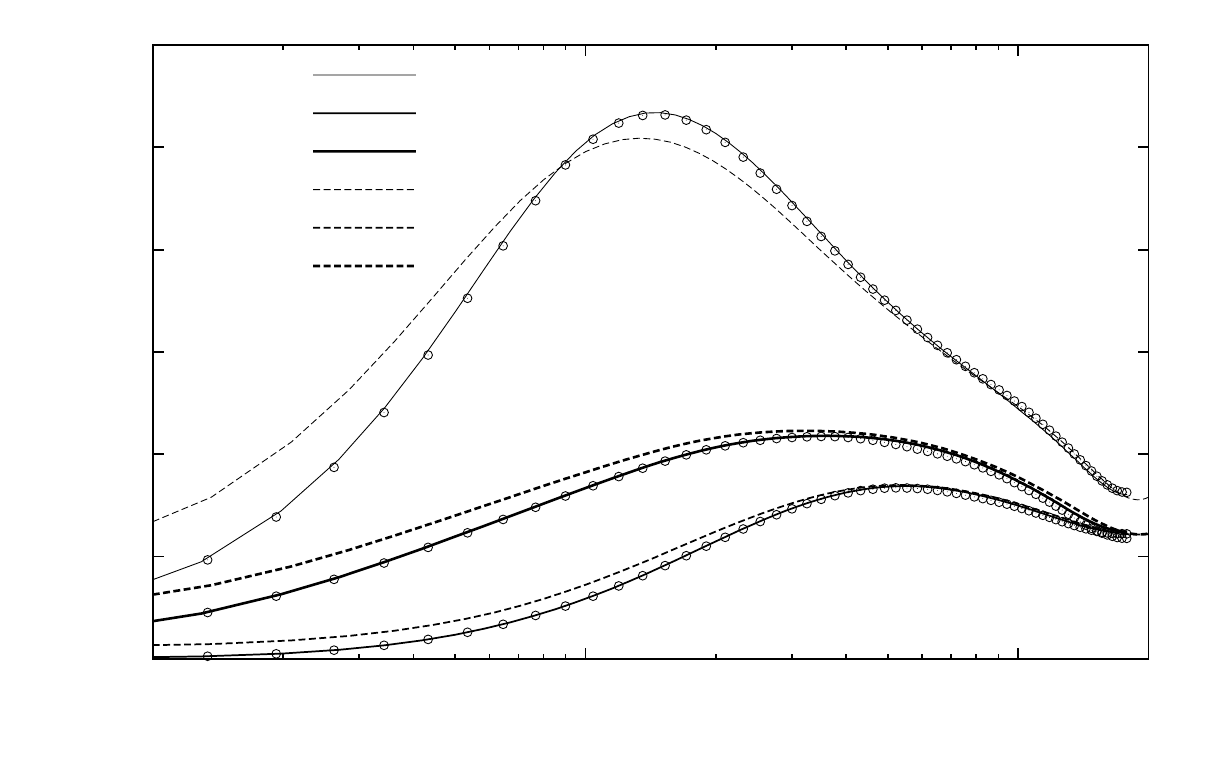}}%
    \gplfronttext
  \end{picture}%
\endgroup

%% file: rms-slab.tex
\begingroup
  \makeatletter
  \providecommand\color[2][]{%
    \GenericError{(gnuplot) \space\space\space\@spaces}{%
      Package color not loaded in conjunction with
      terminal option `colourtext'%
    }{See the gnuplot documentation for explanation.%
    }{Either use 'blacktext' in gnuplot or load the package
      color.sty in LaTeX.}%
    \renewcommand\color[2][]{}%
  }%
  \providecommand\includegraphics[2][]{%
    \GenericError{(gnuplot) \space\space\space\@spaces}{%
      Package graphicx or graphics not loaded%
    }{See the gnuplot documentation for explanation.%
    }{The gnuplot epslatex terminal needs graphicx.sty or graphics.sty.}%
    \renewcommand\includegraphics[2][]{}%
  }%
  \providecommand\rotatebox[2]{#2}%
  \@ifundefined{ifGPcolor}{%
    \newif\ifGPcolor
    \GPcolorfalse
  }{}%
  \@ifundefined{ifGPblacktext}{%
    \newif\ifGPblacktext
    \GPblacktexttrue
  }{}%
  \let\gplgaddtomacro\g@addto@macro
  \gdef\gplbacktext{}%
  \gdef\gplfronttext{}%
  \makeatother
  \ifGPblacktext
    \def\colorrgb#1{}%
    \def\colorgray#1{}%
  \else
    \ifGPcolor
      \def\colorrgb#1{\color[rgb]{#1}}%
      \def\colorgray#1{\color[gray]{#1}}%
      \expandafter\def\csname LTw\endcsname{\color{white}}%
      \expandafter\def\csname LTb\endcsname{\color{black}}%
      \expandafter\def\csname LTa\endcsname{\color{black}}%
      \expandafter\def\csname LT0\endcsname{\color[rgb]{1,0,0}}%
      \expandafter\def\csname LT1\endcsname{\color[rgb]{0,1,0}}%
      \expandafter\def\csname LT2\endcsname{\color[rgb]{0,0,1}}%
      \expandafter\def\csname LT3\endcsname{\color[rgb]{1,0,1}}%
      \expandafter\def\csname LT4\endcsname{\color[rgb]{0,1,1}}%
      \expandafter\def\csname LT5\endcsname{\color[rgb]{1,1,0}}%
      \expandafter\def\csname LT6\endcsname{\color[rgb]{0,0,0}}%
      \expandafter\def\csname LT7\endcsname{\color[rgb]{1,0.3,0}}%
      \expandafter\def\csname LT8\endcsname{\color[rgb]{0.5,0.5,0.5}}%
    \else
      \def\colorrgb#1{\color{black}}%
      \def\colorgray#1{\color[gray]{#1}}%
      \expandafter\def\csname LTw\endcsname{\color{white}}%
      \expandafter\def\csname LTb\endcsname{\color{black}}%
      \expandafter\def\csname LTa\endcsname{\color{black}}%
      \expandafter\def\csname LT0\endcsname{\color{black}}%
      \expandafter\def\csname LT1\endcsname{\color{black}}%
      \expandafter\def\csname LT2\endcsname{\color{black}}%
      \expandafter\def\csname LT3\endcsname{\color{black}}%
      \expandafter\def\csname LT4\endcsname{\color{black}}%
      \expandafter\def\csname LT5\endcsname{\color{black}}%
      \expandafter\def\csname LT6\endcsname{\color{black}}%
      \expandafter\def\csname LT7\endcsname{\color{black}}%
      \expandafter\def\csname LT8\endcsname{\color{black}}%
    \fi
  \fi
  \setlength{\unitlength}{0.0500bp}%
  \begin{picture}(7012.00,4506.00)%
    \gplgaddtomacro\gplbacktext{%
      \csname LTb\endcsname%
      \put(1012,704){\makebox(0,0)[r]{\strut{} 0}}%
      \put(1012,1097){\makebox(0,0)[r]{\strut{} 0.002}}%
      \put(1012,1490){\makebox(0,0)[r]{\strut{} 0.004}}%
      \put(1012,1883){\makebox(0,0)[r]{\strut{} 0.006}}%
      \put(1012,2276){\makebox(0,0)[r]{\strut{} 0.008}}%
      \put(1012,2669){\makebox(0,0)[r]{\strut{} 0.01}}%
      \put(1012,3062){\makebox(0,0)[r]{\strut{} 0.012}}%
      \put(1012,3455){\makebox(0,0)[r]{\strut{} 0.014}}%
      \put(1012,3848){\makebox(0,0)[r]{\strut{} 0.016}}%
      \put(1012,4241){\makebox(0,0)[r]{\strut{} 0.018}}%
      \put(1144,484){\makebox(0,0){\strut{}-0.2}}%
      \put(2512,484){\makebox(0,0){\strut{}-0.15}}%
      \put(3879,484){\makebox(0,0){\strut{}-0.1}}%
      \put(5247,484){\makebox(0,0){\strut{}-0.05}}%
      \put(6615,484){\makebox(0,0){\strut{} 0}}%
      \put(176,2472){\rotatebox{-270}{\makebox(0,0){\strut{}$u_{\textrm{i}_\textrm{rms}}^+$}}}%
      \put(3879,154){\makebox(0,0){\strut{}$y$}}%
    }%
    \gplgaddtomacro\gplfronttext{%
      \csname LTb\endcsname%
      \put(1936,4068){\makebox(0,0)[r]{\strut{}u Por}}%
      \csname LTb\endcsname%
      \put(1936,3848){\makebox(0,0)[r]{\strut{}v Por}}%
      \csname LTb\endcsname%
      \put(1936,3628){\makebox(0,0)[r]{\strut{}w Por}}%
    }%
    \gplbacktext
    \put(0,0){\includegraphics{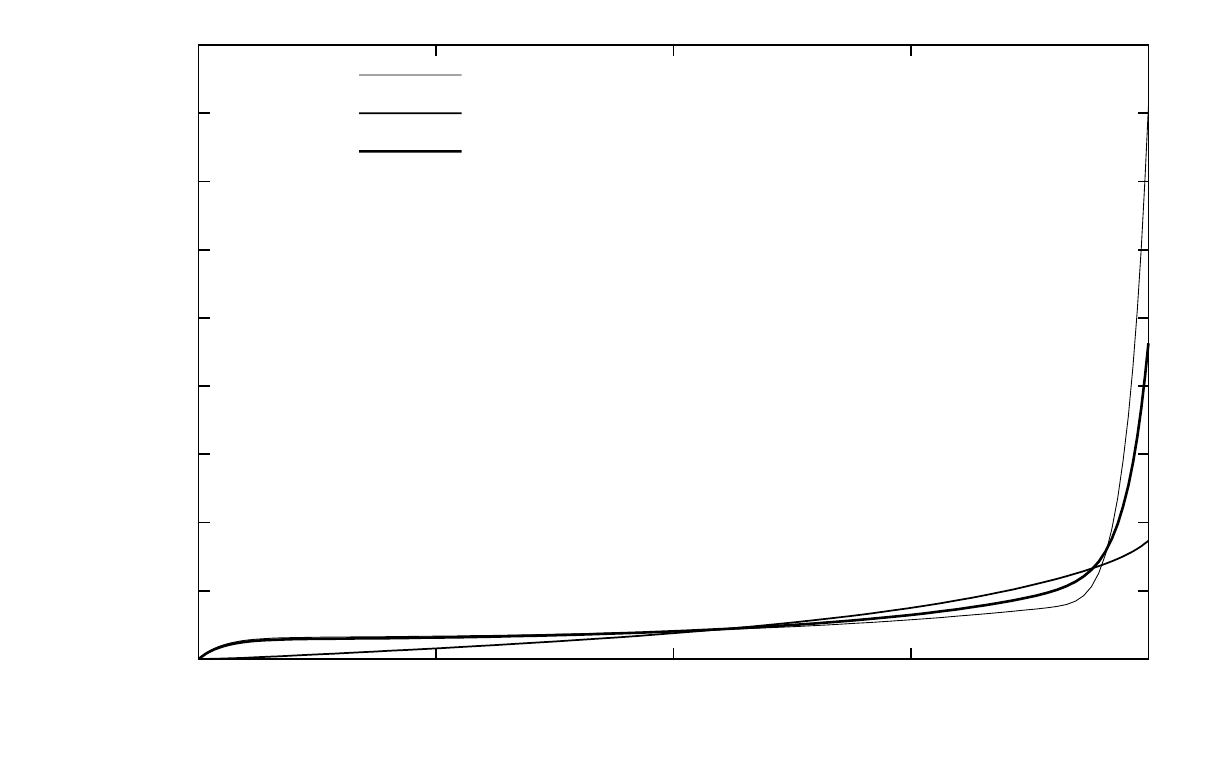}}%
    \gplfronttext
  \end{picture}%
\endgroup

%% file: uv_corrcoeff.tex
\begingroup
  \makeatletter
  \providecommand\color[2][]{%
    \GenericError{(gnuplot) \space\space\space\@spaces}{%
      Package color not loaded in conjunction with
      terminal option `colourtext'%
    }{See the gnuplot documentation for explanation.%
    }{Either use 'blacktext' in gnuplot or load the package
      color.sty in LaTeX.}%
    \renewcommand\color[2][]{}%
  }%
  \providecommand\includegraphics[2][]{%
    \GenericError{(gnuplot) \space\space\space\@spaces}{%
      Package graphicx or graphics not loaded%
    }{See the gnuplot documentation for explanation.%
    }{The gnuplot epslatex terminal needs graphicx.sty or graphics.sty.}%
    \renewcommand\includegraphics[2][]{}%
  }%
  \providecommand\rotatebox[2]{#2}%
  \@ifundefined{ifGPcolor}{%
    \newif\ifGPcolor
    \GPcolorfalse
  }{}%
  \@ifundefined{ifGPblacktext}{%
    \newif\ifGPblacktext
    \GPblacktexttrue
  }{}%
  \let\gplgaddtomacro\g@addto@macro
  \gdef\gplbacktext{}%
  \gdef\gplfronttext{}%
  \makeatother
  \ifGPblacktext
    \def\colorrgb#1{}%
    \def\colorgray#1{}%
  \else
    \ifGPcolor
      \def\colorrgb#1{\color[rgb]{#1}}%
      \def\colorgray#1{\color[gray]{#1}}%
      \expandafter\def\csname LTw\endcsname{\color{white}}%
      \expandafter\def\csname LTb\endcsname{\color{black}}%
      \expandafter\def\csname LTa\endcsname{\color{black}}%
      \expandafter\def\csname LT0\endcsname{\color[rgb]{1,0,0}}%
      \expandafter\def\csname LT1\endcsname{\color[rgb]{0,1,0}}%
      \expandafter\def\csname LT2\endcsname{\color[rgb]{0,0,1}}%
      \expandafter\def\csname LT3\endcsname{\color[rgb]{1,0,1}}%
      \expandafter\def\csname LT4\endcsname{\color[rgb]{0,1,1}}%
      \expandafter\def\csname LT5\endcsname{\color[rgb]{1,1,0}}%
      \expandafter\def\csname LT6\endcsname{\color[rgb]{0,0,0}}%
      \expandafter\def\csname LT7\endcsname{\color[rgb]{1,0.3,0}}%
      \expandafter\def\csname LT8\endcsname{\color[rgb]{0.5,0.5,0.5}}%
    \else
      \def\colorrgb#1{\color{black}}%
      \def\colorgray#1{\color[gray]{#1}}%
      \expandafter\def\csname LTw\endcsname{\color{white}}%
      \expandafter\def\csname LTb\endcsname{\color{black}}%
      \expandafter\def\csname LTa\endcsname{\color{black}}%
      \expandafter\def\csname LT0\endcsname{\color{black}}%
      \expandafter\def\csname LT1\endcsname{\color{black}}%
      \expandafter\def\csname LT2\endcsname{\color{black}}%
      \expandafter\def\csname LT3\endcsname{\color{black}}%
      \expandafter\def\csname LT4\endcsname{\color{black}}%
      \expandafter\def\csname LT5\endcsname{\color{black}}%
      \expandafter\def\csname LT6\endcsname{\color{black}}%
      \expandafter\def\csname LT7\endcsname{\color{black}}%
      \expandafter\def\csname LT8\endcsname{\color{black}}%
    \fi
  \fi
  \setlength{\unitlength}{0.0500bp}%
  \begin{picture}(7012.00,4506.00)%
    \gplgaddtomacro\gplbacktext{%
      \csname LTb\endcsname%
      \put(748,704){\makebox(0,0)[r]{\strut{} 0}}%
      \put(748,1411){\makebox(0,0)[r]{\strut{} 0.1}}%
      \put(748,2119){\makebox(0,0)[r]{\strut{} 0.2}}%
      \put(748,2826){\makebox(0,0)[r]{\strut{} 0.3}}%
      \put(748,3534){\makebox(0,0)[r]{\strut{} 0.4}}%
      \put(748,4241){\makebox(0,0)[r]{\strut{} 0.5}}%
      \put(880,484){\makebox(0,0){\strut{}-0.2}}%
      \put(1836,484){\makebox(0,0){\strut{} 0}}%
      \put(2792,484){\makebox(0,0){\strut{} 0.2}}%
      \put(3748,484){\makebox(0,0){\strut{} 0.4}}%
      \put(4703,484){\makebox(0,0){\strut{} 0.6}}%
      \put(5659,484){\makebox(0,0){\strut{} 0.8}}%
      \put(6615,484){\makebox(0,0){\strut{} 1}}%
      \put(176,2472){\rotatebox{-270}{\makebox(0,0){\strut{}$-C_{uv}$}}}%
      \put(3747,154){\makebox(0,0){\strut{}$y$}}%
    }%
    \gplgaddtomacro\gplfronttext{%
      \csname LTb\endcsname%
      \put(5628,4068){\makebox(0,0)[r]{\strut{}Imp}}%
      \csname LTb\endcsname%
      \put(5628,3848){\makebox(0,0)[r]{\strut{}Por}}%
    }%
    \gplbacktext
    \put(0,0){\includegraphics{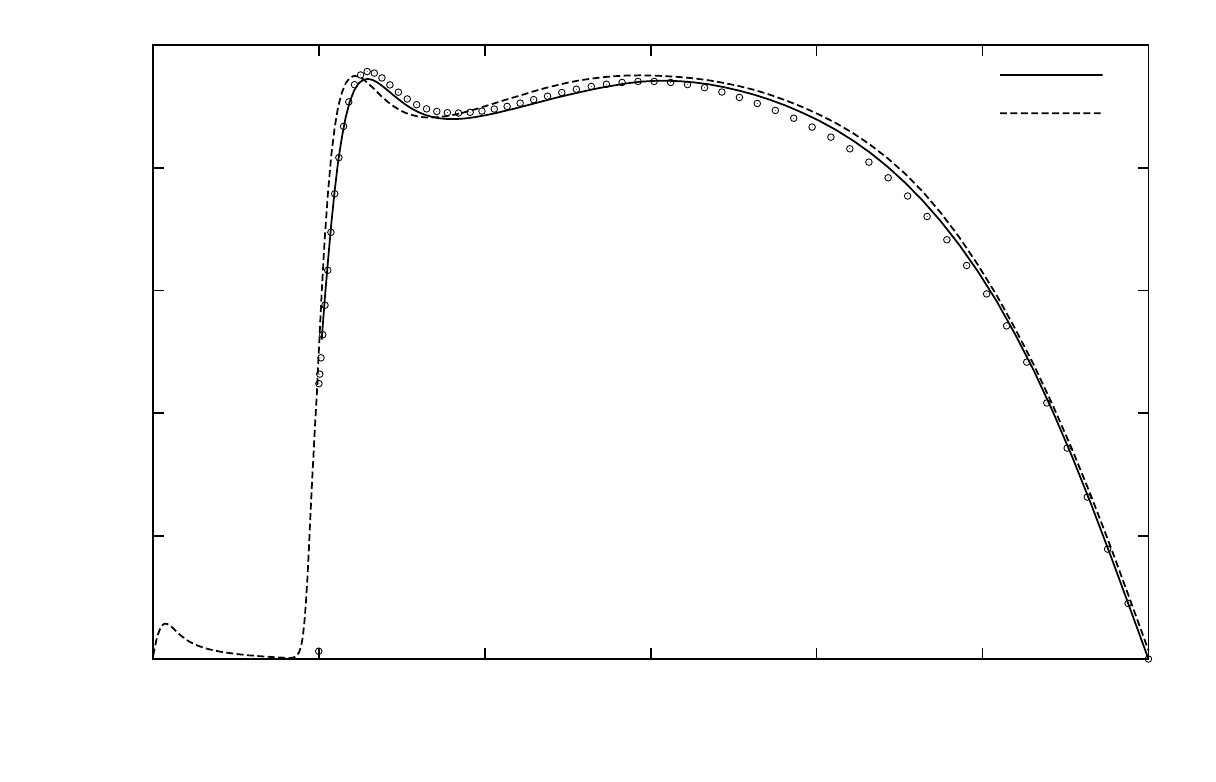}}%
    \gplfronttext
  \end{picture}%
\endgroup

%% file: vort.tex
\begingroup
  \makeatletter
  \providecommand\color[2][]{%
    \GenericError{(gnuplot) \space\space\space\@spaces}{%
      Package color not loaded in conjunction with
      terminal option `colourtext'%
    }{See the gnuplot documentation for explanation.%
    }{Either use 'blacktext' in gnuplot or load the package
      color.sty in LaTeX.}%
    \renewcommand\color[2][]{}%
  }%
  \providecommand\includegraphics[2][]{%
    \GenericError{(gnuplot) \space\space\space\@spaces}{%
      Package graphicx or graphics not loaded%
    }{See the gnuplot documentation for explanation.%
    }{The gnuplot epslatex terminal needs graphicx.sty or graphics.sty.}%
    \renewcommand\includegraphics[2][]{}%
  }%
  \providecommand\rotatebox[2]{#2}%
  \@ifundefined{ifGPcolor}{%
    \newif\ifGPcolor
    \GPcolorfalse
  }{}%
  \@ifundefined{ifGPblacktext}{%
    \newif\ifGPblacktext
    \GPblacktexttrue
  }{}%
  \let\gplgaddtomacro\g@addto@macro
  \gdef\gplbacktext{}%
  \gdef\gplfronttext{}%
  \makeatother
  \ifGPblacktext
    \def\colorrgb#1{}%
    \def\colorgray#1{}%
  \else
    \ifGPcolor
      \def\colorrgb#1{\color[rgb]{#1}}%
      \def\colorgray#1{\color[gray]{#1}}%
      \expandafter\def\csname LTw\endcsname{\color{white}}%
      \expandafter\def\csname LTb\endcsname{\color{black}}%
      \expandafter\def\csname LTa\endcsname{\color{black}}%
      \expandafter\def\csname LT0\endcsname{\color[rgb]{1,0,0}}%
      \expandafter\def\csname LT1\endcsname{\color[rgb]{0,1,0}}%
      \expandafter\def\csname LT2\endcsname{\color[rgb]{0,0,1}}%
      \expandafter\def\csname LT3\endcsname{\color[rgb]{1,0,1}}%
      \expandafter\def\csname LT4\endcsname{\color[rgb]{0,1,1}}%
      \expandafter\def\csname LT5\endcsname{\color[rgb]{1,1,0}}%
      \expandafter\def\csname LT6\endcsname{\color[rgb]{0,0,0}}%
      \expandafter\def\csname LT7\endcsname{\color[rgb]{1,0.3,0}}%
      \expandafter\def\csname LT8\endcsname{\color[rgb]{0.5,0.5,0.5}}%
    \else
      \def\colorrgb#1{\color{black}}%
      \def\colorgray#1{\color[gray]{#1}}%
      \expandafter\def\csname LTw\endcsname{\color{white}}%
      \expandafter\def\csname LTb\endcsname{\color{black}}%
      \expandafter\def\csname LTa\endcsname{\color{black}}%
      \expandafter\def\csname LT0\endcsname{\color{black}}%
      \expandafter\def\csname LT1\endcsname{\color{black}}%
      \expandafter\def\csname LT2\endcsname{\color{black}}%
      \expandafter\def\csname LT3\endcsname{\color{black}}%
      \expandafter\def\csname LT4\endcsname{\color{black}}%
      \expandafter\def\csname LT5\endcsname{\color{black}}%
      \expandafter\def\csname LT6\endcsname{\color{black}}%
      \expandafter\def\csname LT7\endcsname{\color{black}}%
      \expandafter\def\csname LT8\endcsname{\color{black}}%
    \fi
  \fi
  \setlength{\unitlength}{0.0500bp}%
  \begin{picture}(7012.00,4506.00)%
    \gplgaddtomacro\gplbacktext{%
      \csname LTb\endcsname%
      \put(1078,704){\makebox(0,0)[r]{\strut{} 0}}%
      \put(1078,1097){\makebox(0,0)[r]{\strut{} 0.05}}%
      \put(1078,1490){\makebox(0,0)[r]{\strut{} 0.1}}%
      \put(1078,1883){\makebox(0,0)[r]{\strut{} 0.15}}%
      \put(1078,2276){\makebox(0,0)[r]{\strut{} 0.2}}%
      \put(1078,2669){\makebox(0,0)[r]{\strut{} 0.25}}%
      \put(1078,3062){\makebox(0,0)[r]{\strut{} 0.3}}%
      \put(1078,3455){\makebox(0,0)[r]{\strut{} 0.35}}%
      \put(1078,3848){\makebox(0,0)[r]{\strut{} 0.4}}%
      \put(1078,4241){\makebox(0,0)[r]{\strut{} 0.45}}%
      \put(1210,484){\makebox(0,0){\strut{} 1}}%
      \put(3559,484){\makebox(0,0){\strut{} 10}}%
      \put(5908,484){\makebox(0,0){\strut{} 100}}%
      \put(176,2472){\rotatebox{-270}{\makebox(0,0){\strut{}$\omega_\textrm{i,rms}^+$}}}%
      \put(3912,154){\makebox(0,0){\strut{}$y^+$}}%
    }%
    \gplgaddtomacro\gplfronttext{%
      \csname LTb\endcsname%
      \put(5628,4068){\makebox(0,0)[r]{\strut{}$\omega_x$ Imp}}%
      \csname LTb\endcsname%
      \put(5628,3848){\makebox(0,0)[r]{\strut{}$\omega_y$ Imp}}%
      \csname LTb\endcsname%
      \put(5628,3628){\makebox(0,0)[r]{\strut{}$\omega_z$ Imp}}%
      \csname LTb\endcsname%
      \put(5628,3408){\makebox(0,0)[r]{\strut{}$\omega_x$ Por}}%
      \csname LTb\endcsname%
      \put(5628,3188){\makebox(0,0)[r]{\strut{}$\omega_y$ Por}}%
      \csname LTb\endcsname%
      \put(5628,2968){\makebox(0,0)[r]{\strut{}$\omega_z$ Por}}%
    }%
    \gplbacktext
    \put(0,0){\includegraphics{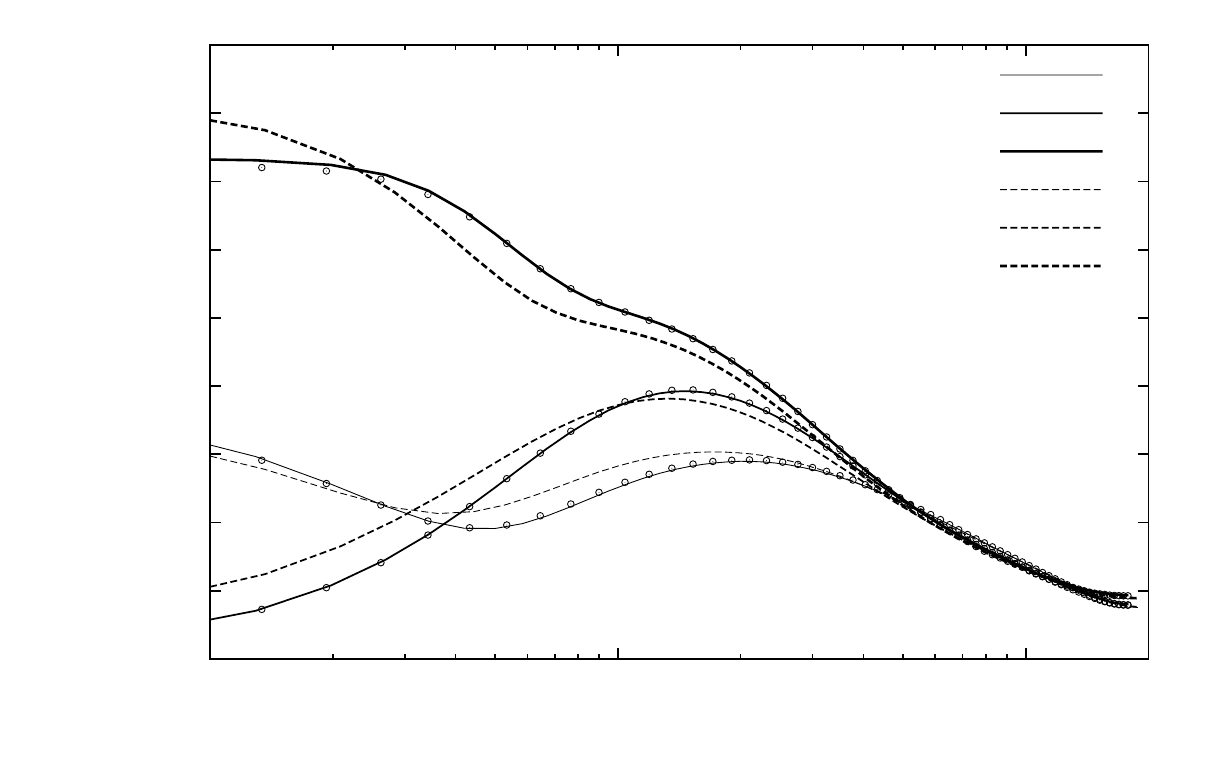}}%
    \gplfronttext
  \end{picture}%
\endgroup

%% file: diss.tex
\begingroup
  \makeatletter
  \providecommand\color[2][]{%
    \GenericError{(gnuplot) \space\space\space\@spaces}{%
      Package color not loaded in conjunction with
      terminal option `colourtext'%
    }{See the gnuplot documentation for explanation.%
    }{Either use 'blacktext' in gnuplot or load the package
      color.sty in LaTeX.}%
    \renewcommand\color[2][]{}%
  }%
  \providecommand\includegraphics[2][]{%
    \GenericError{(gnuplot) \space\space\space\@spaces}{%
      Package graphicx or graphics not loaded%
    }{See the gnuplot documentation for explanation.%
    }{The gnuplot epslatex terminal needs graphicx.sty or graphics.sty.}%
    \renewcommand\includegraphics[2][]{}%
  }%
  \providecommand\rotatebox[2]{#2}%
  \@ifundefined{ifGPcolor}{%
    \newif\ifGPcolor
    \GPcolorfalse
  }{}%
  \@ifundefined{ifGPblacktext}{%
    \newif\ifGPblacktext
    \GPblacktexttrue
  }{}%
  \let\gplgaddtomacro\g@addto@macro
  \gdef\gplbacktext{}%
  \gdef\gplfronttext{}%
  \makeatother
  \ifGPblacktext
    \def\colorrgb#1{}%
    \def\colorgray#1{}%
  \else
    \ifGPcolor
      \def\colorrgb#1{\color[rgb]{#1}}%
      \def\colorgray#1{\color[gray]{#1}}%
      \expandafter\def\csname LTw\endcsname{\color{white}}%
      \expandafter\def\csname LTb\endcsname{\color{black}}%
      \expandafter\def\csname LTa\endcsname{\color{black}}%
      \expandafter\def\csname LT0\endcsname{\color[rgb]{1,0,0}}%
      \expandafter\def\csname LT1\endcsname{\color[rgb]{0,1,0}}%
      \expandafter\def\csname LT2\endcsname{\color[rgb]{0,0,1}}%
      \expandafter\def\csname LT3\endcsname{\color[rgb]{1,0,1}}%
      \expandafter\def\csname LT4\endcsname{\color[rgb]{0,1,1}}%
      \expandafter\def\csname LT5\endcsname{\color[rgb]{1,1,0}}%
      \expandafter\def\csname LT6\endcsname{\color[rgb]{0,0,0}}%
      \expandafter\def\csname LT7\endcsname{\color[rgb]{1,0.3,0}}%
      \expandafter\def\csname LT8\endcsname{\color[rgb]{0.5,0.5,0.5}}%
    \else
      \def\colorrgb#1{\color{black}}%
      \def\colorgray#1{\color[gray]{#1}}%
      \expandafter\def\csname LTw\endcsname{\color{white}}%
      \expandafter\def\csname LTb\endcsname{\color{black}}%
      \expandafter\def\csname LTa\endcsname{\color{black}}%
      \expandafter\def\csname LT0\endcsname{\color{black}}%
      \expandafter\def\csname LT1\endcsname{\color{black}}%
      \expandafter\def\csname LT2\endcsname{\color{black}}%
      \expandafter\def\csname LT3\endcsname{\color{black}}%
      \expandafter\def\csname LT4\endcsname{\color{black}}%
      \expandafter\def\csname LT5\endcsname{\color{black}}%
      \expandafter\def\csname LT6\endcsname{\color{black}}%
      \expandafter\def\csname LT7\endcsname{\color{black}}%
      \expandafter\def\csname LT8\endcsname{\color{black}}%
    \fi
  \fi
  \setlength{\unitlength}{0.0500bp}%
  \begin{picture}(7200.00,4320.00)%
    \gplgaddtomacro\gplbacktext{%
      \csname LTb\endcsname%
      \put(1122,704){\makebox(0,0)[r]{\strut{} 0}}%
      \put(1122,1183){\makebox(0,0)[r]{\strut{} 0.002}}%
      \put(1122,1661){\makebox(0,0)[r]{\strut{} 0.004}}%
      \put(1122,2140){\makebox(0,0)[r]{\strut{} 0.006}}%
      \put(1122,2619){\makebox(0,0)[r]{\strut{} 0.008}}%
      \put(1122,3098){\makebox(0,0)[r]{\strut{} 0.01}}%
      \put(1122,3576){\makebox(0,0)[r]{\strut{} 0.012}}%
      \put(1122,4055){\makebox(0,0)[r]{\strut{} 0.014}}%
      \put(1254,484){\makebox(0,0){\strut{}-0.2}}%
      \put(1590,484){\makebox(0,0){\strut{} 0}}%
      \put(1926,484){\makebox(0,0){\strut{} 0.2}}%
      \put(2262,484){\makebox(0,0){\strut{} 0.4}}%
      \put(2597,484){\makebox(0,0){\strut{} 0.6}}%
      \put(2933,484){\makebox(0,0){\strut{} 0.8}}%
      \put(3269,484){\makebox(0,0){\strut{} 1}}%
      \put(484,2379){\rotatebox{-270}{\makebox(0,0){\strut{}$\epsilon$}}}%
      \put(2261,154){\makebox(0,0){\strut{}$y$}}%
    }%
    \gplgaddtomacro\gplfronttext{%
      \csname LTb\endcsname%
      \put(2282,3882){\makebox(0,0)[r]{\strut{}Imp}}%
      \csname LTb\endcsname%
      \put(2282,3662){\makebox(0,0)[r]{\strut{}Por}}%
    }%
    \gplgaddtomacro\gplbacktext{%
      \csname LTb\endcsname%
      \put(4590,704){\makebox(0,0)[r]{\strut{} 0}}%
      \put(4590,1542){\makebox(0,0)[r]{\strut{} 0.05}}%
      \put(4590,2380){\makebox(0,0)[r]{\strut{} 0.1}}%
      \put(4590,3217){\makebox(0,0)[r]{\strut{} 0.15}}%
      \put(4590,4055){\makebox(0,0)[r]{\strut{} 0.2}}%
      \put(4722,484){\makebox(0,0){\strut{} 1}}%
      \put(5674,484){\makebox(0,0){\strut{} 10}}%
      \put(6626,484){\makebox(0,0){\strut{} 100}}%
      \put(4084,2379){\rotatebox{-270}{\makebox(0,0){\strut{}$\epsilon^+$}}}%
      \put(5795,154){\makebox(0,0){\strut{}$y^+$}}%
    }%
    \gplgaddtomacro\gplfronttext{%
      \csname LTb\endcsname%
      \put(5882,3882){\makebox(0,0)[r]{\strut{}Imp}}%
      \csname LTb\endcsname%
      \put(5882,3662){\makebox(0,0)[r]{\strut{}Por}}%
    }%
    \gplbacktext
    \put(0,0){\includegraphics{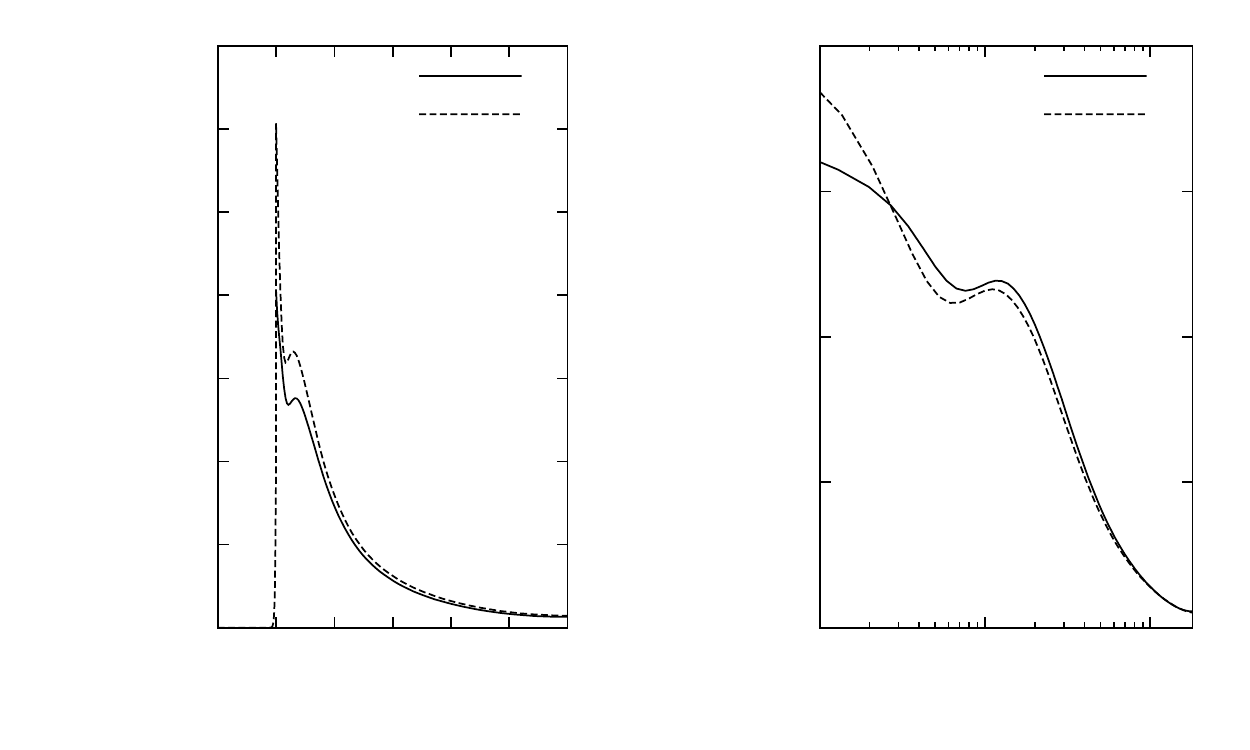}}%
    \gplfronttext
  \end{picture}%
\endgroup

%% file: lumley.tex
\begingroup
  \makeatletter
  \providecommand\color[2][]{%
    \GenericError{(gnuplot) \space\space\space\@spaces}{%
      Package color not loaded in conjunction with
      terminal option `colourtext'%
    }{See the gnuplot documentation for explanation.%
    }{Either use 'blacktext' in gnuplot or load the package
      color.sty in LaTeX.}%
    \renewcommand\color[2][]{}%
  }%
  \providecommand\includegraphics[2][]{%
    \GenericError{(gnuplot) \space\space\space\@spaces}{%
      Package graphicx or graphics not loaded%
    }{See the gnuplot documentation for explanation.%
    }{The gnuplot epslatex terminal needs graphicx.sty or graphics.sty.}%
    \renewcommand\includegraphics[2][]{}%
  }%
  \providecommand\rotatebox[2]{#2}%
  \@ifundefined{ifGPcolor}{%
    \newif\ifGPcolor
    \GPcolorfalse
  }{}%
  \@ifundefined{ifGPblacktext}{%
    \newif\ifGPblacktext
    \GPblacktexttrue
  }{}%
  \let\gplgaddtomacro\g@addto@macro
  \gdef\gplbacktext{}%
  \gdef\gplfronttext{}%
  \makeatother
  \ifGPblacktext
    \def\colorrgb#1{}%
    \def\colorgray#1{}%
  \else
    \ifGPcolor
      \def\colorrgb#1{\color[rgb]{#1}}%
      \def\colorgray#1{\color[gray]{#1}}%
      \expandafter\def\csname LTw\endcsname{\color{white}}%
      \expandafter\def\csname LTb\endcsname{\color{black}}%
      \expandafter\def\csname LTa\endcsname{\color{black}}%
      \expandafter\def\csname LT0\endcsname{\color[rgb]{1,0,0}}%
      \expandafter\def\csname LT1\endcsname{\color[rgb]{0,1,0}}%
      \expandafter\def\csname LT2\endcsname{\color[rgb]{0,0,1}}%
      \expandafter\def\csname LT3\endcsname{\color[rgb]{1,0,1}}%
      \expandafter\def\csname LT4\endcsname{\color[rgb]{0,1,1}}%
      \expandafter\def\csname LT5\endcsname{\color[rgb]{1,1,0}}%
      \expandafter\def\csname LT6\endcsname{\color[rgb]{0,0,0}}%
      \expandafter\def\csname LT7\endcsname{\color[rgb]{1,0.3,0}}%
      \expandafter\def\csname LT8\endcsname{\color[rgb]{0.5,0.5,0.5}}%
    \else
      \def\colorrgb#1{\color{black}}%
      \def\colorgray#1{\color[gray]{#1}}%
      \expandafter\def\csname LTw\endcsname{\color{white}}%
      \expandafter\def\csname LTb\endcsname{\color{black}}%
      \expandafter\def\csname LTa\endcsname{\color{black}}%
      \expandafter\def\csname LT0\endcsname{\color{black}}%
      \expandafter\def\csname LT1\endcsname{\color{black}}%
      \expandafter\def\csname LT2\endcsname{\color{black}}%
      \expandafter\def\csname LT3\endcsname{\color{black}}%
      \expandafter\def\csname LT4\endcsname{\color{black}}%
      \expandafter\def\csname LT5\endcsname{\color{black}}%
      \expandafter\def\csname LT6\endcsname{\color{black}}%
      \expandafter\def\csname LT7\endcsname{\color{black}}%
      \expandafter\def\csname LT8\endcsname{\color{black}}%
    \fi
  \fi
  \setlength{\unitlength}{0.0500bp}%
  \begin{picture}(7012.00,4506.00)%
    \gplgaddtomacro\gplbacktext{%
      \csname LTb\endcsname%
      \put(880,704){\makebox(0,0)[r]{\strut{} 0}}%
      \put(880,1294){\makebox(0,0)[r]{\strut{} 0.05}}%
      \put(880,1883){\makebox(0,0)[r]{\strut{} 0.1}}%
      \put(880,2473){\makebox(0,0)[r]{\strut{} 0.15}}%
      \put(880,3062){\makebox(0,0)[r]{\strut{} 0.2}}%
      \put(880,3652){\makebox(0,0)[r]{\strut{} 0.25}}%
      \put(880,4241){\makebox(0,0)[r]{\strut{} 0.3}}%
      \put(1012,484){\makebox(0,0){\strut{}-0.01}}%
      \put(1946,484){\makebox(0,0){\strut{} 0}}%
      \put(2880,484){\makebox(0,0){\strut{} 0.01}}%
      \put(3813,484){\makebox(0,0){\strut{} 0.02}}%
      \put(4747,484){\makebox(0,0){\strut{} 0.03}}%
      \put(5681,484){\makebox(0,0){\strut{} 0.04}}%
      \put(6615,484){\makebox(0,0){\strut{} 0.05}}%
      \put(176,2472){\rotatebox{-270}{\makebox(0,0){\strut{}$\textrm{-II}$}}}%
      \put(3813,154){\makebox(0,0){\strut{}$\textrm{III}$}}%
      \put(2133,822){\makebox(0,0)[l]{\strut{}$\textrm{iso}$}}%
      \put(6148,3887){\makebox(0,0)[l]{\strut{}$1\textrm{C}$}}%
      \put(1059,2060){\makebox(0,0)[l]{\strut{}$2\textrm{C axi}$}}%
      \put(4280,2473){\makebox(0,0)[l]{\strut{}$\textrm{axi}$}}%
      \put(2880,2590){\makebox(0,0)[l]{\strut{}$2\textrm{C}$}}%
      \put(1087,1270){\makebox(0,0)[l]{\strut{}$\textrm{axi}$}}%
    }%
    \gplgaddtomacro\gplfronttext{%
      \csname LTb\endcsname%
      \put(1540,4068){\makebox(0,0)[r]{\strut{}Imp}}%
      \csname LTb\endcsname%
      \put(1540,3848){\makebox(0,0)[r]{\strut{}Por}}%
    }%
    \gplbacktext
    \put(0,0){\includegraphics{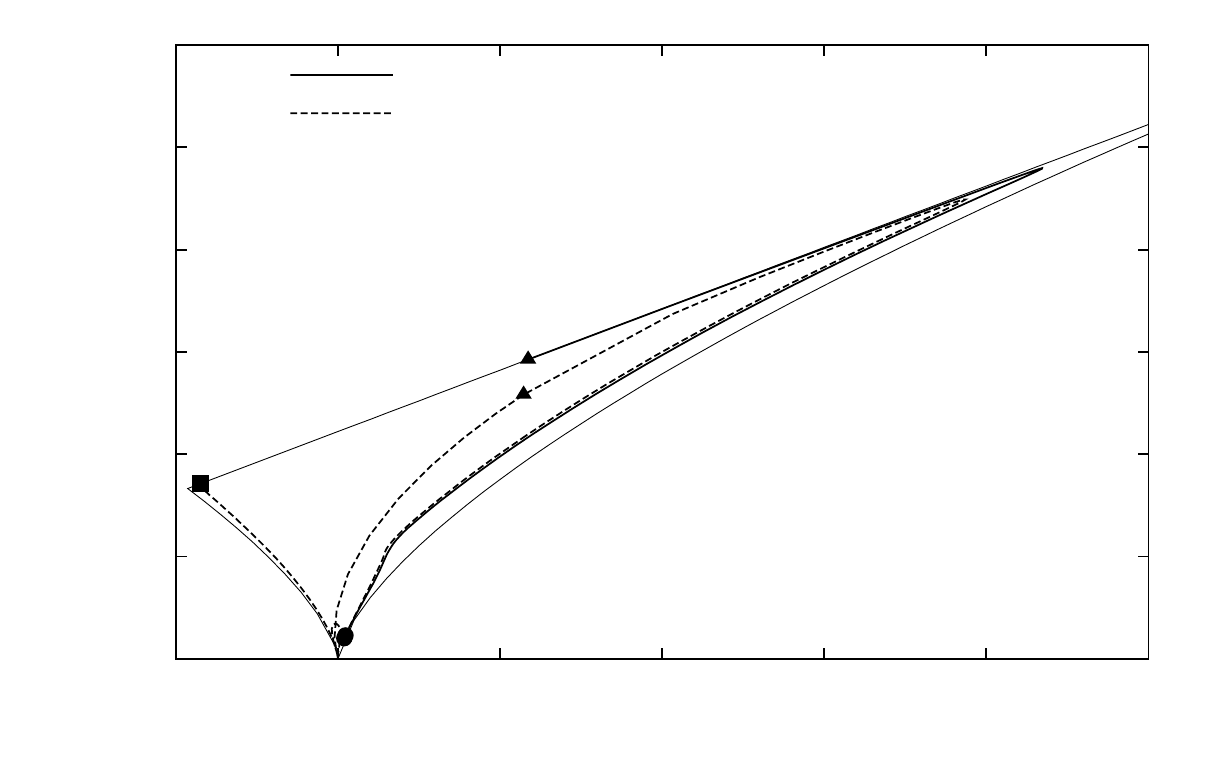}}%
    \gplfronttext
  \end{picture}%
\endgroup

%% file: cor.tex
\begingroup
  \makeatletter
  \providecommand\color[2][]{%
    \GenericError{(gnuplot) \space\space\space\@spaces}{%
      Package color not loaded in conjunction with
      terminal option `colourtext'%
    }{See the gnuplot documentation for explanation.%
    }{Either use 'blacktext' in gnuplot or load the package
      color.sty in LaTeX.}%
    \renewcommand\color[2][]{}%
  }%
  \providecommand\includegraphics[2][]{%
    \GenericError{(gnuplot) \space\space\space\@spaces}{%
      Package graphicx or graphics not loaded%
    }{See the gnuplot documentation for explanation.%
    }{The gnuplot epslatex terminal needs graphicx.sty or graphics.sty.}%
    \renewcommand\includegraphics[2][]{}%
  }%
  \providecommand\rotatebox[2]{#2}%
  \@ifundefined{ifGPcolor}{%
    \newif\ifGPcolor
    \GPcolorfalse
  }{}%
  \@ifundefined{ifGPblacktext}{%
    \newif\ifGPblacktext
    \GPblacktexttrue
  }{}%
  \let\gplgaddtomacro\g@addto@macro
  \gdef\gplbacktext{}%
  \gdef\gplfronttext{}%
  \makeatother
  \ifGPblacktext
    \def\colorrgb#1{}%
    \def\colorgray#1{}%
  \else
    \ifGPcolor
      \def\colorrgb#1{\color[rgb]{#1}}%
      \def\colorgray#1{\color[gray]{#1}}%
      \expandafter\def\csname LTw\endcsname{\color{white}}%
      \expandafter\def\csname LTb\endcsname{\color{black}}%
      \expandafter\def\csname LTa\endcsname{\color{black}}%
      \expandafter\def\csname LT0\endcsname{\color[rgb]{1,0,0}}%
      \expandafter\def\csname LT1\endcsname{\color[rgb]{0,1,0}}%
      \expandafter\def\csname LT2\endcsname{\color[rgb]{0,0,1}}%
      \expandafter\def\csname LT3\endcsname{\color[rgb]{1,0,1}}%
      \expandafter\def\csname LT4\endcsname{\color[rgb]{0,1,1}}%
      \expandafter\def\csname LT5\endcsname{\color[rgb]{1,1,0}}%
      \expandafter\def\csname LT6\endcsname{\color[rgb]{0,0,0}}%
      \expandafter\def\csname LT7\endcsname{\color[rgb]{1,0.3,0}}%
      \expandafter\def\csname LT8\endcsname{\color[rgb]{0.5,0.5,0.5}}%
    \else
      \def\colorrgb#1{\color{black}}%
      \def\colorgray#1{\color[gray]{#1}}%
      \expandafter\def\csname LTw\endcsname{\color{white}}%
      \expandafter\def\csname LTb\endcsname{\color{black}}%
      \expandafter\def\csname LTa\endcsname{\color{black}}%
      \expandafter\def\csname LT0\endcsname{\color{black}}%
      \expandafter\def\csname LT1\endcsname{\color{black}}%
      \expandafter\def\csname LT2\endcsname{\color{black}}%
      \expandafter\def\csname LT3\endcsname{\color{black}}%
      \expandafter\def\csname LT4\endcsname{\color{black}}%
      \expandafter\def\csname LT5\endcsname{\color{black}}%
      \expandafter\def\csname LT6\endcsname{\color{black}}%
      \expandafter\def\csname LT7\endcsname{\color{black}}%
      \expandafter\def\csname LT8\endcsname{\color{black}}%
    \fi
  \fi
  \setlength{\unitlength}{0.0500bp}%
  \begin{picture}(7012.00,5760.00)%
    \gplgaddtomacro\gplbacktext{%
      \csname LTb\endcsname%
      \put(748,3186){\makebox(0,0)[r]{\strut{}-0.4}}%
      \put(748,3433){\makebox(0,0)[r]{\strut{}-0.2}}%
      \put(748,3680){\makebox(0,0)[r]{\strut{} 0}}%
      \put(748,3927){\makebox(0,0)[r]{\strut{} 0.2}}%
      \put(748,4175){\makebox(0,0)[r]{\strut{} 0.4}}%
      \put(748,4422){\makebox(0,0)[r]{\strut{} 0.6}}%
      \put(748,4669){\makebox(0,0)[r]{\strut{} 0.8}}%
      \put(748,4916){\makebox(0,0)[r]{\strut{} 1}}%
      \put(880,2966){\makebox(0,0){\strut{} 0}}%
      \put(1235,2966){\makebox(0,0){\strut{} 1}}%
      \put(1590,2966){\makebox(0,0){\strut{} 2}}%
      \put(1945,2966){\makebox(0,0){\strut{} 3}}%
      \put(2300,2966){\makebox(0,0){\strut{} 4}}%
      \put(2655,2966){\makebox(0,0){\strut{} 5}}%
      \put(3010,2966){\makebox(0,0){\strut{} 6}}%
      \put(176,4051){\rotatebox{-270}{\makebox(0,0){\strut{}$R_{\textrm{ii}}$}}}%
      \put(1994,2746){\makebox(0,0){\strut{}$x$}}%
    }%
    \gplgaddtomacro\gplfronttext{%
      \csname LTb\endcsname%
      \put(2769,5609){\makebox(0,0)[r]{\strut{}u Imp}}%
      \csname LTb\endcsname%
      \put(2769,5389){\makebox(0,0)[r]{\strut{}v Imp}}%
      \csname LTb\endcsname%
      \put(2769,5169){\makebox(0,0)[r]{\strut{}w Imp}}%
      \csname LTb\endcsname%
      \put(4284,5609){\makebox(0,0)[r]{\strut{}u Por}}%
      \csname LTb\endcsname%
      \put(4284,5389){\makebox(0,0)[r]{\strut{}v Por}}%
      \csname LTb\endcsname%
      \put(4284,5169){\makebox(0,0)[r]{\strut{}w Por}}%
    }%
    \gplgaddtomacro\gplbacktext{%
      \csname LTb\endcsname%
      \put(4254,3186){\makebox(0,0)[r]{\strut{}-0.4}}%
      \put(4254,3433){\makebox(0,0)[r]{\strut{}-0.2}}%
      \put(4254,3680){\makebox(0,0)[r]{\strut{} 0}}%
      \put(4254,3927){\makebox(0,0)[r]{\strut{} 0.2}}%
      \put(4254,4175){\makebox(0,0)[r]{\strut{} 0.4}}%
      \put(4254,4422){\makebox(0,0)[r]{\strut{} 0.6}}%
      \put(4254,4669){\makebox(0,0)[r]{\strut{} 0.8}}%
      \put(4254,4916){\makebox(0,0)[r]{\strut{} 1}}%
      \put(4386,2966){\makebox(0,0){\strut{} 0}}%
      \put(4741,2966){\makebox(0,0){\strut{} 1}}%
      \put(5096,2966){\makebox(0,0){\strut{} 2}}%
      \put(5451,2966){\makebox(0,0){\strut{} 3}}%
      \put(5806,2966){\makebox(0,0){\strut{} 4}}%
      \put(6161,2966){\makebox(0,0){\strut{} 5}}%
      \put(6516,2966){\makebox(0,0){\strut{} 6}}%
      \put(3682,4051){\rotatebox{-270}{\makebox(0,0){\strut{}$R_{\textrm{ii}}$}}}%
      \put(5500,2746){\makebox(0,0){\strut{}$x$}}%
    }%
    \gplgaddtomacro\gplfronttext{%
    }%
    \gplgaddtomacro\gplbacktext{%
      \csname LTb\endcsname%
      \put(748,594){\makebox(0,0)[r]{\strut{}-0.4}}%
      \put(748,841){\makebox(0,0)[r]{\strut{}-0.2}}%
      \put(748,1088){\makebox(0,0)[r]{\strut{} 0}}%
      \put(748,1335){\makebox(0,0)[r]{\strut{} 0.2}}%
      \put(748,1583){\makebox(0,0)[r]{\strut{} 0.4}}%
      \put(748,1830){\makebox(0,0)[r]{\strut{} 0.6}}%
      \put(748,2077){\makebox(0,0)[r]{\strut{} 0.8}}%
      \put(748,2324){\makebox(0,0)[r]{\strut{} 1}}%
      \put(880,374){\makebox(0,0){\strut{} 0}}%
      \put(1235,374){\makebox(0,0){\strut{} 0.5}}%
      \put(1590,374){\makebox(0,0){\strut{} 1}}%
      \put(1945,374){\makebox(0,0){\strut{} 1.5}}%
      \put(2300,374){\makebox(0,0){\strut{} 2}}%
      \put(2655,374){\makebox(0,0){\strut{} 2.5}}%
      \put(3010,374){\makebox(0,0){\strut{} 3}}%
      \put(176,1459){\rotatebox{-270}{\makebox(0,0){\strut{}$R_{\textrm{ii}}$}}}%
      \put(1994,154){\makebox(0,0){\strut{}$z$}}%
    }%
    \gplgaddtomacro\gplfronttext{%
    }%
    \gplgaddtomacro\gplbacktext{%
      \csname LTb\endcsname%
      \put(4254,594){\makebox(0,0)[r]{\strut{}-0.4}}%
      \put(4254,841){\makebox(0,0)[r]{\strut{}-0.2}}%
      \put(4254,1088){\makebox(0,0)[r]{\strut{} 0}}%
      \put(4254,1335){\makebox(0,0)[r]{\strut{} 0.2}}%
      \put(4254,1583){\makebox(0,0)[r]{\strut{} 0.4}}%
      \put(4254,1830){\makebox(0,0)[r]{\strut{} 0.6}}%
      \put(4254,2077){\makebox(0,0)[r]{\strut{} 0.8}}%
      \put(4254,2324){\makebox(0,0)[r]{\strut{} 1}}%
      \put(4386,374){\makebox(0,0){\strut{} 0}}%
      \put(4741,374){\makebox(0,0){\strut{} 0.5}}%
      \put(5096,374){\makebox(0,0){\strut{} 1}}%
      \put(5451,374){\makebox(0,0){\strut{} 1.5}}%
      \put(5806,374){\makebox(0,0){\strut{} 2}}%
      \put(6161,374){\makebox(0,0){\strut{} 2.5}}%
      \put(6516,374){\makebox(0,0){\strut{} 3}}%
      \put(3682,1459){\rotatebox{-270}{\makebox(0,0){\strut{}$R_{\textrm{ii}}$}}}%
      \put(5500,154){\makebox(0,0){\strut{}$z$}}%
    }%
    \gplgaddtomacro\gplfronttext{%
    }%
    \gplbacktext
    \put(0,0){\includegraphics{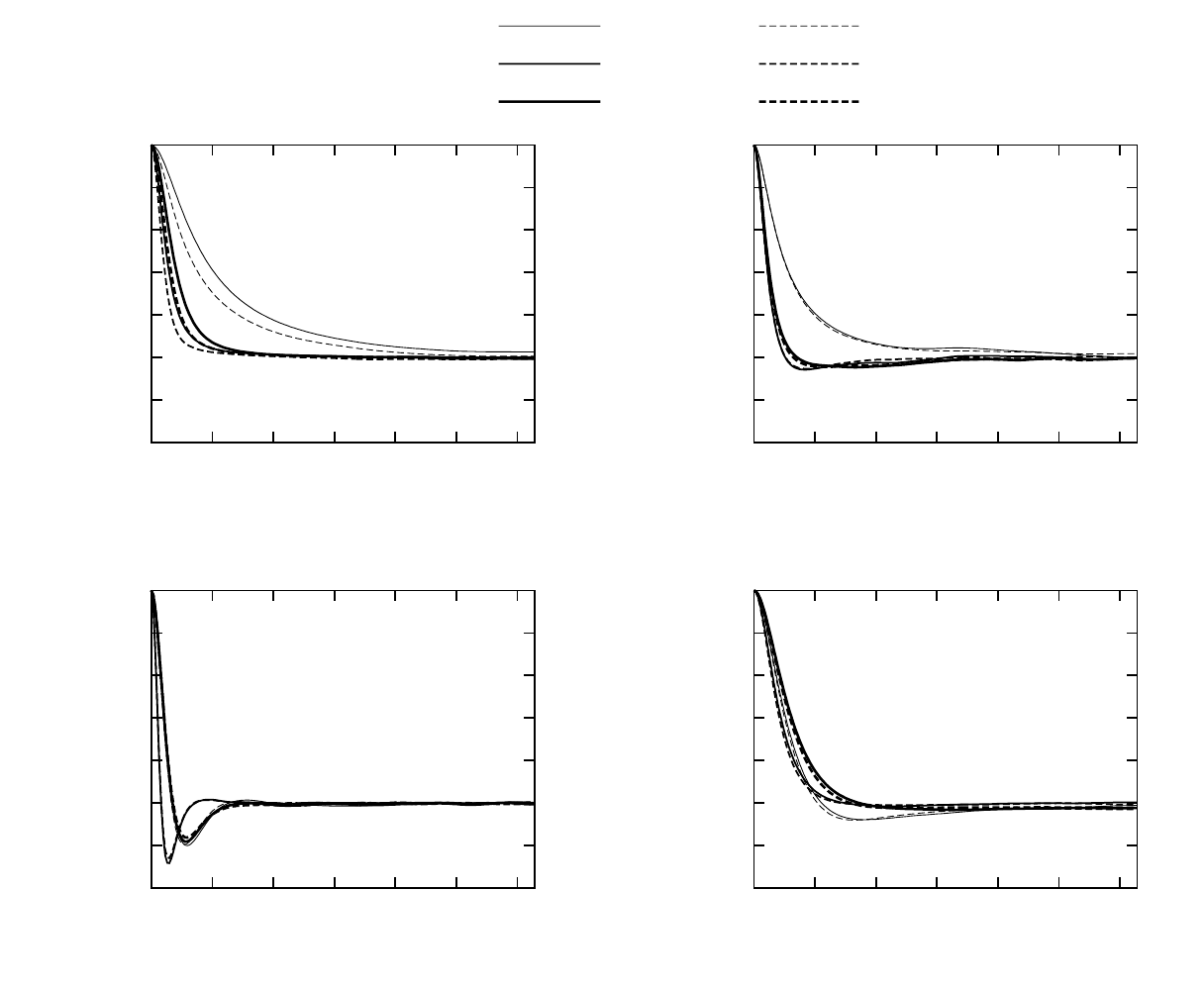}}%
    \gplfronttext
  \end{picture}%
\endgroup

%% file: loglaw_sigma.tex
\begingroup
  \makeatletter
  \providecommand\color[2][]{%
    \GenericError{(gnuplot) \space\space\space\@spaces}{%
      Package color not loaded in conjunction with
      terminal option `colourtext'%
    }{See the gnuplot documentation for explanation.%
    }{Either use 'blacktext' in gnuplot or load the package
      color.sty in LaTeX.}%
    \renewcommand\color[2][]{}%
  }%
  \providecommand\includegraphics[2][]{%
    \GenericError{(gnuplot) \space\space\space\@spaces}{%
      Package graphicx or graphics not loaded%
    }{See the gnuplot documentation for explanation.%
    }{The gnuplot epslatex terminal needs graphicx.sty or graphics.sty.}%
    \renewcommand\includegraphics[2][]{}%
  }%
  \providecommand\rotatebox[2]{#2}%
  \@ifundefined{ifGPcolor}{%
    \newif\ifGPcolor
    \GPcolorfalse
  }{}%
  \@ifundefined{ifGPblacktext}{%
    \newif\ifGPblacktext
    \GPblacktexttrue
  }{}%
  \let\gplgaddtomacro\g@addto@macro
  \gdef\gplbacktext{}%
  \gdef\gplfronttext{}%
  \makeatother
  \ifGPblacktext
    \def\colorrgb#1{}%
    \def\colorgray#1{}%
  \else
    \ifGPcolor
      \def\colorrgb#1{\color[rgb]{#1}}%
      \def\colorgray#1{\color[gray]{#1}}%
      \expandafter\def\csname LTw\endcsname{\color{white}}%
      \expandafter\def\csname LTb\endcsname{\color{black}}%
      \expandafter\def\csname LTa\endcsname{\color{black}}%
      \expandafter\def\csname LT0\endcsname{\color[rgb]{1,0,0}}%
      \expandafter\def\csname LT1\endcsname{\color[rgb]{0,1,0}}%
      \expandafter\def\csname LT2\endcsname{\color[rgb]{0,0,1}}%
      \expandafter\def\csname LT3\endcsname{\color[rgb]{1,0,1}}%
      \expandafter\def\csname LT4\endcsname{\color[rgb]{0,1,1}}%
      \expandafter\def\csname LT5\endcsname{\color[rgb]{1,1,0}}%
      \expandafter\def\csname LT6\endcsname{\color[rgb]{0,0,0}}%
      \expandafter\def\csname LT7\endcsname{\color[rgb]{1,0.3,0}}%
      \expandafter\def\csname LT8\endcsname{\color[rgb]{0.5,0.5,0.5}}%
    \else
      \def\colorrgb#1{\color{black}}%
      \def\colorgray#1{\color[gray]{#1}}%
      \expandafter\def\csname LTw\endcsname{\color{white}}%
      \expandafter\def\csname LTb\endcsname{\color{black}}%
      \expandafter\def\csname LTa\endcsname{\color{black}}%
      \expandafter\def\csname LT0\endcsname{\color{black}}%
      \expandafter\def\csname LT1\endcsname{\color{black}}%
      \expandafter\def\csname LT2\endcsname{\color{black}}%
      \expandafter\def\csname LT3\endcsname{\color{black}}%
      \expandafter\def\csname LT4\endcsname{\color{black}}%
      \expandafter\def\csname LT5\endcsname{\color{black}}%
      \expandafter\def\csname LT6\endcsname{\color{black}}%
      \expandafter\def\csname LT7\endcsname{\color{black}}%
      \expandafter\def\csname LT8\endcsname{\color{black}}%
    \fi
  \fi
  \setlength{\unitlength}{0.0500bp}%
  \begin{picture}(7012.00,4506.00)%
    \gplgaddtomacro\gplbacktext{%
      \csname LTb\endcsname%
      \put(594,704){\makebox(0,0)[r]{\strut{} 0}}%
      \put(594,1588){\makebox(0,0)[r]{\strut{} 5}}%
      \put(594,2473){\makebox(0,0)[r]{\strut{} 10}}%
      \put(594,3357){\makebox(0,0)[r]{\strut{} 15}}%
      \put(594,4241){\makebox(0,0)[r]{\strut{} 20}}%
      \put(726,484){\makebox(0,0){\strut{} 1}}%
      \put(3103,484){\makebox(0,0){\strut{} 10}}%
      \put(5481,484){\makebox(0,0){\strut{} 100}}%
      \put(220,2472){\rotatebox{-270}{\makebox(0,0){\strut{}$(\overline{u}-U_i)^+$}}}%
      \put(3670,154){\makebox(0,0){\strut{}$y^+$}}%
    }%
    \gplgaddtomacro\gplfronttext{%
      \csname LTb\endcsname%
      \put(2046,4068){\makebox(0,0)[r]{\strut{}Imp}}%
      \csname LTb\endcsname%
      \put(2046,3848){\makebox(0,0)[r]{\strut{}$\sigma=0.00025$}}%
      \csname LTb\endcsname%
      \put(2046,3628){\makebox(0,0)[r]{\strut{}$\sigma=0.00050$}}%
      \csname LTb\endcsname%
      \put(2046,3408){\makebox(0,0)[r]{\strut{}$\sigma=0.00100$}}%
      \csname LTb\endcsname%
      \put(2046,3188){\makebox(0,0)[r]{\strut{}$\sigma=0.00200$}}%
      \csname LTb\endcsname%
      \put(2046,2968){\makebox(0,0)[r]{\strut{}$\sigma=0.00400$}}%
    }%
    \gplbacktext
    \put(0,0){\includegraphics{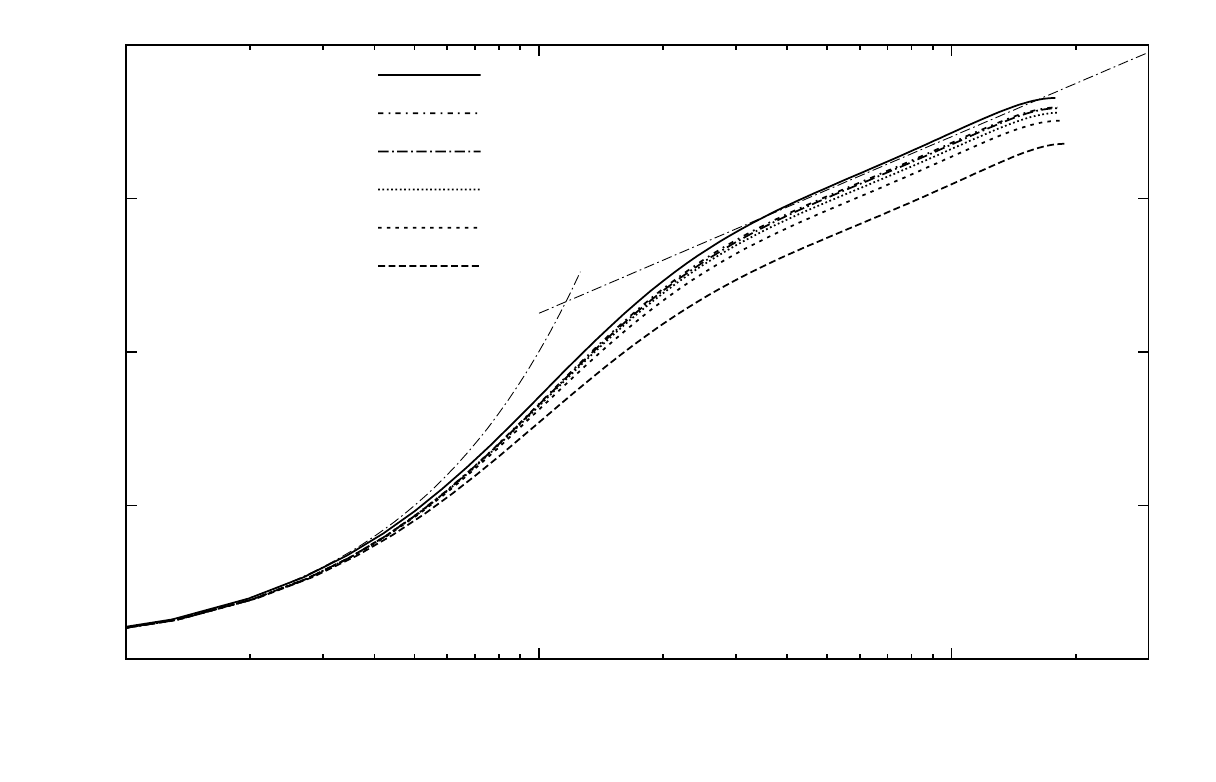}}%
    \gplfronttext
  \end{picture}%
\endgroup

%% file: rms_sigma.tex
\begingroup
  \makeatletter
  \providecommand\color[2][]{%
    \GenericError{(gnuplot) \space\space\space\@spaces}{%
      Package color not loaded in conjunction with
      terminal option `colourtext'%
    }{See the gnuplot documentation for explanation.%
    }{Either use 'blacktext' in gnuplot or load the package
      color.sty in LaTeX.}%
    \renewcommand\color[2][]{}%
  }%
  \providecommand\includegraphics[2][]{%
    \GenericError{(gnuplot) \space\space\space\@spaces}{%
      Package graphicx or graphics not loaded%
    }{See the gnuplot documentation for explanation.%
    }{The gnuplot epslatex terminal needs graphicx.sty or graphics.sty.}%
    \renewcommand\includegraphics[2][]{}%
  }%
  \providecommand\rotatebox[2]{#2}%
  \@ifundefined{ifGPcolor}{%
    \newif\ifGPcolor
    \GPcolorfalse
  }{}%
  \@ifundefined{ifGPblacktext}{%
    \newif\ifGPblacktext
    \GPblacktexttrue
  }{}%
  \let\gplgaddtomacro\g@addto@macro
  \gdef\gplbacktext{}%
  \gdef\gplfronttext{}%
  \makeatother
  \ifGPblacktext
    \def\colorrgb#1{}%
    \def\colorgray#1{}%
  \else
    \ifGPcolor
      \def\colorrgb#1{\color[rgb]{#1}}%
      \def\colorgray#1{\color[gray]{#1}}%
      \expandafter\def\csname LTw\endcsname{\color{white}}%
      \expandafter\def\csname LTb\endcsname{\color{black}}%
      \expandafter\def\csname LTa\endcsname{\color{black}}%
      \expandafter\def\csname LT0\endcsname{\color[rgb]{1,0,0}}%
      \expandafter\def\csname LT1\endcsname{\color[rgb]{0,1,0}}%
      \expandafter\def\csname LT2\endcsname{\color[rgb]{0,0,1}}%
      \expandafter\def\csname LT3\endcsname{\color[rgb]{1,0,1}}%
      \expandafter\def\csname LT4\endcsname{\color[rgb]{0,1,1}}%
      \expandafter\def\csname LT5\endcsname{\color[rgb]{1,1,0}}%
      \expandafter\def\csname LT6\endcsname{\color[rgb]{0,0,0}}%
      \expandafter\def\csname LT7\endcsname{\color[rgb]{1,0.3,0}}%
      \expandafter\def\csname LT8\endcsname{\color[rgb]{0.5,0.5,0.5}}%
    \else
      \def\colorrgb#1{\color{black}}%
      \def\colorgray#1{\color[gray]{#1}}%
      \expandafter\def\csname LTw\endcsname{\color{white}}%
      \expandafter\def\csname LTb\endcsname{\color{black}}%
      \expandafter\def\csname LTa\endcsname{\color{black}}%
      \expandafter\def\csname LT0\endcsname{\color{black}}%
      \expandafter\def\csname LT1\endcsname{\color{black}}%
      \expandafter\def\csname LT2\endcsname{\color{black}}%
      \expandafter\def\csname LT3\endcsname{\color{black}}%
      \expandafter\def\csname LT4\endcsname{\color{black}}%
      \expandafter\def\csname LT5\endcsname{\color{black}}%
      \expandafter\def\csname LT6\endcsname{\color{black}}%
      \expandafter\def\csname LT7\endcsname{\color{black}}%
      \expandafter\def\csname LT8\endcsname{\color{black}}%
    \fi
  \fi
  \setlength{\unitlength}{0.0500bp}%
  \begin{picture}(7012.00,7012.00)%
    \gplgaddtomacro\gplbacktext{%
      \csname LTb\endcsname%
      \put(726,3859){\makebox(0,0)[r]{\strut{} 0}}%
      \put(726,4229){\makebox(0,0)[r]{\strut{} 0.5}}%
      \put(726,4599){\makebox(0,0)[r]{\strut{} 1}}%
      \put(726,4970){\makebox(0,0)[r]{\strut{} 1.5}}%
      \put(726,5340){\makebox(0,0)[r]{\strut{} 2}}%
      \put(726,5710){\makebox(0,0)[r]{\strut{} 2.5}}%
      \put(726,6080){\makebox(0,0)[r]{\strut{} 3}}%
      \put(858,3639){\makebox(0,0){\strut{} 1}}%
      \put(1836,3639){\makebox(0,0){\strut{} 10}}%
      \put(2815,3639){\makebox(0,0){\strut{} 100}}%
      \put(352,4969){\rotatebox{-270}{\makebox(0,0){\strut{}$u_\textrm{rms}/u_\tau$}}}%
      \put(1983,3309){\makebox(0,0){\strut{}$y^+$}}%
    }%
    \gplgaddtomacro\gplfronttext{%
      \csname LTb\endcsname%
      \put(2604,6562){\makebox(0,0)[r]{\strut{}Imp}}%
      \csname LTb\endcsname%
      \put(2604,6342){\makebox(0,0)[r]{\strut{}$\sigma=0.004$}}%
      \csname LTb\endcsname%
      \put(4383,6562){\makebox(0,0)[r]{\strut{}$\sigma=0.002$}}%
      \csname LTb\endcsname%
      \put(4383,6342){\makebox(0,0)[r]{\strut{}$\sigma=0.001$}}%
    }%
    \gplgaddtomacro\gplbacktext{%
      \csname LTb\endcsname%
      \put(4232,3859){\makebox(0,0)[r]{\strut{} 0}}%
      \put(4232,4229){\makebox(0,0)[r]{\strut{} 0.5}}%
      \put(4232,4599){\makebox(0,0)[r]{\strut{} 1}}%
      \put(4232,4970){\makebox(0,0)[r]{\strut{} 1.5}}%
      \put(4232,5340){\makebox(0,0)[r]{\strut{} 2}}%
      \put(4232,5710){\makebox(0,0)[r]{\strut{} 2.5}}%
      \put(4232,6080){\makebox(0,0)[r]{\strut{} 3}}%
      \put(4364,3639){\makebox(0,0){\strut{} 1}}%
      \put(5342,3639){\makebox(0,0){\strut{} 10}}%
      \put(6321,3639){\makebox(0,0){\strut{} 100}}%
      \put(3858,4969){\rotatebox{-270}{\makebox(0,0){\strut{}$v_\textrm{rms}/u_\tau$}}}%
      \put(5489,3309){\makebox(0,0){\strut{}$y^+$}}%
    }%
    \gplgaddtomacro\gplfronttext{%
    }%
    \gplgaddtomacro\gplbacktext{%
      \csname LTb\endcsname%
      \put(726,704){\makebox(0,0)[r]{\strut{} 0}}%
      \put(726,1074){\makebox(0,0)[r]{\strut{} 0.5}}%
      \put(726,1444){\makebox(0,0)[r]{\strut{} 1}}%
      \put(726,1815){\makebox(0,0)[r]{\strut{} 1.5}}%
      \put(726,2185){\makebox(0,0)[r]{\strut{} 2}}%
      \put(726,2555){\makebox(0,0)[r]{\strut{} 2.5}}%
      \put(726,2925){\makebox(0,0)[r]{\strut{} 3}}%
      \put(858,484){\makebox(0,0){\strut{} 1}}%
      \put(1836,484){\makebox(0,0){\strut{} 10}}%
      \put(2815,484){\makebox(0,0){\strut{} 100}}%
      \put(352,1814){\rotatebox{-270}{\makebox(0,0){\strut{}$w_\textrm{rms}/u_\tau$}}}%
      \put(1983,154){\makebox(0,0){\strut{}$y^+$}}%
    }%
    \gplgaddtomacro\gplfronttext{%
    }%
    \gplgaddtomacro\gplbacktext{%
      \csname LTb\endcsname%
      \put(4232,704){\makebox(0,0)[r]{\strut{} 0}}%
      \put(4232,1148){\makebox(0,0)[r]{\strut{} 0.1}}%
      \put(4232,1592){\makebox(0,0)[r]{\strut{} 0.2}}%
      \put(4232,2037){\makebox(0,0)[r]{\strut{} 0.3}}%
      \put(4232,2481){\makebox(0,0)[r]{\strut{} 0.4}}%
      \put(4232,2925){\makebox(0,0)[r]{\strut{} 0.5}}%
      \put(4364,484){\makebox(0,0){\strut{}-0.2}}%
      \put(4739,484){\makebox(0,0){\strut{} 0}}%
      \put(5114,484){\makebox(0,0){\strut{} 0.2}}%
      \put(5490,484){\makebox(0,0){\strut{} 0.4}}%
      \put(5865,484){\makebox(0,0){\strut{} 0.6}}%
      \put(6240,484){\makebox(0,0){\strut{} 0.8}}%
      \put(6615,484){\makebox(0,0){\strut{} 1}}%
      \put(3858,1814){\rotatebox{-270}{\makebox(0,0){\strut{}$-C_{uv}$}}}%
      \put(5489,154){\makebox(0,0){\strut{}$y$}}%
    }%
    \gplgaddtomacro\gplfronttext{%
    }%
    \gplbacktext
    \put(0,0){\includegraphics{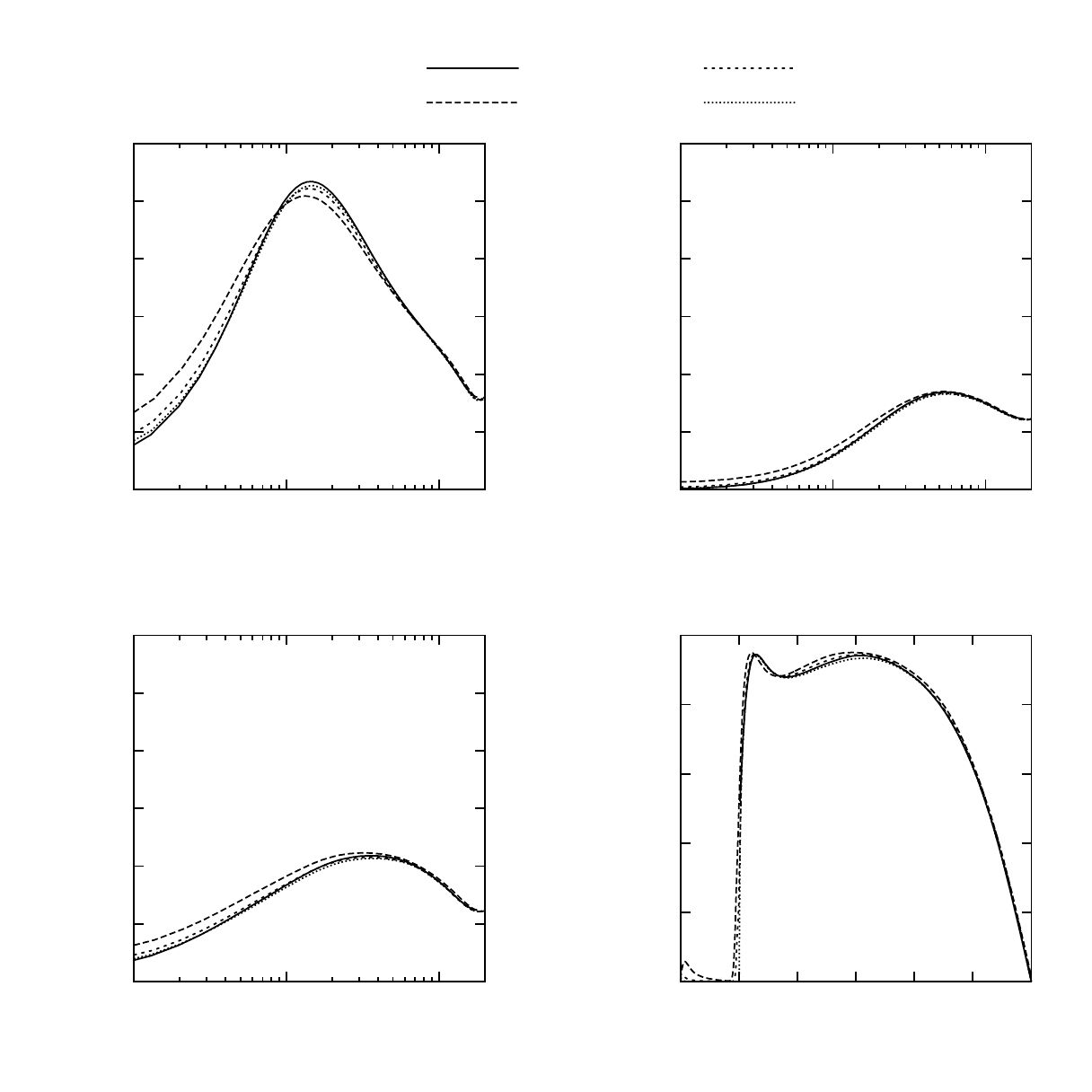}}%
    \gplfronttext
  \end{picture}%
\endgroup

%% file: rmsinterface-sigma.tex
\begingroup
  \makeatletter
  \providecommand\color[2][]{%
    \GenericError{(gnuplot) \space\space\space\@spaces}{%
      Package color not loaded in conjunction with
      terminal option `colourtext'%
    }{See the gnuplot documentation for explanation.%
    }{Either use 'blacktext' in gnuplot or load the package
      color.sty in LaTeX.}%
    \renewcommand\color[2][]{}%
  }%
  \providecommand\includegraphics[2][]{%
    \GenericError{(gnuplot) \space\space\space\@spaces}{%
      Package graphicx or graphics not loaded%
    }{See the gnuplot documentation for explanation.%
    }{The gnuplot epslatex terminal needs graphicx.sty or graphics.sty.}%
    \renewcommand\includegraphics[2][]{}%
  }%
  \providecommand\rotatebox[2]{#2}%
  \@ifundefined{ifGPcolor}{%
    \newif\ifGPcolor
    \GPcolorfalse
  }{}%
  \@ifundefined{ifGPblacktext}{%
    \newif\ifGPblacktext
    \GPblacktexttrue
  }{}%
  \let\gplgaddtomacro\g@addto@macro
  \gdef\gplbacktext{}%
  \gdef\gplfronttext{}%
  \makeatother
  \ifGPblacktext
    \def\colorrgb#1{}%
    \def\colorgray#1{}%
  \else
    \ifGPcolor
      \def\colorrgb#1{\color[rgb]{#1}}%
      \def\colorgray#1{\color[gray]{#1}}%
      \expandafter\def\csname LTw\endcsname{\color{white}}%
      \expandafter\def\csname LTb\endcsname{\color{black}}%
      \expandafter\def\csname LTa\endcsname{\color{black}}%
      \expandafter\def\csname LT0\endcsname{\color[rgb]{1,0,0}}%
      \expandafter\def\csname LT1\endcsname{\color[rgb]{0,1,0}}%
      \expandafter\def\csname LT2\endcsname{\color[rgb]{0,0,1}}%
      \expandafter\def\csname LT3\endcsname{\color[rgb]{1,0,1}}%
      \expandafter\def\csname LT4\endcsname{\color[rgb]{0,1,1}}%
      \expandafter\def\csname LT5\endcsname{\color[rgb]{1,1,0}}%
      \expandafter\def\csname LT6\endcsname{\color[rgb]{0,0,0}}%
      \expandafter\def\csname LT7\endcsname{\color[rgb]{1,0.3,0}}%
      \expandafter\def\csname LT8\endcsname{\color[rgb]{0.5,0.5,0.5}}%
    \else
      \def\colorrgb#1{\color{black}}%
      \def\colorgray#1{\color[gray]{#1}}%
      \expandafter\def\csname LTw\endcsname{\color{white}}%
      \expandafter\def\csname LTb\endcsname{\color{black}}%
      \expandafter\def\csname LTa\endcsname{\color{black}}%
      \expandafter\def\csname LT0\endcsname{\color{black}}%
      \expandafter\def\csname LT1\endcsname{\color{black}}%
      \expandafter\def\csname LT2\endcsname{\color{black}}%
      \expandafter\def\csname LT3\endcsname{\color{black}}%
      \expandafter\def\csname LT4\endcsname{\color{black}}%
      \expandafter\def\csname LT5\endcsname{\color{black}}%
      \expandafter\def\csname LT6\endcsname{\color{black}}%
      \expandafter\def\csname LT7\endcsname{\color{black}}%
      \expandafter\def\csname LT8\endcsname{\color{black}}%
    \fi
  \fi
  \setlength{\unitlength}{0.0500bp}%
  \begin{picture}(7012.00,4506.00)%
    \gplgaddtomacro\gplbacktext{%
      \csname LTb\endcsname%
      \put(1012,704){\makebox(0,0)[r]{\strut{} 0}}%
      \put(1012,1097){\makebox(0,0)[r]{\strut{} 0.002}}%
      \put(1012,1490){\makebox(0,0)[r]{\strut{} 0.004}}%
      \put(1012,1883){\makebox(0,0)[r]{\strut{} 0.006}}%
      \put(1012,2276){\makebox(0,0)[r]{\strut{} 0.008}}%
      \put(1012,2669){\makebox(0,0)[r]{\strut{} 0.01}}%
      \put(1012,3062){\makebox(0,0)[r]{\strut{} 0.012}}%
      \put(1012,3455){\makebox(0,0)[r]{\strut{} 0.014}}%
      \put(1012,3848){\makebox(0,0)[r]{\strut{} 0.016}}%
      \put(1012,4241){\makebox(0,0)[r]{\strut{} 0.018}}%
      \put(1144,484){\makebox(0,0){\strut{} 0}}%
      \put(1828,484){\makebox(0,0){\strut{} 0.0005}}%
      \put(2512,484){\makebox(0,0){\strut{} 0.001}}%
      \put(3196,484){\makebox(0,0){\strut{} 0.0015}}%
      \put(3880,484){\makebox(0,0){\strut{} 0.002}}%
      \put(4563,484){\makebox(0,0){\strut{} 0.0025}}%
      \put(5247,484){\makebox(0,0){\strut{} 0.003}}%
      \put(5931,484){\makebox(0,0){\strut{} 0.0035}}%
      \put(6615,484){\makebox(0,0){\strut{} 0.004}}%
      \put(176,2472){\rotatebox{-270}{\makebox(0,0){\strut{}$u_{\textrm{i}_\textrm{rms}}^+$}}}%
      \put(3879,154){\makebox(0,0){\strut{}$\sigma$}}%
    }%
    \gplgaddtomacro\gplfronttext{%
      \csname LTb\endcsname%
      \put(1408,4068){\makebox(0,0)[r]{\strut{}$u$}}%
      \csname LTb\endcsname%
      \put(1408,3848){\makebox(0,0)[r]{\strut{}$v$}}%
      \csname LTb\endcsname%
      \put(1408,3628){\makebox(0,0)[r]{\strut{}$w$}}%
    }%
    \gplbacktext
    \put(0,0){\includegraphics{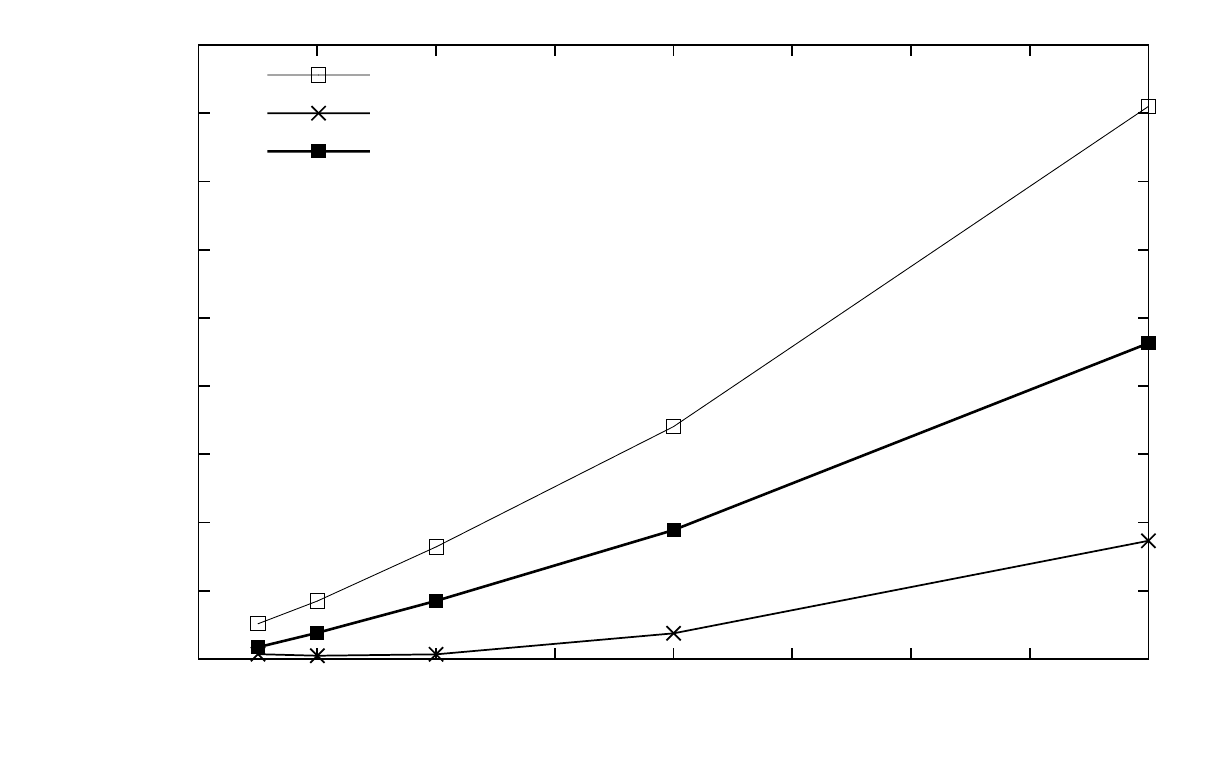}}%
    \gplfronttext
  \end{picture}%
\endgroup

%% file: cor_sigma.tex
\begingroup
  \makeatletter
  \providecommand\color[2][]{%
    \GenericError{(gnuplot) \space\space\space\@spaces}{%
      Package color not loaded in conjunction with
      terminal option `colourtext'%
    }{See the gnuplot documentation for explanation.%
    }{Either use 'blacktext' in gnuplot or load the package
      color.sty in LaTeX.}%
    \renewcommand\color[2][]{}%
  }%
  \providecommand\includegraphics[2][]{%
    \GenericError{(gnuplot) \space\space\space\@spaces}{%
      Package graphicx or graphics not loaded%
    }{See the gnuplot documentation for explanation.%
    }{The gnuplot epslatex terminal needs graphicx.sty or graphics.sty.}%
    \renewcommand\includegraphics[2][]{}%
  }%
  \providecommand\rotatebox[2]{#2}%
  \@ifundefined{ifGPcolor}{%
    \newif\ifGPcolor
    \GPcolorfalse
  }{}%
  \@ifundefined{ifGPblacktext}{%
    \newif\ifGPblacktext
    \GPblacktexttrue
  }{}%
  \let\gplgaddtomacro\g@addto@macro
  \gdef\gplbacktext{}%
  \gdef\gplfronttext{}%
  \makeatother
  \ifGPblacktext
    \def\colorrgb#1{}%
    \def\colorgray#1{}%
  \else
    \ifGPcolor
      \def\colorrgb#1{\color[rgb]{#1}}%
      \def\colorgray#1{\color[gray]{#1}}%
      \expandafter\def\csname LTw\endcsname{\color{white}}%
      \expandafter\def\csname LTb\endcsname{\color{black}}%
      \expandafter\def\csname LTa\endcsname{\color{black}}%
      \expandafter\def\csname LT0\endcsname{\color[rgb]{1,0,0}}%
      \expandafter\def\csname LT1\endcsname{\color[rgb]{0,1,0}}%
      \expandafter\def\csname LT2\endcsname{\color[rgb]{0,0,1}}%
      \expandafter\def\csname LT3\endcsname{\color[rgb]{1,0,1}}%
      \expandafter\def\csname LT4\endcsname{\color[rgb]{0,1,1}}%
      \expandafter\def\csname LT5\endcsname{\color[rgb]{1,1,0}}%
      \expandafter\def\csname LT6\endcsname{\color[rgb]{0,0,0}}%
      \expandafter\def\csname LT7\endcsname{\color[rgb]{1,0.3,0}}%
      \expandafter\def\csname LT8\endcsname{\color[rgb]{0.5,0.5,0.5}}%
    \else
      \def\colorrgb#1{\color{black}}%
      \def\colorgray#1{\color[gray]{#1}}%
      \expandafter\def\csname LTw\endcsname{\color{white}}%
      \expandafter\def\csname LTb\endcsname{\color{black}}%
      \expandafter\def\csname LTa\endcsname{\color{black}}%
      \expandafter\def\csname LT0\endcsname{\color{black}}%
      \expandafter\def\csname LT1\endcsname{\color{black}}%
      \expandafter\def\csname LT2\endcsname{\color{black}}%
      \expandafter\def\csname LT3\endcsname{\color{black}}%
      \expandafter\def\csname LT4\endcsname{\color{black}}%
      \expandafter\def\csname LT5\endcsname{\color{black}}%
      \expandafter\def\csname LT6\endcsname{\color{black}}%
      \expandafter\def\csname LT7\endcsname{\color{black}}%
      \expandafter\def\csname LT8\endcsname{\color{black}}%
    \fi
  \fi
  \setlength{\unitlength}{0.0500bp}%
  \begin{picture}(7012.00,5760.00)%
    \gplgaddtomacro\gplbacktext{%
      \csname LTb\endcsname%
      \put(726,3343){\makebox(0,0)[r]{\strut{} 0}}%
      \put(726,3658){\makebox(0,0)[r]{\strut{} 0.2}}%
      \put(726,3972){\makebox(0,0)[r]{\strut{} 0.4}}%
      \put(726,4287){\makebox(0,0)[r]{\strut{} 0.6}}%
      \put(726,4601){\makebox(0,0)[r]{\strut{} 0.8}}%
      \put(726,4916){\makebox(0,0)[r]{\strut{} 1}}%
      \put(858,2966){\makebox(0,0){\strut{} 0}}%
      \put(1216,2966){\makebox(0,0){\strut{} 1}}%
      \put(1575,2966){\makebox(0,0){\strut{} 2}}%
      \put(1933,2966){\makebox(0,0){\strut{} 3}}%
      \put(2292,2966){\makebox(0,0){\strut{} 4}}%
      \put(2650,2966){\makebox(0,0){\strut{} 5}}%
      \put(3009,2966){\makebox(0,0){\strut{} 6}}%
      \put(220,4051){\rotatebox{-270}{\makebox(0,0){\strut{}$R_{11}$}}}%
      \put(1983,2746){\makebox(0,0){\strut{}$x$}}%
    }%
    \gplgaddtomacro\gplfronttext{%
      \csname LTb\endcsname%
      \put(5380,1661){\makebox(0,0)[r]{\strut{}Imp}}%
      \csname LTb\endcsname%
      \put(5380,1441){\makebox(0,0)[r]{\strut{}$\sigma=0.001$}}%
      \csname LTb\endcsname%
      \put(5380,1221){\makebox(0,0)[r]{\strut{}$\sigma=0.002$}}%
      \csname LTb\endcsname%
      \put(5380,1001){\makebox(0,0)[r]{\strut{}$\sigma=0.004$}}%
    }%
    \gplgaddtomacro\gplbacktext{%
      \csname LTb\endcsname%
      \put(4232,3343){\makebox(0,0)[r]{\strut{} 0}}%
      \put(4232,3658){\makebox(0,0)[r]{\strut{} 0.2}}%
      \put(4232,3972){\makebox(0,0)[r]{\strut{} 0.4}}%
      \put(4232,4287){\makebox(0,0)[r]{\strut{} 0.6}}%
      \put(4232,4601){\makebox(0,0)[r]{\strut{} 0.8}}%
      \put(4232,4916){\makebox(0,0)[r]{\strut{} 1}}%
      \put(4364,2966){\makebox(0,0){\strut{} 0}}%
      \put(4722,2966){\makebox(0,0){\strut{} 1}}%
      \put(5081,2966){\makebox(0,0){\strut{} 2}}%
      \put(5439,2966){\makebox(0,0){\strut{} 3}}%
      \put(5798,2966){\makebox(0,0){\strut{} 4}}%
      \put(6156,2966){\makebox(0,0){\strut{} 5}}%
      \put(6515,2966){\makebox(0,0){\strut{} 6}}%
      \put(3726,4051){\rotatebox{-270}{\makebox(0,0){\strut{}$R_{22}$}}}%
      \put(5489,2746){\makebox(0,0){\strut{}$x$}}%
    }%
    \gplgaddtomacro\gplfronttext{%
    }%
    \gplgaddtomacro\gplbacktext{%
      \csname LTb\endcsname%
      \put(726,751){\makebox(0,0)[r]{\strut{} 0}}%
      \put(726,1066){\makebox(0,0)[r]{\strut{} 0.2}}%
      \put(726,1380){\makebox(0,0)[r]{\strut{} 0.4}}%
      \put(726,1695){\makebox(0,0)[r]{\strut{} 0.6}}%
      \put(726,2009){\makebox(0,0)[r]{\strut{} 0.8}}%
      \put(726,2324){\makebox(0,0)[r]{\strut{} 1}}%
      \put(858,374){\makebox(0,0){\strut{} 0}}%
      \put(1216,374){\makebox(0,0){\strut{} 0.5}}%
      \put(1575,374){\makebox(0,0){\strut{} 1}}%
      \put(1933,374){\makebox(0,0){\strut{} 1.5}}%
      \put(2292,374){\makebox(0,0){\strut{} 2}}%
      \put(2650,374){\makebox(0,0){\strut{} 2.5}}%
      \put(3009,374){\makebox(0,0){\strut{} 3}}%
      \put(220,1459){\rotatebox{-270}{\makebox(0,0){\strut{}$R_{33}$}}}%
      \put(1983,154){\makebox(0,0){\strut{}$x$}}%
    }%
    \gplgaddtomacro\gplfronttext{%
    }%
    \gplbacktext
    \put(0,0){\includegraphics{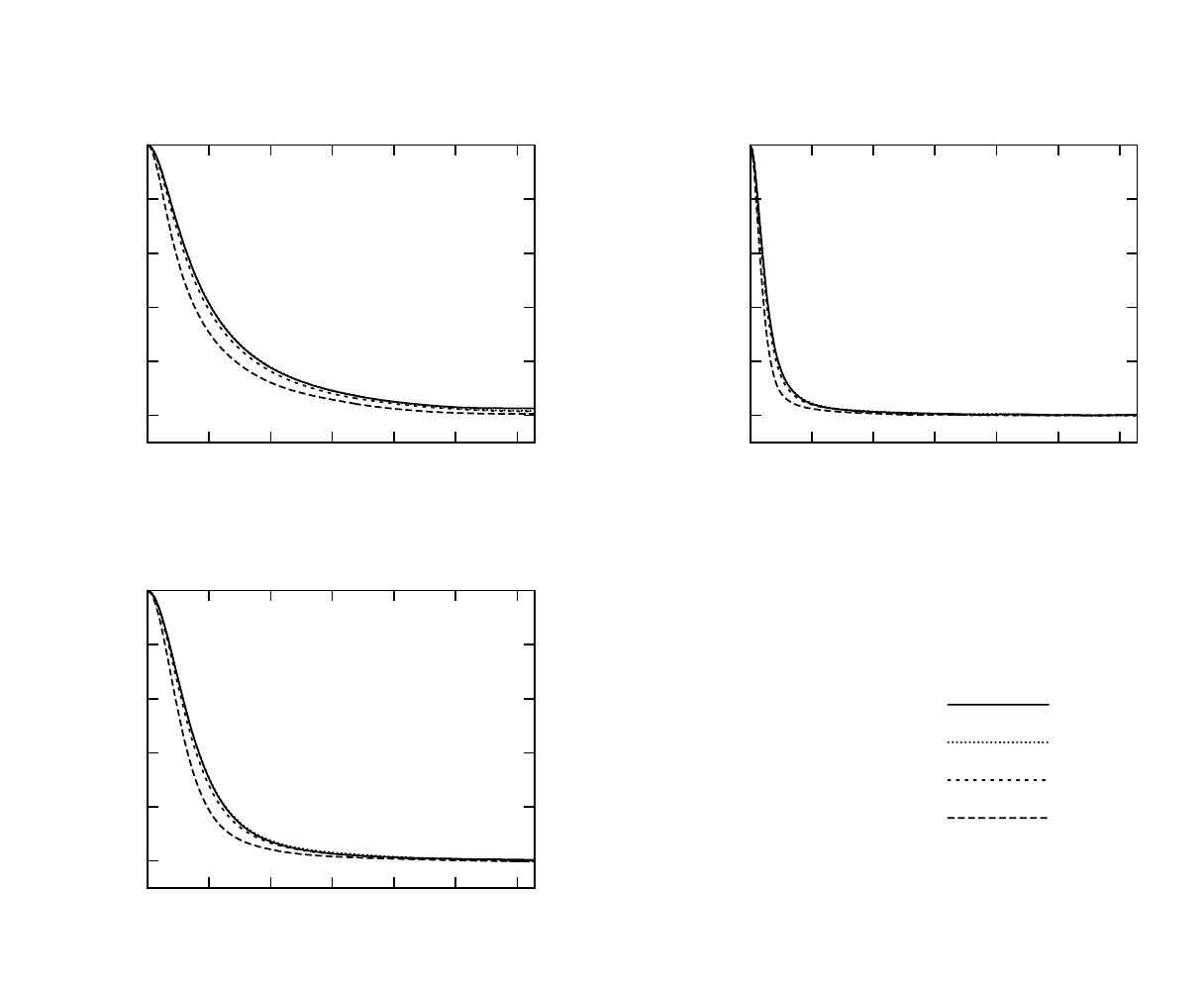}}%
    \gplfronttext
  \end{picture}%
\endgroup

%% file: loglaw_epsilon.tex
\begingroup
  \makeatletter
  \providecommand\color[2][]{%
    \GenericError{(gnuplot) \space\space\space\@spaces}{%
      Package color not loaded in conjunction with
      terminal option `colourtext'%
    }{See the gnuplot documentation for explanation.%
    }{Either use 'blacktext' in gnuplot or load the package
      color.sty in LaTeX.}%
    \renewcommand\color[2][]{}%
  }%
  \providecommand\includegraphics[2][]{%
    \GenericError{(gnuplot) \space\space\space\@spaces}{%
      Package graphicx or graphics not loaded%
    }{See the gnuplot documentation for explanation.%
    }{The gnuplot epslatex terminal needs graphicx.sty or graphics.sty.}%
    \renewcommand\includegraphics[2][]{}%
  }%
  \providecommand\rotatebox[2]{#2}%
  \@ifundefined{ifGPcolor}{%
    \newif\ifGPcolor
    \GPcolorfalse
  }{}%
  \@ifundefined{ifGPblacktext}{%
    \newif\ifGPblacktext
    \GPblacktexttrue
  }{}%
  \let\gplgaddtomacro\g@addto@macro
  \gdef\gplbacktext{}%
  \gdef\gplfronttext{}%
  \makeatother
  \ifGPblacktext
    \def\colorrgb#1{}%
    \def\colorgray#1{}%
  \else
    \ifGPcolor
      \def\colorrgb#1{\color[rgb]{#1}}%
      \def\colorgray#1{\color[gray]{#1}}%
      \expandafter\def\csname LTw\endcsname{\color{white}}%
      \expandafter\def\csname LTb\endcsname{\color{black}}%
      \expandafter\def\csname LTa\endcsname{\color{black}}%
      \expandafter\def\csname LT0\endcsname{\color[rgb]{1,0,0}}%
      \expandafter\def\csname LT1\endcsname{\color[rgb]{0,1,0}}%
      \expandafter\def\csname LT2\endcsname{\color[rgb]{0,0,1}}%
      \expandafter\def\csname LT3\endcsname{\color[rgb]{1,0,1}}%
      \expandafter\def\csname LT4\endcsname{\color[rgb]{0,1,1}}%
      \expandafter\def\csname LT5\endcsname{\color[rgb]{1,1,0}}%
      \expandafter\def\csname LT6\endcsname{\color[rgb]{0,0,0}}%
      \expandafter\def\csname LT7\endcsname{\color[rgb]{1,0.3,0}}%
      \expandafter\def\csname LT8\endcsname{\color[rgb]{0.5,0.5,0.5}}%
    \else
      \def\colorrgb#1{\color{black}}%
      \def\colorgray#1{\color[gray]{#1}}%
      \expandafter\def\csname LTw\endcsname{\color{white}}%
      \expandafter\def\csname LTb\endcsname{\color{black}}%
      \expandafter\def\csname LTa\endcsname{\color{black}}%
      \expandafter\def\csname LT0\endcsname{\color{black}}%
      \expandafter\def\csname LT1\endcsname{\color{black}}%
      \expandafter\def\csname LT2\endcsname{\color{black}}%
      \expandafter\def\csname LT3\endcsname{\color{black}}%
      \expandafter\def\csname LT4\endcsname{\color{black}}%
      \expandafter\def\csname LT5\endcsname{\color{black}}%
      \expandafter\def\csname LT6\endcsname{\color{black}}%
      \expandafter\def\csname LT7\endcsname{\color{black}}%
      \expandafter\def\csname LT8\endcsname{\color{black}}%
    \fi
  \fi
  \setlength{\unitlength}{0.0500bp}%
  \begin{picture}(7012.00,4506.00)%
    \gplgaddtomacro\gplbacktext{%
      \csname LTb\endcsname%
      \put(594,704){\makebox(0,0)[r]{\strut{} 0}}%
      \put(594,1588){\makebox(0,0)[r]{\strut{} 5}}%
      \put(594,2473){\makebox(0,0)[r]{\strut{} 10}}%
      \put(594,3357){\makebox(0,0)[r]{\strut{} 15}}%
      \put(594,4241){\makebox(0,0)[r]{\strut{} 20}}%
      \put(726,484){\makebox(0,0){\strut{} 1}}%
      \put(3103,484){\makebox(0,0){\strut{} 10}}%
      \put(5481,484){\makebox(0,0){\strut{} 100}}%
      \put(220,2472){\rotatebox{-270}{\makebox(0,0){\strut{}$(\overline{u}-U_i)^+$}}}%
      \put(3670,154){\makebox(0,0){\strut{}$y^+$}}%
    }%
    \gplgaddtomacro\gplfronttext{%
      \csname LTb\endcsname%
      \put(1518,4068){\makebox(0,0)[r]{\strut{}Imp}}%
      \csname LTb\endcsname%
      \put(1518,3848){\makebox(0,0)[r]{\strut{}$\varepsilon=0.3$}}%
      \csname LTb\endcsname%
      \put(1518,3628){\makebox(0,0)[r]{\strut{}$\varepsilon=0.6$}}%
      \csname LTb\endcsname%
      \put(1518,3408){\makebox(0,0)[r]{\strut{}$\varepsilon=0.9$}}%
    }%
    \gplbacktext
    \put(0,0){\includegraphics{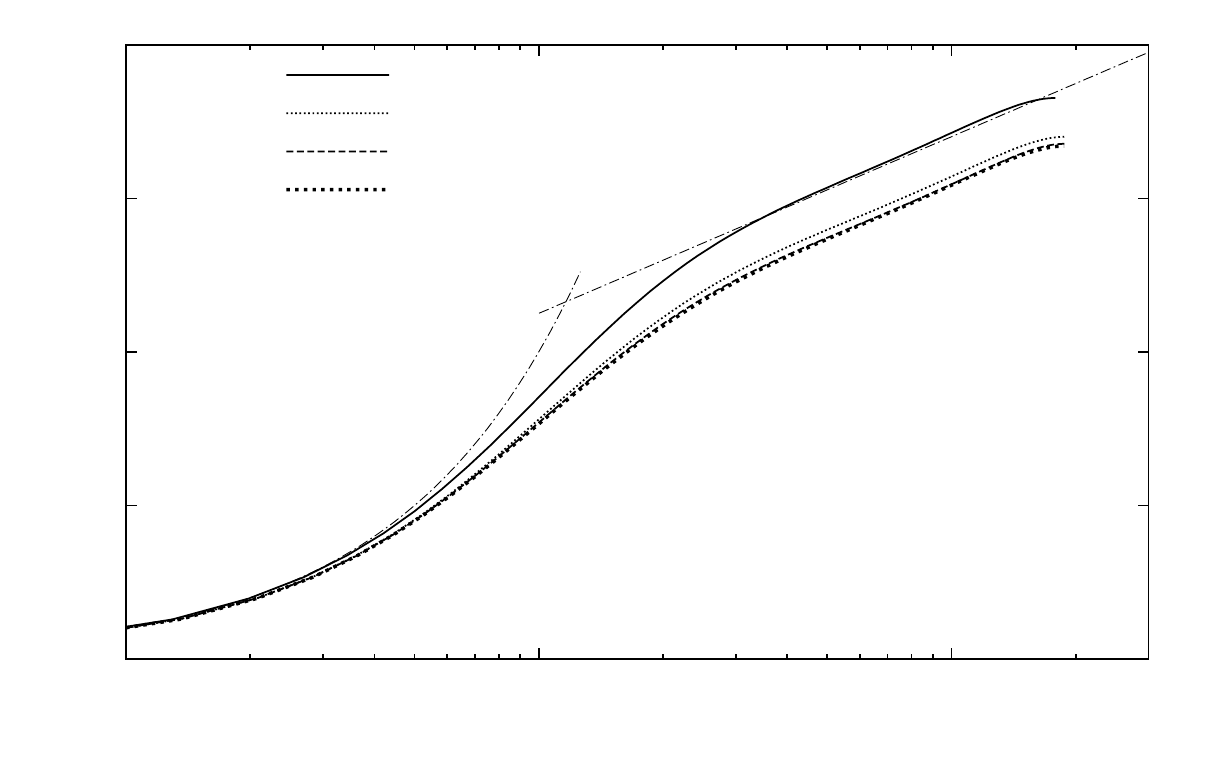}}%
    \gplfronttext
  \end{picture}%
\endgroup

%% file: loglaw_tau.tex
\begingroup
  \makeatletter
  \providecommand\color[2][]{%
    \GenericError{(gnuplot) \space\space\space\@spaces}{%
      Package color not loaded in conjunction with
      terminal option `colourtext'%
    }{See the gnuplot documentation for explanation.%
    }{Either use 'blacktext' in gnuplot or load the package
      color.sty in LaTeX.}%
    \renewcommand\color[2][]{}%
  }%
  \providecommand\includegraphics[2][]{%
    \GenericError{(gnuplot) \space\space\space\@spaces}{%
      Package graphicx or graphics not loaded%
    }{See the gnuplot documentation for explanation.%
    }{The gnuplot epslatex terminal needs graphicx.sty or graphics.sty.}%
    \renewcommand\includegraphics[2][]{}%
  }%
  \providecommand\rotatebox[2]{#2}%
  \@ifundefined{ifGPcolor}{%
    \newif\ifGPcolor
    \GPcolorfalse
  }{}%
  \@ifundefined{ifGPblacktext}{%
    \newif\ifGPblacktext
    \GPblacktexttrue
  }{}%
  \let\gplgaddtomacro\g@addto@macro
  \gdef\gplbacktext{}%
  \gdef\gplfronttext{}%
  \makeatother
  \ifGPblacktext
    \def\colorrgb#1{}%
    \def\colorgray#1{}%
  \else
    \ifGPcolor
      \def\colorrgb#1{\color[rgb]{#1}}%
      \def\colorgray#1{\color[gray]{#1}}%
      \expandafter\def\csname LTw\endcsname{\color{white}}%
      \expandafter\def\csname LTb\endcsname{\color{black}}%
      \expandafter\def\csname LTa\endcsname{\color{black}}%
      \expandafter\def\csname LT0\endcsname{\color[rgb]{1,0,0}}%
      \expandafter\def\csname LT1\endcsname{\color[rgb]{0,1,0}}%
      \expandafter\def\csname LT2\endcsname{\color[rgb]{0,0,1}}%
      \expandafter\def\csname LT3\endcsname{\color[rgb]{1,0,1}}%
      \expandafter\def\csname LT4\endcsname{\color[rgb]{0,1,1}}%
      \expandafter\def\csname LT5\endcsname{\color[rgb]{1,1,0}}%
      \expandafter\def\csname LT6\endcsname{\color[rgb]{0,0,0}}%
      \expandafter\def\csname LT7\endcsname{\color[rgb]{1,0.3,0}}%
      \expandafter\def\csname LT8\endcsname{\color[rgb]{0.5,0.5,0.5}}%
    \else
      \def\colorrgb#1{\color{black}}%
      \def\colorgray#1{\color[gray]{#1}}%
      \expandafter\def\csname LTw\endcsname{\color{white}}%
      \expandafter\def\csname LTb\endcsname{\color{black}}%
      \expandafter\def\csname LTa\endcsname{\color{black}}%
      \expandafter\def\csname LT0\endcsname{\color{black}}%
      \expandafter\def\csname LT1\endcsname{\color{black}}%
      \expandafter\def\csname LT2\endcsname{\color{black}}%
      \expandafter\def\csname LT3\endcsname{\color{black}}%
      \expandafter\def\csname LT4\endcsname{\color{black}}%
      \expandafter\def\csname LT5\endcsname{\color{black}}%
      \expandafter\def\csname LT6\endcsname{\color{black}}%
      \expandafter\def\csname LT7\endcsname{\color{black}}%
      \expandafter\def\csname LT8\endcsname{\color{black}}%
    \fi
  \fi
  \setlength{\unitlength}{0.0500bp}%
  \begin{picture}(7012.00,4506.00)%
    \gplgaddtomacro\gplbacktext{%
      \csname LTb\endcsname%
      \put(726,704){\makebox(0,0)[r]{\strut{} 0}}%
      \put(726,1176){\makebox(0,0)[r]{\strut{} 0.2}}%
      \put(726,1647){\makebox(0,0)[r]{\strut{} 0.4}}%
      \put(726,2119){\makebox(0,0)[r]{\strut{} 0.6}}%
      \put(726,2590){\makebox(0,0)[r]{\strut{} 0.8}}%
      \put(726,3062){\makebox(0,0)[r]{\strut{} 1}}%
      \put(726,3534){\makebox(0,0)[r]{\strut{} 1.2}}%
      \put(726,4005){\makebox(0,0)[r]{\strut{} 1.4}}%
      \put(858,484){\makebox(0,0){\strut{}-0.2}}%
      \put(1233,484){\makebox(0,0){\strut{} 0}}%
      \put(1608,484){\makebox(0,0){\strut{} 0.2}}%
      \put(1984,484){\makebox(0,0){\strut{} 0.4}}%
      \put(2359,484){\makebox(0,0){\strut{} 0.6}}%
      \put(2734,484){\makebox(0,0){\strut{} 0.8}}%
      \put(3109,484){\makebox(0,0){\strut{} 1}}%
      \put(220,2472){\rotatebox{-270}{\makebox(0,0){\strut{}$\overline{u}$}}}%
      \put(1983,154){\makebox(0,0){\strut{}$y$}}%
    }%
    \gplgaddtomacro\gplfronttext{%
      \csname LTb\endcsname%
      \put(2122,1757){\makebox(0,0)[r]{\strut{}Imp}}%
      \csname LTb\endcsname%
      \put(2122,1537){\makebox(0,0)[r]{\strut{}$\tau=-1$}}%
      \csname LTb\endcsname%
      \put(2122,1317){\makebox(0,0)[r]{\strut{}$\tau=0$}}%
      \csname LTb\endcsname%
      \put(2122,1097){\makebox(0,0)[r]{\strut{}$\tau=0.5$}}%
      \csname LTb\endcsname%
      \put(2122,877){\makebox(0,0)[r]{\strut{}$\tau=+1$}}%
    }%
    \gplgaddtomacro\gplbacktext{%
      \csname LTb\endcsname%
      \put(4100,704){\makebox(0,0)[r]{\strut{} 0}}%
      \put(4100,1588){\makebox(0,0)[r]{\strut{} 5}}%
      \put(4100,2473){\makebox(0,0)[r]{\strut{} 10}}%
      \put(4100,3357){\makebox(0,0)[r]{\strut{} 15}}%
      \put(4100,4241){\makebox(0,0)[r]{\strut{} 20}}%
      \put(4232,484){\makebox(0,0){\strut{} 1}}%
      \put(5194,484){\makebox(0,0){\strut{} 10}}%
      \put(6156,484){\makebox(0,0){\strut{} 100}}%
      \put(3726,2472){\rotatebox{-270}{\makebox(0,0){\strut{}$(\overline{u}-U_i)^+$}}}%
      \put(5423,154){\makebox(0,0){\strut{}$y^+$}}%
    }%
    \gplgaddtomacro\gplfronttext{%
      \csname LTb\endcsname%
      \put(5628,1757){\makebox(0,0)[r]{\strut{}Imp}}%
      \csname LTb\endcsname%
      \put(5628,1537){\makebox(0,0)[r]{\strut{}$\tau=-1$}}%
      \csname LTb\endcsname%
      \put(5628,1317){\makebox(0,0)[r]{\strut{}$\tau=0$}}%
      \csname LTb\endcsname%
      \put(5628,1097){\makebox(0,0)[r]{\strut{}$\tau=0.5$}}%
      \csname LTb\endcsname%
      \put(5628,877){\makebox(0,0)[r]{\strut{}$\tau=+1$}}%
    }%
    \gplbacktext
    \put(0,0){\includegraphics{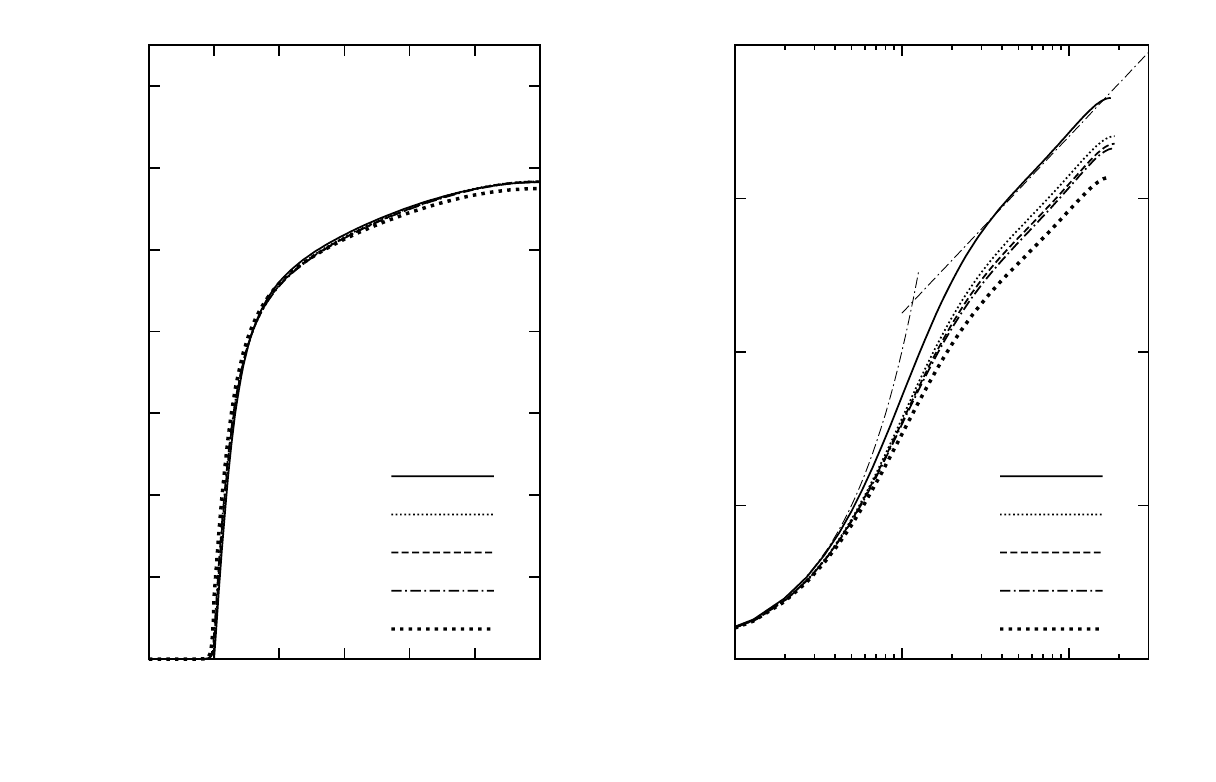}}%
    \gplfronttext
  \end{picture}%
\endgroup

%% file: rms_tau.tex
\begingroup
  \makeatletter
  \providecommand\color[2][]{%
    \GenericError{(gnuplot) \space\space\space\@spaces}{%
      Package color not loaded in conjunction with
      terminal option `colourtext'%
    }{See the gnuplot documentation for explanation.%
    }{Either use 'blacktext' in gnuplot or load the package
      color.sty in LaTeX.}%
    \renewcommand\color[2][]{}%
  }%
  \providecommand\includegraphics[2][]{%
    \GenericError{(gnuplot) \space\space\space\@spaces}{%
      Package graphicx or graphics not loaded%
    }{See the gnuplot documentation for explanation.%
    }{The gnuplot epslatex terminal needs graphicx.sty or graphics.sty.}%
    \renewcommand\includegraphics[2][]{}%
  }%
  \providecommand\rotatebox[2]{#2}%
  \@ifundefined{ifGPcolor}{%
    \newif\ifGPcolor
    \GPcolorfalse
  }{}%
  \@ifundefined{ifGPblacktext}{%
    \newif\ifGPblacktext
    \GPblacktexttrue
  }{}%
  \let\gplgaddtomacro\g@addto@macro
  \gdef\gplbacktext{}%
  \gdef\gplfronttext{}%
  \makeatother
  \ifGPblacktext
    \def\colorrgb#1{}%
    \def\colorgray#1{}%
  \else
    \ifGPcolor
      \def\colorrgb#1{\color[rgb]{#1}}%
      \def\colorgray#1{\color[gray]{#1}}%
      \expandafter\def\csname LTw\endcsname{\color{white}}%
      \expandafter\def\csname LTb\endcsname{\color{black}}%
      \expandafter\def\csname LTa\endcsname{\color{black}}%
      \expandafter\def\csname LT0\endcsname{\color[rgb]{1,0,0}}%
      \expandafter\def\csname LT1\endcsname{\color[rgb]{0,1,0}}%
      \expandafter\def\csname LT2\endcsname{\color[rgb]{0,0,1}}%
      \expandafter\def\csname LT3\endcsname{\color[rgb]{1,0,1}}%
      \expandafter\def\csname LT4\endcsname{\color[rgb]{0,1,1}}%
      \expandafter\def\csname LT5\endcsname{\color[rgb]{1,1,0}}%
      \expandafter\def\csname LT6\endcsname{\color[rgb]{0,0,0}}%
      \expandafter\def\csname LT7\endcsname{\color[rgb]{1,0.3,0}}%
      \expandafter\def\csname LT8\endcsname{\color[rgb]{0.5,0.5,0.5}}%
    \else
      \def\colorrgb#1{\color{black}}%
      \def\colorgray#1{\color[gray]{#1}}%
      \expandafter\def\csname LTw\endcsname{\color{white}}%
      \expandafter\def\csname LTb\endcsname{\color{black}}%
      \expandafter\def\csname LTa\endcsname{\color{black}}%
      \expandafter\def\csname LT0\endcsname{\color{black}}%
      \expandafter\def\csname LT1\endcsname{\color{black}}%
      \expandafter\def\csname LT2\endcsname{\color{black}}%
      \expandafter\def\csname LT3\endcsname{\color{black}}%
      \expandafter\def\csname LT4\endcsname{\color{black}}%
      \expandafter\def\csname LT5\endcsname{\color{black}}%
      \expandafter\def\csname LT6\endcsname{\color{black}}%
      \expandafter\def\csname LT7\endcsname{\color{black}}%
      \expandafter\def\csname LT8\endcsname{\color{black}}%
    \fi
  \fi
  \setlength{\unitlength}{0.0500bp}%
  \begin{picture}(7012.00,7012.00)%
    \gplgaddtomacro\gplbacktext{%
      \csname LTb\endcsname%
      \put(726,3859){\makebox(0,0)[r]{\strut{} 0}}%
      \put(726,4229){\makebox(0,0)[r]{\strut{} 0.5}}%
      \put(726,4599){\makebox(0,0)[r]{\strut{} 1}}%
      \put(726,4970){\makebox(0,0)[r]{\strut{} 1.5}}%
      \put(726,5340){\makebox(0,0)[r]{\strut{} 2}}%
      \put(726,5710){\makebox(0,0)[r]{\strut{} 2.5}}%
      \put(726,6080){\makebox(0,0)[r]{\strut{} 3}}%
      \put(858,3639){\makebox(0,0){\strut{} 1}}%
      \put(1836,3639){\makebox(0,0){\strut{} 10}}%
      \put(2815,3639){\makebox(0,0){\strut{} 100}}%
      \put(352,4969){\rotatebox{-270}{\makebox(0,0){\strut{}$u_\textrm{rms}/u_\tau$}}}%
      \put(1983,3309){\makebox(0,0){\strut{}$y^+$}}%
    }%
    \gplgaddtomacro\gplfronttext{%
      \csname LTb\endcsname%
      \put(1812,6562){\makebox(0,0)[r]{\strut{}Imp}}%
      \csname LTb\endcsname%
      \put(1812,6342){\makebox(0,0)[r]{\strut{}$\tau=-1$}}%
      \csname LTb\endcsname%
      \put(3327,6562){\makebox(0,0)[r]{\strut{}$\tau=0$}}%
      \csname LTb\endcsname%
      \put(3327,6342){\makebox(0,0)[r]{\strut{}$\tau=0.5$}}%
      \csname LTb\endcsname%
      \put(4842,6562){\makebox(0,0)[r]{\strut{}$\tau=+1$}}%
    }%
    \gplgaddtomacro\gplbacktext{%
      \csname LTb\endcsname%
      \put(4232,3859){\makebox(0,0)[r]{\strut{} 0}}%
      \put(4232,4229){\makebox(0,0)[r]{\strut{} 0.5}}%
      \put(4232,4599){\makebox(0,0)[r]{\strut{} 1}}%
      \put(4232,4970){\makebox(0,0)[r]{\strut{} 1.5}}%
      \put(4232,5340){\makebox(0,0)[r]{\strut{} 2}}%
      \put(4232,5710){\makebox(0,0)[r]{\strut{} 2.5}}%
      \put(4232,6080){\makebox(0,0)[r]{\strut{} 3}}%
      \put(4364,3639){\makebox(0,0){\strut{} 1}}%
      \put(5342,3639){\makebox(0,0){\strut{} 10}}%
      \put(6321,3639){\makebox(0,0){\strut{} 100}}%
      \put(3858,4969){\rotatebox{-270}{\makebox(0,0){\strut{}$v_\textrm{rms}/u_\tau$}}}%
      \put(5489,3309){\makebox(0,0){\strut{}$y^+$}}%
    }%
    \gplgaddtomacro\gplfronttext{%
    }%
    \gplgaddtomacro\gplbacktext{%
      \csname LTb\endcsname%
      \put(726,704){\makebox(0,0)[r]{\strut{} 0}}%
      \put(726,1074){\makebox(0,0)[r]{\strut{} 0.5}}%
      \put(726,1444){\makebox(0,0)[r]{\strut{} 1}}%
      \put(726,1815){\makebox(0,0)[r]{\strut{} 1.5}}%
      \put(726,2185){\makebox(0,0)[r]{\strut{} 2}}%
      \put(726,2555){\makebox(0,0)[r]{\strut{} 2.5}}%
      \put(726,2925){\makebox(0,0)[r]{\strut{} 3}}%
      \put(858,484){\makebox(0,0){\strut{} 1}}%
      \put(1836,484){\makebox(0,0){\strut{} 10}}%
      \put(2815,484){\makebox(0,0){\strut{} 100}}%
      \put(352,1814){\rotatebox{-270}{\makebox(0,0){\strut{}$w_\textrm{rms}/u_\tau$}}}%
      \put(1983,154){\makebox(0,0){\strut{}$y^+$}}%
    }%
    \gplgaddtomacro\gplfronttext{%
    }%
    \gplgaddtomacro\gplbacktext{%
      \csname LTb\endcsname%
      \put(4232,704){\makebox(0,0)[r]{\strut{} 0}}%
      \put(4232,1148){\makebox(0,0)[r]{\strut{} 0.1}}%
      \put(4232,1592){\makebox(0,0)[r]{\strut{} 0.2}}%
      \put(4232,2037){\makebox(0,0)[r]{\strut{} 0.3}}%
      \put(4232,2481){\makebox(0,0)[r]{\strut{} 0.4}}%
      \put(4232,2925){\makebox(0,0)[r]{\strut{} 0.5}}%
      \put(4364,484){\makebox(0,0){\strut{}-0.2}}%
      \put(4739,484){\makebox(0,0){\strut{} 0}}%
      \put(5114,484){\makebox(0,0){\strut{} 0.2}}%
      \put(5490,484){\makebox(0,0){\strut{} 0.4}}%
      \put(5865,484){\makebox(0,0){\strut{} 0.6}}%
      \put(6240,484){\makebox(0,0){\strut{} 0.8}}%
      \put(6615,484){\makebox(0,0){\strut{} 1}}%
      \put(3858,1814){\rotatebox{-270}{\makebox(0,0){\strut{}$-C_{uv}$}}}%
      \put(5489,154){\makebox(0,0){\strut{}$y$}}%
    }%
    \gplgaddtomacro\gplfronttext{%
    }%
    \gplbacktext
    \put(0,0){\includegraphics{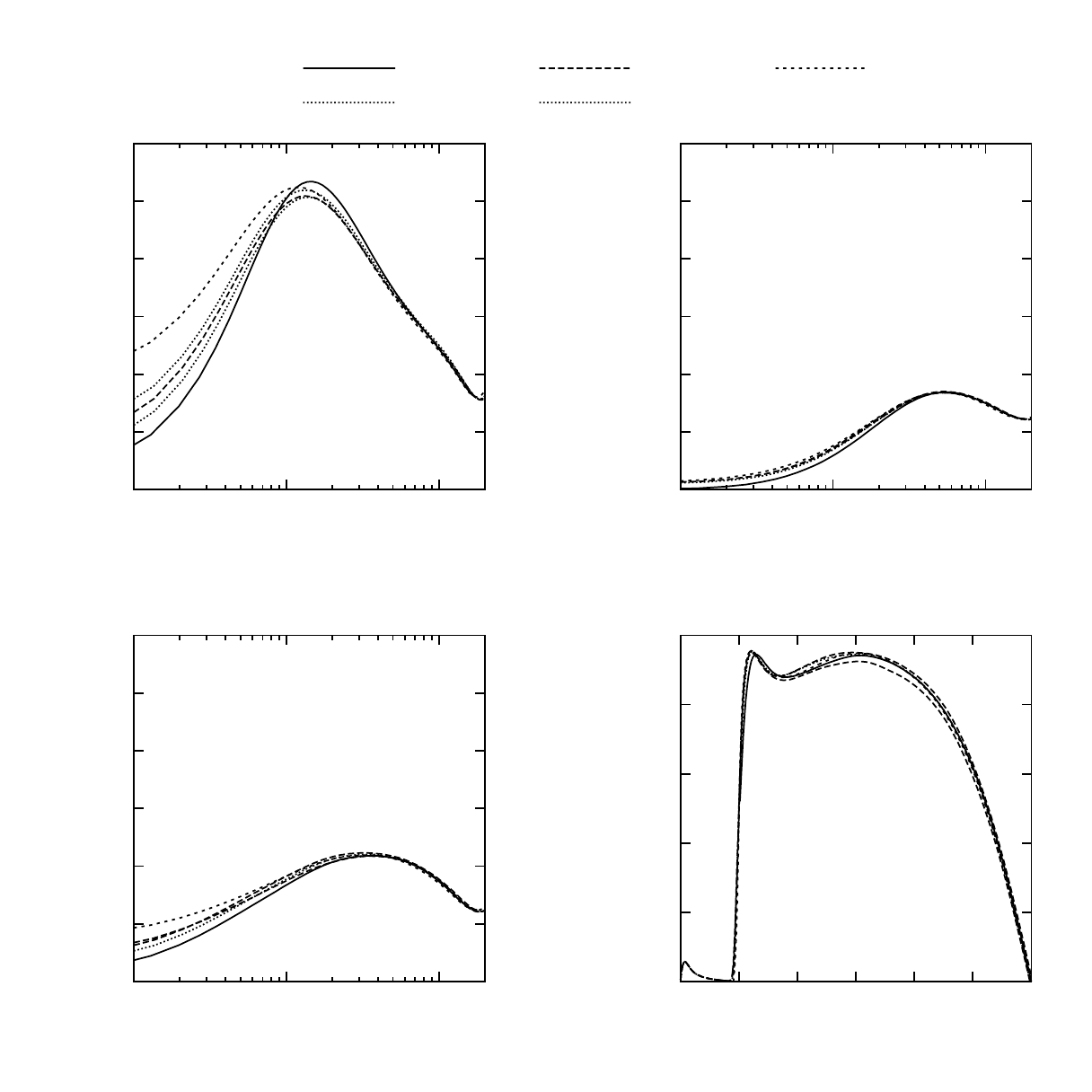}}%
    \gplfronttext
  \end{picture}%
\endgroup

%% file: cor_tau.tex
\begingroup
  \makeatletter
  \providecommand\color[2][]{%
    \GenericError{(gnuplot) \space\space\space\@spaces}{%
      Package color not loaded in conjunction with
      terminal option `colourtext'%
    }{See the gnuplot documentation for explanation.%
    }{Either use 'blacktext' in gnuplot or load the package
      color.sty in LaTeX.}%
    \renewcommand\color[2][]{}%
  }%
  \providecommand\includegraphics[2][]{%
    \GenericError{(gnuplot) \space\space\space\@spaces}{%
      Package graphicx or graphics not loaded%
    }{See the gnuplot documentation for explanation.%
    }{The gnuplot epslatex terminal needs graphicx.sty or graphics.sty.}%
    \renewcommand\includegraphics[2][]{}%
  }%
  \providecommand\rotatebox[2]{#2}%
  \@ifundefined{ifGPcolor}{%
    \newif\ifGPcolor
    \GPcolorfalse
  }{}%
  \@ifundefined{ifGPblacktext}{%
    \newif\ifGPblacktext
    \GPblacktexttrue
  }{}%
  \let\gplgaddtomacro\g@addto@macro
  \gdef\gplbacktext{}%
  \gdef\gplfronttext{}%
  \makeatother
  \ifGPblacktext
    \def\colorrgb#1{}%
    \def\colorgray#1{}%
  \else
    \ifGPcolor
      \def\colorrgb#1{\color[rgb]{#1}}%
      \def\colorgray#1{\color[gray]{#1}}%
      \expandafter\def\csname LTw\endcsname{\color{white}}%
      \expandafter\def\csname LTb\endcsname{\color{black}}%
      \expandafter\def\csname LTa\endcsname{\color{black}}%
      \expandafter\def\csname LT0\endcsname{\color[rgb]{1,0,0}}%
      \expandafter\def\csname LT1\endcsname{\color[rgb]{0,1,0}}%
      \expandafter\def\csname LT2\endcsname{\color[rgb]{0,0,1}}%
      \expandafter\def\csname LT3\endcsname{\color[rgb]{1,0,1}}%
      \expandafter\def\csname LT4\endcsname{\color[rgb]{0,1,1}}%
      \expandafter\def\csname LT5\endcsname{\color[rgb]{1,1,0}}%
      \expandafter\def\csname LT6\endcsname{\color[rgb]{0,0,0}}%
      \expandafter\def\csname LT7\endcsname{\color[rgb]{1,0.3,0}}%
      \expandafter\def\csname LT8\endcsname{\color[rgb]{0.5,0.5,0.5}}%
    \else
      \def\colorrgb#1{\color{black}}%
      \def\colorgray#1{\color[gray]{#1}}%
      \expandafter\def\csname LTw\endcsname{\color{white}}%
      \expandafter\def\csname LTb\endcsname{\color{black}}%
      \expandafter\def\csname LTa\endcsname{\color{black}}%
      \expandafter\def\csname LT0\endcsname{\color{black}}%
      \expandafter\def\csname LT1\endcsname{\color{black}}%
      \expandafter\def\csname LT2\endcsname{\color{black}}%
      \expandafter\def\csname LT3\endcsname{\color{black}}%
      \expandafter\def\csname LT4\endcsname{\color{black}}%
      \expandafter\def\csname LT5\endcsname{\color{black}}%
      \expandafter\def\csname LT6\endcsname{\color{black}}%
      \expandafter\def\csname LT7\endcsname{\color{black}}%
      \expandafter\def\csname LT8\endcsname{\color{black}}%
    \fi
  \fi
  \setlength{\unitlength}{0.0500bp}%
  \begin{picture}(7012.00,5760.00)%
    \gplgaddtomacro\gplbacktext{%
      \csname LTb\endcsname%
      \put(726,3343){\makebox(0,0)[r]{\strut{} 0}}%
      \put(726,3658){\makebox(0,0)[r]{\strut{} 0.2}}%
      \put(726,3972){\makebox(0,0)[r]{\strut{} 0.4}}%
      \put(726,4287){\makebox(0,0)[r]{\strut{} 0.6}}%
      \put(726,4601){\makebox(0,0)[r]{\strut{} 0.8}}%
      \put(726,4916){\makebox(0,0)[r]{\strut{} 1}}%
      \put(858,2966){\makebox(0,0){\strut{} 0}}%
      \put(1216,2966){\makebox(0,0){\strut{} 1}}%
      \put(1575,2966){\makebox(0,0){\strut{} 2}}%
      \put(1933,2966){\makebox(0,0){\strut{} 3}}%
      \put(2292,2966){\makebox(0,0){\strut{} 4}}%
      \put(2650,2966){\makebox(0,0){\strut{} 5}}%
      \put(3009,2966){\makebox(0,0){\strut{} 6}}%
      \put(220,4051){\rotatebox{-270}{\makebox(0,0){\strut{}$R_{11}$}}}%
      \put(1983,2746){\makebox(0,0){\strut{}$x$}}%
    }%
    \gplgaddtomacro\gplfronttext{%
      \csname LTb\endcsname%
      \put(5380,1661){\makebox(0,0)[r]{\strut{}Imp}}%
      \csname LTb\endcsname%
      \put(5380,1441){\makebox(0,0)[r]{\strut{}$\tau=-1$}}%
      \csname LTb\endcsname%
      \put(5380,1221){\makebox(0,0)[r]{\strut{}$\tau=0$}}%
      \csname LTb\endcsname%
      \put(5380,1001){\makebox(0,0)[r]{\strut{}$\tau=0.5$}}%
      \csname LTb\endcsname%
      \put(5380,781){\makebox(0,0)[r]{\strut{}$\tau=+1$}}%
    }%
    \gplgaddtomacro\gplbacktext{%
      \csname LTb\endcsname%
      \put(4232,3343){\makebox(0,0)[r]{\strut{} 0}}%
      \put(4232,3658){\makebox(0,0)[r]{\strut{} 0.2}}%
      \put(4232,3972){\makebox(0,0)[r]{\strut{} 0.4}}%
      \put(4232,4287){\makebox(0,0)[r]{\strut{} 0.6}}%
      \put(4232,4601){\makebox(0,0)[r]{\strut{} 0.8}}%
      \put(4232,4916){\makebox(0,0)[r]{\strut{} 1}}%
      \put(4364,2966){\makebox(0,0){\strut{} 0}}%
      \put(4722,2966){\makebox(0,0){\strut{} 1}}%
      \put(5081,2966){\makebox(0,0){\strut{} 2}}%
      \put(5439,2966){\makebox(0,0){\strut{} 3}}%
      \put(5798,2966){\makebox(0,0){\strut{} 4}}%
      \put(6156,2966){\makebox(0,0){\strut{} 5}}%
      \put(6515,2966){\makebox(0,0){\strut{} 6}}%
      \put(3726,4051){\rotatebox{-270}{\makebox(0,0){\strut{}$R_{22}$}}}%
      \put(5489,2746){\makebox(0,0){\strut{}$x$}}%
    }%
    \gplgaddtomacro\gplfronttext{%
    }%
    \gplgaddtomacro\gplbacktext{%
      \csname LTb\endcsname%
      \put(726,751){\makebox(0,0)[r]{\strut{} 0}}%
      \put(726,1066){\makebox(0,0)[r]{\strut{} 0.2}}%
      \put(726,1380){\makebox(0,0)[r]{\strut{} 0.4}}%
      \put(726,1695){\makebox(0,0)[r]{\strut{} 0.6}}%
      \put(726,2009){\makebox(0,0)[r]{\strut{} 0.8}}%
      \put(726,2324){\makebox(0,0)[r]{\strut{} 1}}%
      \put(858,374){\makebox(0,0){\strut{} 0}}%
      \put(1216,374){\makebox(0,0){\strut{} 0.5}}%
      \put(1575,374){\makebox(0,0){\strut{} 1}}%
      \put(1933,374){\makebox(0,0){\strut{} 1.5}}%
      \put(2292,374){\makebox(0,0){\strut{} 2}}%
      \put(2650,374){\makebox(0,0){\strut{} 2.5}}%
      \put(3009,374){\makebox(0,0){\strut{} 3}}%
      \put(220,1459){\rotatebox{-270}{\makebox(0,0){\strut{}$R_{33}$}}}%
      \put(1983,154){\makebox(0,0){\strut{}$x$}}%
    }%
    \gplgaddtomacro\gplfronttext{%
    }%
    \gplbacktext
    \put(0,0){\includegraphics{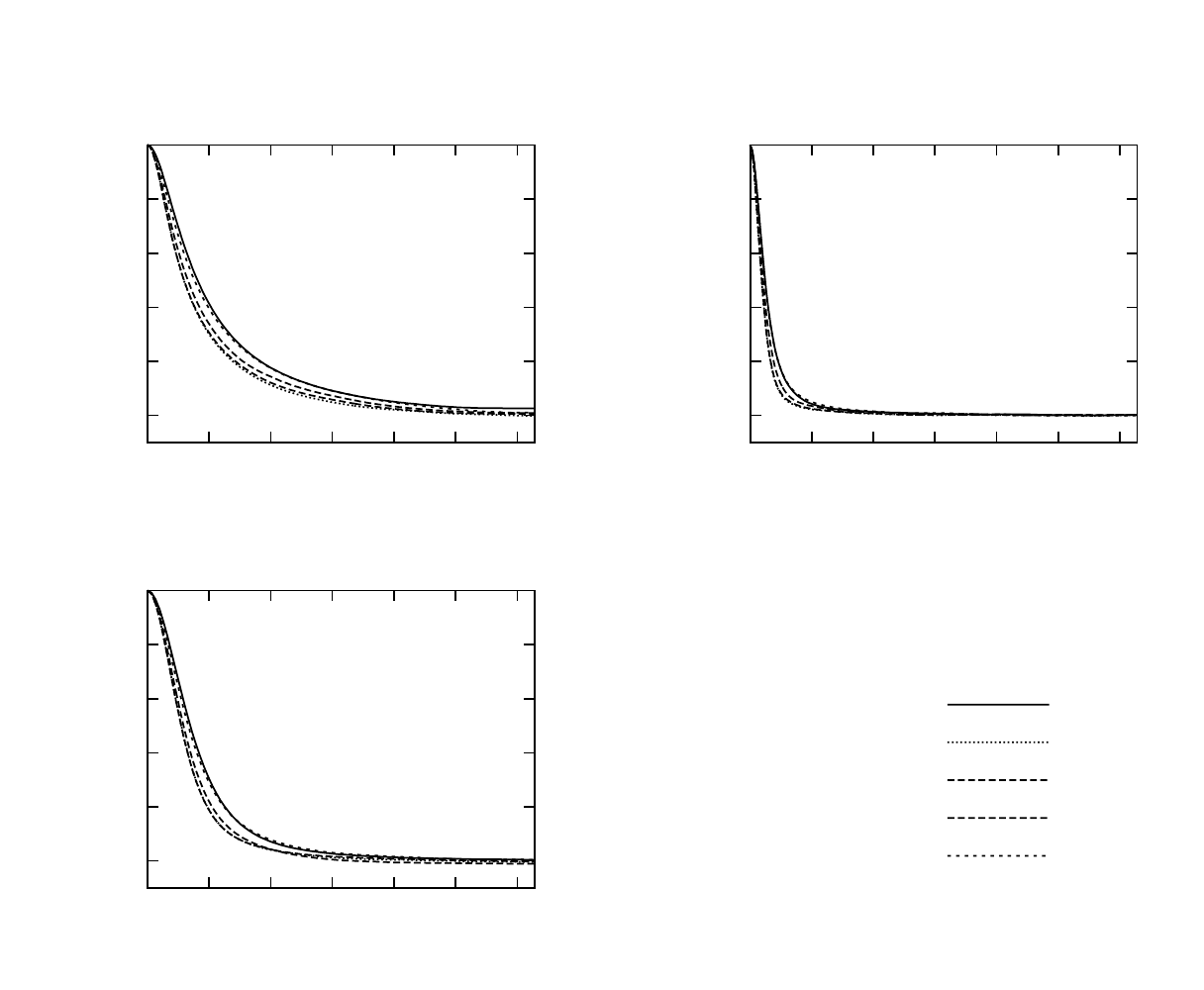}}%
    \gplfronttext
  \end{picture}%
\endgroup

%% file: diss_tau.tex
\begingroup
  \makeatletter
  \providecommand\color[2][]{%
    \GenericError{(gnuplot) \space\space\space\@spaces}{%
      Package color not loaded in conjunction with
      terminal option `colourtext'%
    }{See the gnuplot documentation for explanation.%
    }{Either use 'blacktext' in gnuplot or load the package
      color.sty in LaTeX.}%
    \renewcommand\color[2][]{}%
  }%
  \providecommand\includegraphics[2][]{%
    \GenericError{(gnuplot) \space\space\space\@spaces}{%
      Package graphicx or graphics not loaded%
    }{See the gnuplot documentation for explanation.%
    }{The gnuplot epslatex terminal needs graphicx.sty or graphics.sty.}%
    \renewcommand\includegraphics[2][]{}%
  }%
  \providecommand\rotatebox[2]{#2}%
  \@ifundefined{ifGPcolor}{%
    \newif\ifGPcolor
    \GPcolorfalse
  }{}%
  \@ifundefined{ifGPblacktext}{%
    \newif\ifGPblacktext
    \GPblacktexttrue
  }{}%
  \let\gplgaddtomacro\g@addto@macro
  \gdef\gplbacktext{}%
  \gdef\gplfronttext{}%
  \makeatother
  \ifGPblacktext
    \def\colorrgb#1{}%
    \def\colorgray#1{}%
  \else
    \ifGPcolor
      \def\colorrgb#1{\color[rgb]{#1}}%
      \def\colorgray#1{\color[gray]{#1}}%
      \expandafter\def\csname LTw\endcsname{\color{white}}%
      \expandafter\def\csname LTb\endcsname{\color{black}}%
      \expandafter\def\csname LTa\endcsname{\color{black}}%
      \expandafter\def\csname LT0\endcsname{\color[rgb]{1,0,0}}%
      \expandafter\def\csname LT1\endcsname{\color[rgb]{0,1,0}}%
      \expandafter\def\csname LT2\endcsname{\color[rgb]{0,0,1}}%
      \expandafter\def\csname LT3\endcsname{\color[rgb]{1,0,1}}%
      \expandafter\def\csname LT4\endcsname{\color[rgb]{0,1,1}}%
      \expandafter\def\csname LT5\endcsname{\color[rgb]{1,1,0}}%
      \expandafter\def\csname LT6\endcsname{\color[rgb]{0,0,0}}%
      \expandafter\def\csname LT7\endcsname{\color[rgb]{1,0.3,0}}%
      \expandafter\def\csname LT8\endcsname{\color[rgb]{0.5,0.5,0.5}}%
    \else
      \def\colorrgb#1{\color{black}}%
      \def\colorgray#1{\color[gray]{#1}}%
      \expandafter\def\csname LTw\endcsname{\color{white}}%
      \expandafter\def\csname LTb\endcsname{\color{black}}%
      \expandafter\def\csname LTa\endcsname{\color{black}}%
      \expandafter\def\csname LT0\endcsname{\color{black}}%
      \expandafter\def\csname LT1\endcsname{\color{black}}%
      \expandafter\def\csname LT2\endcsname{\color{black}}%
      \expandafter\def\csname LT3\endcsname{\color{black}}%
      \expandafter\def\csname LT4\endcsname{\color{black}}%
      \expandafter\def\csname LT5\endcsname{\color{black}}%
      \expandafter\def\csname LT6\endcsname{\color{black}}%
      \expandafter\def\csname LT7\endcsname{\color{black}}%
      \expandafter\def\csname LT8\endcsname{\color{black}}%
    \fi
  \fi
  \setlength{\unitlength}{0.0500bp}%
  \begin{picture}(7012.00,4506.00)%
    \gplgaddtomacro\gplbacktext{%
      \csname LTb\endcsname%
      \put(858,704){\makebox(0,0)[r]{\strut{} 0}}%
      \put(858,1588){\makebox(0,0)[r]{\strut{} 0.05}}%
      \put(858,2473){\makebox(0,0)[r]{\strut{} 0.1}}%
      \put(858,3357){\makebox(0,0)[r]{\strut{} 0.15}}%
      \put(858,4241){\makebox(0,0)[r]{\strut{} 0.2}}%
      \put(990,484){\makebox(0,0){\strut{} 1}}%
      \put(3484,484){\makebox(0,0){\strut{} 10}}%
      \put(5978,484){\makebox(0,0){\strut{} 100}}%
      \put(352,2472){\rotatebox{-270}{\makebox(0,0){\strut{}$\epsilon^+$}}}%
      \put(3802,154){\makebox(0,0){\strut{}$y^+$}}%
    }%
    \gplgaddtomacro\gplfronttext{%
      \csname LTb\endcsname%
      \put(5628,4068){\makebox(0,0)[r]{\strut{}$\tau=-1.0$}}%
      \csname LTb\endcsname%
      \put(5628,3848){\makebox(0,0)[r]{\strut{}$\tau=0.0$}}%
      \csname LTb\endcsname%
      \put(5628,3628){\makebox(0,0)[r]{\strut{}$\tau=0.5$}}%
      \csname LTb\endcsname%
      \put(5628,3408){\makebox(0,0)[r]{\strut{}$\tau=1.0$}}%
    }%
    \gplbacktext
    \put(0,0){\includegraphics{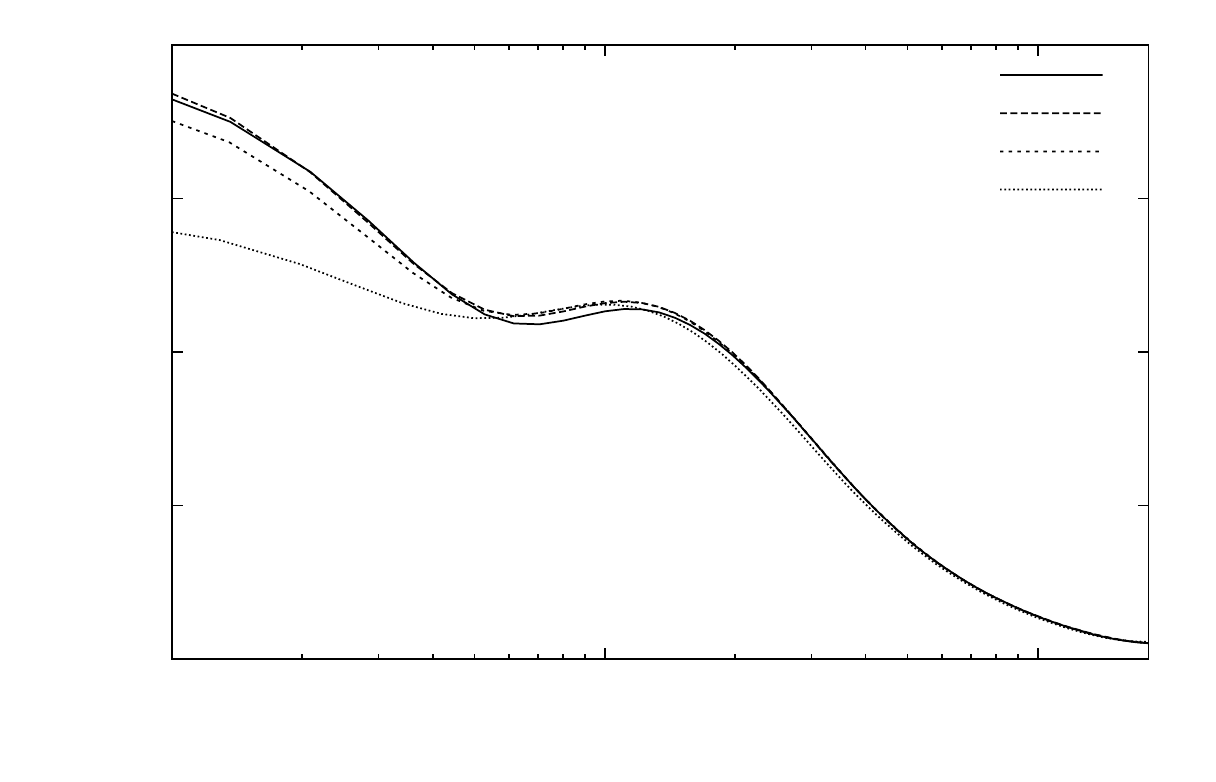}}%
    \gplfronttext
  \end{picture}%
\endgroup

%% file: lumley_tau.tex
\begingroup
  \makeatletter
  \providecommand\color[2][]{%
    \GenericError{(gnuplot) \space\space\space\@spaces}{%
      Package color not loaded in conjunction with
      terminal option `colourtext'%
    }{See the gnuplot documentation for explanation.%
    }{Either use 'blacktext' in gnuplot or load the package
      color.sty in LaTeX.}%
    \renewcommand\color[2][]{}%
  }%
  \providecommand\includegraphics[2][]{%
    \GenericError{(gnuplot) \space\space\space\@spaces}{%
      Package graphicx or graphics not loaded%
    }{See the gnuplot documentation for explanation.%
    }{The gnuplot epslatex terminal needs graphicx.sty or graphics.sty.}%
    \renewcommand\includegraphics[2][]{}%
  }%
  \providecommand\rotatebox[2]{#2}%
  \@ifundefined{ifGPcolor}{%
    \newif\ifGPcolor
    \GPcolorfalse
  }{}%
  \@ifundefined{ifGPblacktext}{%
    \newif\ifGPblacktext
    \GPblacktexttrue
  }{}%
  \let\gplgaddtomacro\g@addto@macro
  \gdef\gplbacktext{}%
  \gdef\gplfronttext{}%
  \makeatother
  \ifGPblacktext
    \def\colorrgb#1{}%
    \def\colorgray#1{}%
  \else
    \ifGPcolor
      \def\colorrgb#1{\color[rgb]{#1}}%
      \def\colorgray#1{\color[gray]{#1}}%
      \expandafter\def\csname LTw\endcsname{\color{white}}%
      \expandafter\def\csname LTb\endcsname{\color{black}}%
      \expandafter\def\csname LTa\endcsname{\color{black}}%
      \expandafter\def\csname LT0\endcsname{\color[rgb]{1,0,0}}%
      \expandafter\def\csname LT1\endcsname{\color[rgb]{0,1,0}}%
      \expandafter\def\csname LT2\endcsname{\color[rgb]{0,0,1}}%
      \expandafter\def\csname LT3\endcsname{\color[rgb]{1,0,1}}%
      \expandafter\def\csname LT4\endcsname{\color[rgb]{0,1,1}}%
      \expandafter\def\csname LT5\endcsname{\color[rgb]{1,1,0}}%
      \expandafter\def\csname LT6\endcsname{\color[rgb]{0,0,0}}%
      \expandafter\def\csname LT7\endcsname{\color[rgb]{1,0.3,0}}%
      \expandafter\def\csname LT8\endcsname{\color[rgb]{0.5,0.5,0.5}}%
    \else
      \def\colorrgb#1{\color{black}}%
      \def\colorgray#1{\color[gray]{#1}}%
      \expandafter\def\csname LTw\endcsname{\color{white}}%
      \expandafter\def\csname LTb\endcsname{\color{black}}%
      \expandafter\def\csname LTa\endcsname{\color{black}}%
      \expandafter\def\csname LT0\endcsname{\color{black}}%
      \expandafter\def\csname LT1\endcsname{\color{black}}%
      \expandafter\def\csname LT2\endcsname{\color{black}}%
      \expandafter\def\csname LT3\endcsname{\color{black}}%
      \expandafter\def\csname LT4\endcsname{\color{black}}%
      \expandafter\def\csname LT5\endcsname{\color{black}}%
      \expandafter\def\csname LT6\endcsname{\color{black}}%
      \expandafter\def\csname LT7\endcsname{\color{black}}%
      \expandafter\def\csname LT8\endcsname{\color{black}}%
    \fi
  \fi
  \setlength{\unitlength}{0.0500bp}%
  \begin{picture}(7012.00,4506.00)%
    \gplgaddtomacro\gplbacktext{%
      \csname LTb\endcsname%
      \put(1078,704){\makebox(0,0)[r]{\strut{} 0}}%
      \put(1078,1294){\makebox(0,0)[r]{\strut{} 0.05}}%
      \put(1078,1883){\makebox(0,0)[r]{\strut{} 0.1}}%
      \put(1078,2473){\makebox(0,0)[r]{\strut{} 0.15}}%
      \put(1078,3062){\makebox(0,0)[r]{\strut{} 0.2}}%
      \put(1078,3652){\makebox(0,0)[r]{\strut{} 0.25}}%
      \put(1078,4241){\makebox(0,0)[r]{\strut{} 0.3}}%
      \put(1210,484){\makebox(0,0){\strut{}-0.01}}%
      \put(2111,484){\makebox(0,0){\strut{} 0}}%
      \put(3012,484){\makebox(0,0){\strut{} 0.01}}%
      \put(3912,484){\makebox(0,0){\strut{} 0.02}}%
      \put(4813,484){\makebox(0,0){\strut{} 0.03}}%
      \put(5714,484){\makebox(0,0){\strut{} 0.04}}%
      \put(6615,484){\makebox(0,0){\strut{} 0.05}}%
      \put(176,2472){\rotatebox{-270}{\makebox(0,0){\strut{}$\textrm{-II}$}}}%
      \put(3912,154){\makebox(0,0){\strut{}$\textrm{III}$}}%
    }%
    \gplgaddtomacro\gplfronttext{%
      \csname LTb\endcsname%
      \put(2134,4068){\makebox(0,0)[r]{\strut{}$\tau=-1.0$}}%
      \csname LTb\endcsname%
      \put(2134,3848){\makebox(0,0)[r]{\strut{}$\tau=0.0$}}%
      \csname LTb\endcsname%
      \put(2134,3628){\makebox(0,0)[r]{\strut{}$\tau=0.5$}}%
      \csname LTb\endcsname%
      \put(2134,3408){\makebox(0,0)[r]{\strut{}$\tau=1.0$}}%
    }%
    \gplbacktext
    \put(0,0){\includegraphics{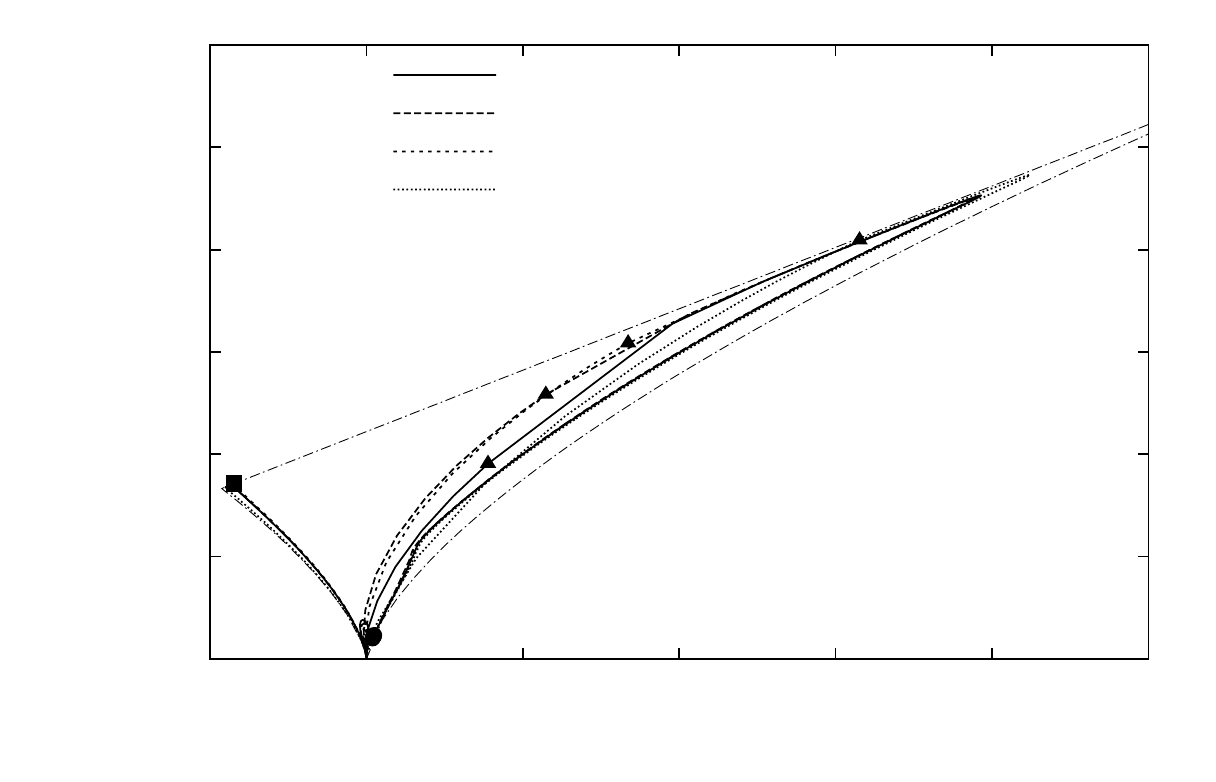}}%
    \gplfronttext
  \end{picture}%
\endgroup

%% file: loglaw_hp.tex
\begingroup
  \makeatletter
  \providecommand\color[2][]{%
    \GenericError{(gnuplot) \space\space\space\@spaces}{%
      Package color not loaded in conjunction with
      terminal option `colourtext'%
    }{See the gnuplot documentation for explanation.%
    }{Either use 'blacktext' in gnuplot or load the package
      color.sty in LaTeX.}%
    \renewcommand\color[2][]{}%
  }%
  \providecommand\includegraphics[2][]{%
    \GenericError{(gnuplot) \space\space\space\@spaces}{%
      Package graphicx or graphics not loaded%
    }{See the gnuplot documentation for explanation.%
    }{The gnuplot epslatex terminal needs graphicx.sty or graphics.sty.}%
    \renewcommand\includegraphics[2][]{}%
  }%
  \providecommand\rotatebox[2]{#2}%
  \@ifundefined{ifGPcolor}{%
    \newif\ifGPcolor
    \GPcolorfalse
  }{}%
  \@ifundefined{ifGPblacktext}{%
    \newif\ifGPblacktext
    \GPblacktexttrue
  }{}%
  \let\gplgaddtomacro\g@addto@macro
  \gdef\gplbacktext{}%
  \gdef\gplfronttext{}%
  \makeatother
  \ifGPblacktext
    \def\colorrgb#1{}%
    \def\colorgray#1{}%
  \else
    \ifGPcolor
      \def\colorrgb#1{\color[rgb]{#1}}%
      \def\colorgray#1{\color[gray]{#1}}%
      \expandafter\def\csname LTw\endcsname{\color{white}}%
      \expandafter\def\csname LTb\endcsname{\color{black}}%
      \expandafter\def\csname LTa\endcsname{\color{black}}%
      \expandafter\def\csname LT0\endcsname{\color[rgb]{1,0,0}}%
      \expandafter\def\csname LT1\endcsname{\color[rgb]{0,1,0}}%
      \expandafter\def\csname LT2\endcsname{\color[rgb]{0,0,1}}%
      \expandafter\def\csname LT3\endcsname{\color[rgb]{1,0,1}}%
      \expandafter\def\csname LT4\endcsname{\color[rgb]{0,1,1}}%
      \expandafter\def\csname LT5\endcsname{\color[rgb]{1,1,0}}%
      \expandafter\def\csname LT6\endcsname{\color[rgb]{0,0,0}}%
      \expandafter\def\csname LT7\endcsname{\color[rgb]{1,0.3,0}}%
      \expandafter\def\csname LT8\endcsname{\color[rgb]{0.5,0.5,0.5}}%
    \else
      \def\colorrgb#1{\color{black}}%
      \def\colorgray#1{\color[gray]{#1}}%
      \expandafter\def\csname LTw\endcsname{\color{white}}%
      \expandafter\def\csname LTb\endcsname{\color{black}}%
      \expandafter\def\csname LTa\endcsname{\color{black}}%
      \expandafter\def\csname LT0\endcsname{\color{black}}%
      \expandafter\def\csname LT1\endcsname{\color{black}}%
      \expandafter\def\csname LT2\endcsname{\color{black}}%
      \expandafter\def\csname LT3\endcsname{\color{black}}%
      \expandafter\def\csname LT4\endcsname{\color{black}}%
      \expandafter\def\csname LT5\endcsname{\color{black}}%
      \expandafter\def\csname LT6\endcsname{\color{black}}%
      \expandafter\def\csname LT7\endcsname{\color{black}}%
      \expandafter\def\csname LT8\endcsname{\color{black}}%
    \fi
  \fi
  \setlength{\unitlength}{0.0500bp}%
  \begin{picture}(7012.00,4506.00)%
    \gplgaddtomacro\gplbacktext{%
      \csname LTb\endcsname%
      \put(594,704){\makebox(0,0)[r]{\strut{} 0}}%
      \put(594,1588){\makebox(0,0)[r]{\strut{} 5}}%
      \put(594,2473){\makebox(0,0)[r]{\strut{} 10}}%
      \put(594,3357){\makebox(0,0)[r]{\strut{} 15}}%
      \put(594,4241){\makebox(0,0)[r]{\strut{} 20}}%
      \put(726,484){\makebox(0,0){\strut{} 1}}%
      \put(3103,484){\makebox(0,0){\strut{} 10}}%
      \put(5481,484){\makebox(0,0){\strut{} 100}}%
      \put(220,2472){\rotatebox{-270}{\makebox(0,0){\strut{}$(\overline{u}-U_i)^+$}}}%
      \put(3670,154){\makebox(0,0){\strut{}$y^+$}}%
    }%
    \gplgaddtomacro\gplfronttext{%
      \csname LTb\endcsname%
      \put(1650,4068){\makebox(0,0)[r]{\strut{}Imp}}%
      \csname LTb\endcsname%
      \put(1650,3848){\makebox(0,0)[r]{\strut{}$h_p=2.0$}}%
      \csname LTb\endcsname%
      \put(1650,3628){\makebox(0,0)[r]{\strut{}$h_p=0.2$}}%
    }%
    \gplbacktext
    \put(0,0){\includegraphics{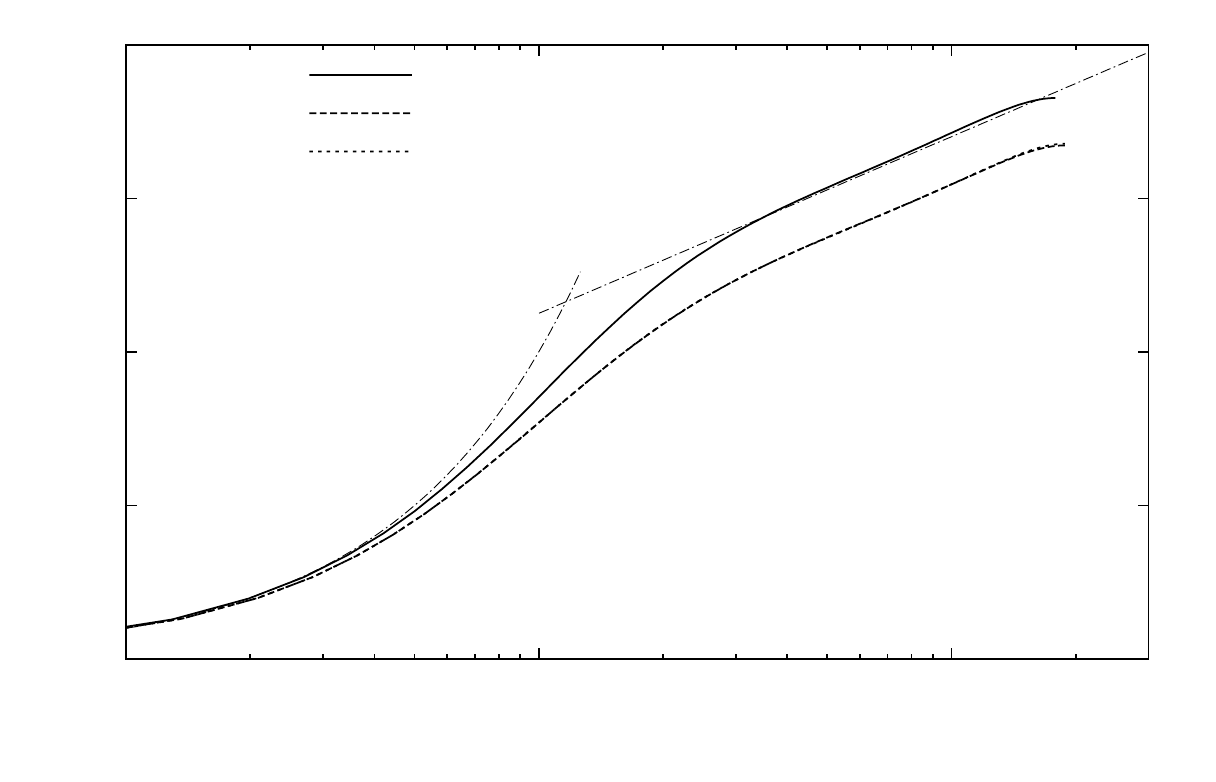}}%
    \gplfronttext
  \end{picture}%
\endgroup

%% file: loglaw_re.tex
\begingroup
  \makeatletter
  \providecommand\color[2][]{%
    \GenericError{(gnuplot) \space\space\space\@spaces}{%
      Package color not loaded in conjunction with
      terminal option `colourtext'%
    }{See the gnuplot documentation for explanation.%
    }{Either use 'blacktext' in gnuplot or load the package
      color.sty in LaTeX.}%
    \renewcommand\color[2][]{}%
  }%
  \providecommand\includegraphics[2][]{%
    \GenericError{(gnuplot) \space\space\space\@spaces}{%
      Package graphicx or graphics not loaded%
    }{See the gnuplot documentation for explanation.%
    }{The gnuplot epslatex terminal needs graphicx.sty or graphics.sty.}%
    \renewcommand\includegraphics[2][]{}%
  }%
  \providecommand\rotatebox[2]{#2}%
  \@ifundefined{ifGPcolor}{%
    \newif\ifGPcolor
    \GPcolorfalse
  }{}%
  \@ifundefined{ifGPblacktext}{%
    \newif\ifGPblacktext
    \GPblacktexttrue
  }{}%
  \let\gplgaddtomacro\g@addto@macro
  \gdef\gplbacktext{}%
  \gdef\gplfronttext{}%
  \makeatother
  \ifGPblacktext
    \def\colorrgb#1{}%
    \def\colorgray#1{}%
  \else
    \ifGPcolor
      \def\colorrgb#1{\color[rgb]{#1}}%
      \def\colorgray#1{\color[gray]{#1}}%
      \expandafter\def\csname LTw\endcsname{\color{white}}%
      \expandafter\def\csname LTb\endcsname{\color{black}}%
      \expandafter\def\csname LTa\endcsname{\color{black}}%
      \expandafter\def\csname LT0\endcsname{\color[rgb]{1,0,0}}%
      \expandafter\def\csname LT1\endcsname{\color[rgb]{0,1,0}}%
      \expandafter\def\csname LT2\endcsname{\color[rgb]{0,0,1}}%
      \expandafter\def\csname LT3\endcsname{\color[rgb]{1,0,1}}%
      \expandafter\def\csname LT4\endcsname{\color[rgb]{0,1,1}}%
      \expandafter\def\csname LT5\endcsname{\color[rgb]{1,1,0}}%
      \expandafter\def\csname LT6\endcsname{\color[rgb]{0,0,0}}%
      \expandafter\def\csname LT7\endcsname{\color[rgb]{1,0.3,0}}%
      \expandafter\def\csname LT8\endcsname{\color[rgb]{0.5,0.5,0.5}}%
    \else
      \def\colorrgb#1{\color{black}}%
      \def\colorgray#1{\color[gray]{#1}}%
      \expandafter\def\csname LTw\endcsname{\color{white}}%
      \expandafter\def\csname LTb\endcsname{\color{black}}%
      \expandafter\def\csname LTa\endcsname{\color{black}}%
      \expandafter\def\csname LT0\endcsname{\color{black}}%
      \expandafter\def\csname LT1\endcsname{\color{black}}%
      \expandafter\def\csname LT2\endcsname{\color{black}}%
      \expandafter\def\csname LT3\endcsname{\color{black}}%
      \expandafter\def\csname LT4\endcsname{\color{black}}%
      \expandafter\def\csname LT5\endcsname{\color{black}}%
      \expandafter\def\csname LT6\endcsname{\color{black}}%
      \expandafter\def\csname LT7\endcsname{\color{black}}%
      \expandafter\def\csname LT8\endcsname{\color{black}}%
    \fi
  \fi
  \setlength{\unitlength}{0.0500bp}%
  \begin{picture}(7012.00,4506.00)%
    \gplgaddtomacro\gplbacktext{%
      \csname LTb\endcsname%
      \put(594,704){\makebox(0,0)[r]{\strut{} 0}}%
      \put(594,1588){\makebox(0,0)[r]{\strut{} 5}}%
      \put(594,2473){\makebox(0,0)[r]{\strut{} 10}}%
      \put(594,3357){\makebox(0,0)[r]{\strut{} 15}}%
      \put(594,4241){\makebox(0,0)[r]{\strut{} 20}}%
      \put(726,484){\makebox(0,0){\strut{} 1}}%
      \put(2989,484){\makebox(0,0){\strut{} 10}}%
      \put(5252,484){\makebox(0,0){\strut{} 100}}%
      \put(220,2472){\rotatebox{-270}{\makebox(0,0){\strut{}$(\overline{u}-U_i)^+$}}}%
      \put(3670,154){\makebox(0,0){\strut{}$y^+$}}%
    }%
    \gplgaddtomacro\gplfronttext{%
      \csname LTb\endcsname%
      \put(3366,4068){\makebox(0,0)[r]{\strut{}$Re=2800$, $\sigma=0.004$ SD}}%
      \csname LTb\endcsname%
      \put(3366,3848){\makebox(0,0)[r]{\strut{}$Re=2800$, $\sigma=0.002$ SD}}%
      \csname LTb\endcsname%
      \put(3366,3628){\makebox(0,0)[r]{\strut{}$Re=6250$, $\sigma=0.002$ SD}}%
    }%
    \gplbacktext
    \put(0,0){\includegraphics{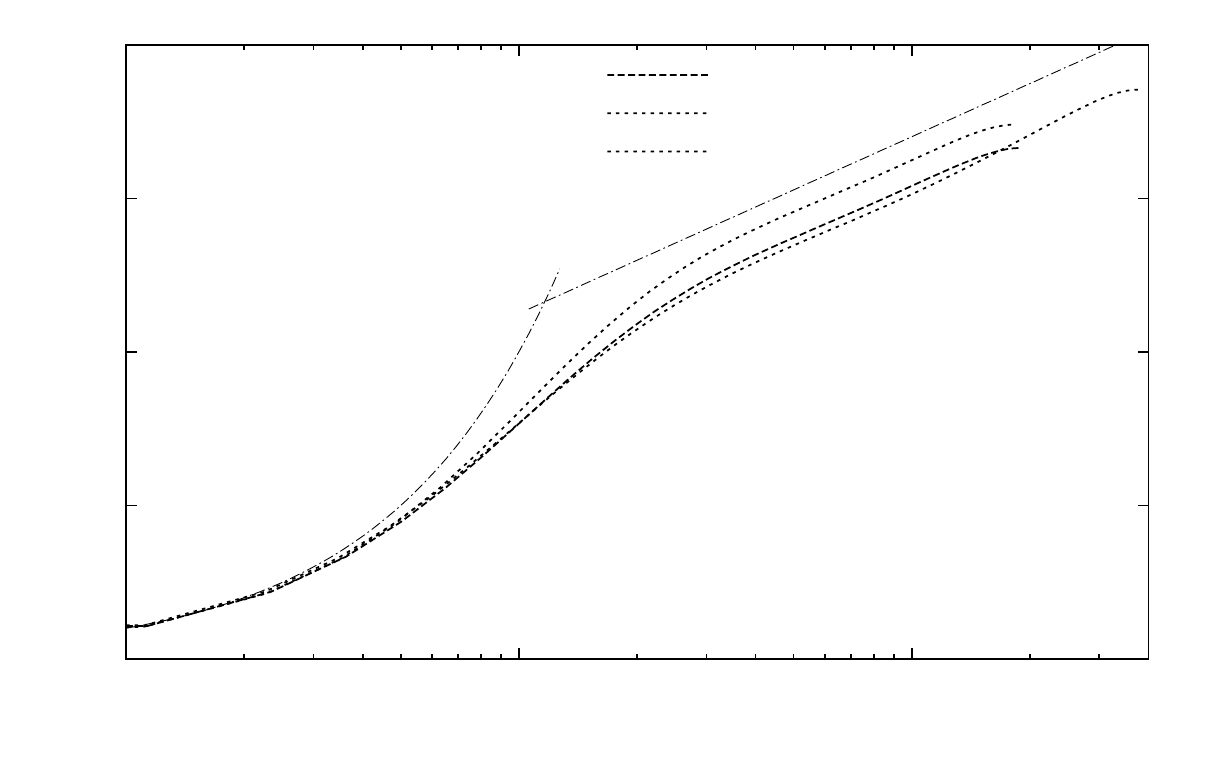}}%
    \gplfronttext
  \end{picture}%
\endgroup